\documentclass[aps,prd,groupedaddress,preprintnumbers,%
              showpacs,showkeys,floatfix]{revtex4-1}       
\usepackage[latin1]{inputenc}                    
\usepackage{graphicx}                            
\usepackage{epsfig}
\usepackage{epsf}
\usepackage{latexsym}                            
\usepackage{amsfonts}                            
\usepackage{amssymb}                             
\usepackage{amsmath}                             
\usepackage[mathscr]{eucal}                      
\usepackage{dcolumn}                             
\usepackage{multirow}
\usepackage{bm}                                  
\usepackage{hyperref}                            
\usepackage[usenames]{color}
\usepackage{array}
\usepackage{appendix}
\usepackage{bbold}

\graphicspath{{.}{Graphics/}}                    

\newcommand{\be}{\begin{equation}}
\newcommand{\ee}{\end{equation}}
\newcommand{\bea}{\begin{eqnarray}}
\newcommand{\eea}{\end{eqnarray}}

\def\eq#1{Eq.~(\ref{#1})}
\def\fig#1{Fig. \ref{#1}}
\def\tbl#1{Table \ref{#1}}
\def \3{\ss }

\newcommand{\tr}{\operatorname{Tr}}
\newcommand{\re}{\operatorname{Re}}

\def\slash#1{\mbox{$\not\!\! #1$}}

\newcommand{\beqn}{\begin{eqnarray}}
\newcommand{\eeqn}{\end{eqnarray}}

\newcommand{\idnty}{\hbox{1$\!\!$1}}

\newcommand{\reci}[1]{\frac{1}{#1}}
\newcommand{\eps}{\epsilon_{abc}}
\newcommand{\con}[3]   {\left( {#1}_a^T C\gamma_5     {#2}_b \right) {#3}_c}
\newcommand{\conm}[3]{\left( {#1}_a^T C\gamma_\mu {#2}_b \right) {#3}_c}
\newcommand{\cone}[3]{\eps \left( {#1}_a^T C\gamma_5 {#2}_b \right) {#3}_c}
\newcommand{\conme}[3]{\eps \left({#1}_a^T \gamma_\mu {#2}_b \right) {#3}_c}

\def\cyp{a}
\def\cyi{b}

\begin{document}

\begin{titlepage}
  {\vspace{-0.5cm} \normalsize
  \hfill \parbox{60mm}{
}}\\[10mm]
  \begin{center}
    \begin{LARGE}
      \textbf{Axial charges of hyperons and charmed baryons using $N_f=2+1+1$ twisted
       mass fermions} \\ [1ex]{\includegraphics[width=0.15\linewidth]{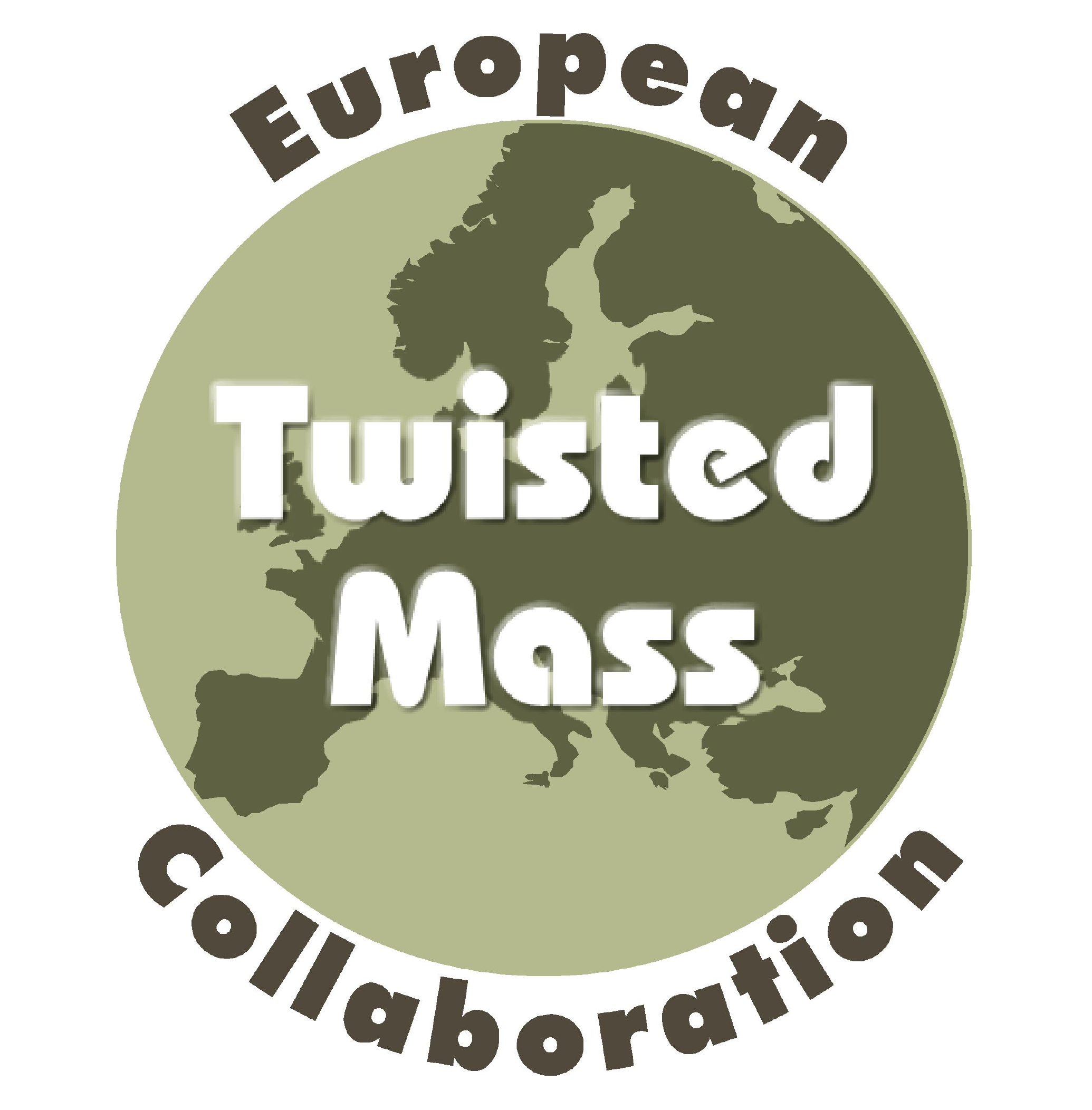}\\[1ex]ETM Collaboration}
    \end{LARGE}
  \end{center}

 \vspace{.5cm}

 \vspace{-0.8cm}
  \baselineskip 20pt plus 2pt minus 2pt
  \begin{center}
    \textbf{
      C.~Alexandrou$^{(\cyp, \cyi)}$,
      K.~Hadjiyiannakou$^{(\cyp, \cyi)}$,
      C. Kallidonis$^{(\cyi)}$, 
}
  \end{center}

  \begin{center}
    \begin{footnotesize}
      \noindent 	
 	$^{(\cyp)}$ Department of Physics, University of Cyprus, P.O. Box 20537,
 	1678 Nicosia, Cyprus\\	
 	$^{(\cyi)}$ Computation-based Science and Technology Research Center, The Cyprus Institute, 20 Kavafi Str., Nicosia 2121, Cyprus \\
     \vspace{0.2cm}
    \end{footnotesize}
  \end{center}
  
  \begin{abstract}

The axial couplings of the low lying baryons are evaluated 
using a total of five ensembles of dynamical twisted mass fermion gauge configurations. The simulations are performed  using  the Iwasaki gauge action and   two degenerate flavors of light quarks,  and a strange and a charm quark fixed to approximately their physical values at two values of the coupling constant. The   lattice spacings, determined using the nucleon mass,  are $a=0.082$~fm and $a=0.065$~fm and the simulations cover a pion mass in the range of about 210~MeV to 430~MeV.   
We study the dependence of the axial couplings on the pion mass in the range of about 210~MeV to 430~MeV as well as the $SU(3)$ breaking effects as we decrease the light quark mass towards its physical value.  
  
\begin{center}
\today
\end{center}
 \end{abstract}
\maketitle 
\end{titlepage}


\section{Introduction}

The axial charges of hyperons are important parameters of low energy effective field theories.
The  nucleon axial charge, the value of which is well known experimentally, is a crucial parameter
entering  in the description of many observables computed within chiral effective theories. It describes  neutron $\beta$-decay and sheds light on  spontaneous chiral symmetry breaking. As a well-measured quantity, it has been traditionally used as a benchmark quantity for lattice QCD computations and it has been extensively studied by  many lattice QCD  collaborations,  including using simulations with a physical value of the  pion mass~\cite{Abdel-Rehim:2015owa,Bali:2014nma}. For recent reviews see Refs.~\citep{Alexandrou:2014xca,Alexandrou:2015yqa,Alexandrou:2015xts,Constantinou:2015agp}. In addition, the quark axial charge $g^q_A$ probes the intrinsic quark spin contribution
to the total spin of a quark in the nucleon, and has been studied both theoretically and experimentally  for a number of years. 

While there has been an extensive work for the nucleon axial charge, the axial charges of hyperons or charmed baryons are less well studied. The knowledge of these axial charges is very important  allowing us to examine the validity of SU(3) relations among them as a function of the pion mass. They are also important parameters for chiral expansions of  baryonic quantities.  Their experimental determination is difficult because most baryons are very short-lived as for instance the $\Delta$, which decays in $~10^{-23} s$. Therefore,  lattice QCD can provide valuable information on these quantities and in general into the structure of these baryons.

In this work, we study the axial charges of hyperons and charmed baryons using  twisted mass fermions with two light quark doublets as well as a strange and a charm  quark with mass fixed to their physical values, denoted as $N_f=2+1+1$ ensembles.  
Results are obtained for the axial charges  of the two 20-plets of spin-1/2 and spin-3/2  baryons that arise when considering the two light, the strange and charm quarks. Five ensembles of twisted mass fermions  are analyzed spanning a pion
mass range between 210~MeV  and 450~MeV allowing us to examine the dependence of the axial charges on the pion mass, which is found similar to the one observed for the nucleon axial charge within this pion mass range.

\section{Lattice setup and simulation details}

In this work, we analyze five ensembles of gauge configurations produced by the European Twisted Mass  Collaboration (ETMC)~\cite{Baron:2008xa,Baron:2009zq}, with $N_f=2+1+1$ maximally twisted quark flavours.
In summary,  these gauge fields are produced using as a gauge action  the Iwasaki improved
gauge action~\cite{Weisz:1982zw,Iwasaki:1985we,Iwasaki:1996sn}, which includes besides the
plaquette term $U^{1\times1}_{x,\mu,\nu}$ also rectangular $(1\times2)$ Wilson 
loops $U^{1\times2}_{x,\mu,\nu}$ given by
\begin{equation}
  \label{eq:Sg}
    S_g =  \frac{\beta}{3}\sum_x\Biggl(  b_0\sum_{\substack{
      \mu,\nu=1\\1\leq\mu<\nu}}^4\left \{1-\re\tr(U^{1\times1}_{x,\mu,\nu})\right \}\Bigr. 
     \Bigl.+
    b_1\sum_{\substack{\mu,\nu=1\\\mu\neq\nu}}^4\left \{1
    -\re\tr(U^{1\times2}_{x,\mu,\nu})\right \}\Biggr)\,.  
\end{equation}
 $\beta=6/g_0^2$ is the bare inverse coupling,  $b_1=-1/12$ and the
(proper) normalization condition $b_0=1-8b_1$. We note that for $b_1=0$ this
action becomes the usual Wilson plaquette gauge action.

The twisted mass Wilson action used for the light degenerate doublet of quarks ($u$,$d$) is given by~\cite{Frezzotti:2003ni,Frezzotti:2000nk}
\be
S_F^{(l)}\left[\chi^{(l)},\overline{\chi}^{(l)},U \right]= a^4\sum_x  \overline{\chi}^{(l)}(x)\bigl(D_W[U] + m_{0,l} + i \mu_l \gamma_5\tau^3  \bigr ) \chi^{(l)}(x)
\label{S_tml}
\ee
with $\tau^3$ the third Pauli matrix acting in  flavour space, $m_{0,l}$ the bare untwisted light quark mass, $\mu_l$ the bare twisted light mass. The massless Wilson-Dirac operator is given by 
\be \label{wilson_term}
D_W[U] = \frac{1}{2} \gamma_{\mu}(\nabla_{\mu} + \nabla_{\mu}^{*})
-\frac{ar}{2} \nabla_{\mu}
\nabla^*_{\mu} 
\ee
where
\be
\nabla_\mu \psi(x)= \frac{1}{a}\biggl[U^\dagger_\mu(x)\psi(x+a\hat{\mu})-\psi(x)\biggr]
\hspace*{0.5cm} {\rm and}\hspace*{0.5cm} 
\nabla^*_{\mu}\psi(x)=-\frac{1}{a}\biggl[U_{\mu}(x-a\hat{\mu})\psi(x-a\hat{\mu})-\psi(x)\biggr]
\quad .
\ee
The quark fields denoted by $\chi^{(l)}$ in \eq{S_tml} are in the so-called ``twisted basis". The fields in the ``physical basis", denoted by $\psi^{(l)}$, are obtained at maximal twist by the transformation

\be
\psi^{(l)}(x)=\reci{\sqrt{2}}\left(\idnty+ i \tau^3\gamma_5\right) \chi^{(l)}(x),\qquad
\overline{\psi}^{(l)}(x)=\overline{\chi}^{(l)}(x) \reci{\sqrt{2}}\left(\idnty + i \tau^3\gamma_5\right)
\quad.
\ee

In addition to the light sector, a twisted heavy mass-split doublet $\chi^{(h)} = \left(\chi_c,\chi_s \right)$ for the strange and charm quarks is introduced, described by the action~\cite{Frezzotti:2004wz,Frezzotti:2003xj}
\be
S_F^{(h)}\left[\chi^{(h)},\overline{\chi}^{(h)},U \right]= a^4\sum_x  \overline{\chi}^{(h)}(x)\bigl(D_W[U] + m_{0,h} + i\mu_\sigma \gamma_5\tau^1 + \tau^3\mu_\delta  \bigr ) \chi^{(h)}(x)
\label{S_tmh}
\ee
where $m_{0,h}$ is the bare untwisted quark mass for the heavy doublet, $\mu_\sigma$ is the bare twisted mass along the $\tau^1$ direction and $\mu_\delta$ is the mass splitting in the $\tau^3$ direction. The quark fields for the heavy quarks in the physical basis are obtained from the twisted basis through the transformation

\be
\psi^{(h)}(x)=\reci{\sqrt{2}}\left(\idnty+ i \tau^1\gamma_5\right) \chi^{(h)}(x),\qquad
\overline{\psi}^{(h)}(x)=\overline{\chi}^{(h)}(x) \reci{\sqrt{2}}\left(\idnty + i \tau^1\gamma_5\right)
\quad.
\ee 

In this paper, unless otherwise stated, the quark fields will be understood as ``physical fields", $\psi$, in particular when we define the interpolating fields of the baryons.

The form of the fermion action in \eq{S_tml} breaks parity and isospin at non-vanishing lattice spacing, as it is also apparent from the form of the Wilson term in \eq{wilson_term}. In particular, the isospin breaking in physical observables is a cut-off effect of ${\cal O}(a^2)$~\cite{Frezzotti:2003ni}. For the masses of baryon isospin multiplets such isospin breaking effects have been found to be small for the ensembles considered in this work~\cite{Alexandrou:2014sha}.

Maximally twisted Wilson quarks are obtained by setting the untwisted quark mass $m_0$ to its critical value $m_{\rm cr}$, while the twisted quark mass parameter $\mu$ is kept non-vanishing to give a mass to the pions. A crucial advantage of the twisted mass formulation is
the fact that, by tuning the bare untwisted quark mass $m_0$ to its critical value
 $m_{\rm cr}$, all physical observables are automatically 
${\cal O}(a)$ improved~\cite{Frezzotti:2003ni,Frezzotti:2003xj}. 
In practice, we implement
maximal twist of Wilson quarks by tuning to zero the bare untwisted 
quark mass, commonly called PCAC mass, $m_{\rm PCAC}$ \cite{Boucaud:2008xu,Frezzotti:2005gi}, which is proportional to
$m_0 - m_{\rm cr}$ up to ${\cal O}(a)$ corrections. 

The gauge configurations analyzed in this work correspond to two lattice volumes and four values of the pion mass for $\beta=1.95$ and one volume and one pion mass  for $\beta=2.10$. The corresponding lattice spacings are  respectively   $a_{\beta=1.95}=0.0820(10)$~fm and $a_{\beta=2.10}=0.0644(7)$ determined from the nucleon mass~\cite{Alexandrou:2014sha}.

For the  heavy quark sector we  use Osterwalder-Seiler
valence strange and charm quarks.  Osterwalder-Seiler fermions are doublets 
like the the u- and d- doublet, i.e.$\chi^s = (s^+, s^-)$ and $\chi^c = (c^+, c^-)$, having an action that is the same as for the
doublet of light quarks, but with $\mu_l$ in Eq.~(\ref{S_tml}) replaced with the tuned value of the bare twisted
mass of the strange or charm valence quark. Taking $m_0$ to be equal to the critical mass determined in the light sector,
the $O(a)$ improvement in any observable still applies. One can equally work with $s^+$ ($c^+$) or $s^-$ ($c^-$)
of the strange (charm) doublets. In the continuum limit both choices are equivalent and in this work we opt for $s^+$ and $c^+$. Since our interest in this work is the baryon spectrum we
choose to tune
the strange and charm quark masses to reproduce the physical masses of the $\Omega^-$ and $\Lambda_c^+$ baryons, respectively. More details on the tuning procedure can be found in Ref.~\cite{Alexandrou:2014sha}. In~\tbl{Table:params} we summarize the parameters of the simulations used in this work, including the $\beta$ value, the spatial lattice extent in lattice units $L/a$, the value of  the bare twisted light quark mass as well as the pion masses.

\begin{table}[h]
\begin{center}
\renewcommand{\arraystretch}{1.2}
\renewcommand{\tabcolsep}{5.5pt}
\begin{tabular}{c|lcccc}
\hline\hline
\multicolumn{5}{c}{ $\beta=1.95$,  $a_{\beta=1.95}=0.0823(10)$~fm, ${r_0/a}=5.710(41)$ }\\
\hline
\multirow{4}{*}{$32^3\times 64$, $L=2.6$~fm} &    $a\mu_l   $  & 0.0025 & 0.0035 & 0.0055   & 0.0075  \\
 									         & No. of Confs  &  199   &  200   &  200     &  200    \\
                                             & $m_\pi$~(GeV) & 0.256  & 0.302  &  0.373   &  0.432 	\\
				                             &   $m_\pi L$   &  3.42  &  4.03  &  4.97    &  5.77    \\
\hline\hline
\multicolumn{5}{c}{ $\beta=2.10$, $a_{\beta=2.10}=0.0646(7)$~fm  ${r_0/a}=7.538(58)$}\\
\hline
\multirow{4}{*}{$48^3\times 96$, $L=3.1$~fm}  &    $a\mu_l$    & 0.0015   \\
                                              & No. of Confs &  200     \\
                              			   	  &$m_\pi$~(GeV) & 0.213    \\
                        			          &   $m_\pi L$  & 3.35     \\ 
\hline \hline
\end{tabular}
\caption{Input parameters ($\beta,L,\mu_l$) of our lattice simulations and corresponding lattice spacing ($a$) and pion mass ($m_{\pi}$). The lattice spacings are determined using the nucleon mass.}
\label{Table:params}
\end{center}
\end{table} 

In the following we will refer to the ensembles  with $\beta=1.95$ as the B-ensembles, and to the ensemble with $\beta=2.10$ as the D-ensemble. We also use the notation B$xx.yy$ or D$xx.yy$ where $xx$ denotes the $a\mu$ value and $yy$ denotes the spatial extent of the lattice, $L/a$, e.g. B25.32 refers to our ensemble with $\beta=1.95$, $a\mu=0.0025$ and $L/a=32$.


\section{Lattice evaluation}

\subsection{Matrix element Decomposition}
We consider the 40 diagonal baryon matrix elements of the axial vector operator $A^\mu(x)=\bar q(x) \gamma^\mu\gamma^5 q(x)$, where $q(x)$ denotes a quark field of a given flavor. For  baryons containing up and down quarks we consider the isovector combination where disconnected contributions vanish in the continuum limit, namely $A^\mu(x)=\bar u(x) \gamma^\mu\gamma^5 u(x)-\bar d(x) \gamma^\mu\gamma^5 d(x)$. The  isoscalar matrix elements of these baryons receive disconnected contributions.  While there has been a big progress in developing  techniques  to  compute them~\cite{Alexandrou:2012py,Alexandrou:2012zz}, the computational resources required are typically two orders of magnitude larger than those required for the connected. The disconnected contribution to the isoscalar axial charge of the nucleon has been computed for the B55.32 ensemble that corresponds to a pion mass $m_\pi=373$~MeV~\cite{Abdel-Rehim:2013wlz,Alexandrou:2013wca}. It has also been computed  for an ensemble of $N_f=2$ clover fermions with pion mass $285$~MeV~\cite{QCDSF:2011aa}. In both calculations they were found to
 be about 10\% of the connected  isoscalar axial charge. Preliminary results at the physical value of the pion mass increase the  value of the disconnected contribution to $g_A^{u+d}$to about 20\% the value of the connected  $g_A^{u+d}$. For the same ensemble the strange axial charge is found to be $g^s_A\sim -0.04(1)$ while the charm axial charge is consistent with zero~\cite{Abdel-Rehim:2015lha}. Given the large computational effort needed to obtain a reliable signal, the computation of disconnected contributions to the matrix elements of the isoscalar current $\bar u(x) \gamma^\mu\gamma^5 u(x)+\bar d(x) \gamma^\mu\gamma^5 d(x)$ and to the strange $\bar s(x) \gamma^\mu\gamma^5 s(x)$ and charm axial $\bar c(x) \gamma^\mu\gamma^5 c(x)$ currents  are neglected in the current work. Instead in 
this first study of the hyperon and charmed baryon  axial charges, we compute
the dominant  connected contributions as well as combinations where the disconnected
contributions cancel in the flavor symmetric limit. Preliminary results on these quantities were
presented in Ref.~\citep{Alexandrou:2014vya}. 

For spin-1/2 baryons the matrix element of the axial-vector current in Euclidean space can be expressed as
\be\label{eq:ac_spin12}
    \langle B(p_f,s_f)|A^\mu|B(p_i,s_i)\rangle =\bar{u}(p_f,s_f) \mathcal{O}^\mu u(p_i,s_i) = \bar{u}_B(p_f,s_f) 
    \left[\gamma^\mu G_A^B(Q^2)-\frac{i Q^\mu}{2m_B} G_p^B(Q^2)\right]\gamma^5 u_B(p_i,s_i)\,,
\ee
where $p_f$, $s_f$ ($p_i,s_i$) are the momentum and spin of the final (initial) spin-1/2 baryonic state ($B$), $q^2=(p_f-p_i)^2=-Q^2$ is the momentum transfer and $u_B$ represents a Dirac (spin-1/2) spinor. For a Dirac spinor we have
\be\label{Eq: spinor_sum_rule_dirac}
\Lambda_{B_{1/2}} = \sum_{s=-1/2}^{1/2} u_B(p,s)\bar{u}_B(p,s) = \frac{-i\slash{p} + M_B}{2 M_B}.
\ee
The corresponding equation in Euclidean space for spin-3/2 baryons reads
\be\label{Eq:axial_spin32}
    \langle B(p_f,s_f)|A^\mu|B(p_i,s_i)\rangle =\bar{v}^{\sigma}_B(p_f,s_f) \mathcal{O}^{\sigma\tau;\mu}v^{\tau}_B(p_i,s_i)\,,    
\ee
where now $v^{\mu}_B$ represents a Rarita-Schwinger spin-3/2 spinor, with
\be\label{Eq:axial_spin32_cont}
    \mathcal{O}^{\sigma\tau;\mu} =  \left[ \delta^{\sigma\tau}\left(g_1^B(Q^2)\gamma^\mu\gamma^5 -i g_3^B(Q^2)\frac{q^\mu}{2M_B}\gamma^5\right) - \frac{q^\sigma q^\tau}{4M_B^2}\left(h_1^B(Q^2)\gamma^\mu\gamma^5 - ih_3^B(Q^2)\frac{q_\mu}{2M_B}\gamma^5\right)\right].  
\ee
The Rarita-Schwinger spinors satisfy the spin sum relation given by \citep{Alexandrou:2013joa}
\be\label{Eq:spinor32}
\Lambda_{B_{3/2}}^{ \sigma\tau} \equiv \sum_{s=-3/2}^{3/2} v^{\sigma}_B(p,s)\bar{v}^{\tau}_B(p,s) = - \frac{-i\slash{p}+M_B}{2M_B} \left(  \delta^{\sigma\tau} - \frac{\gamma^\sigma\gamma^\tau}{3} + \frac{2p^\sigma p^\tau}{3 M_B^2} -i \frac{p^\sigma \gamma^\tau - p^\tau\gamma^\sigma}{3M_B} \right).
\ee
For both spin-1/2 and spin-3/2 baryons the axial charge is obtained from the forward matrix element i.e. setting  $Q^2=0$ in Eqs.~\ref{eq:ac_spin12} and \ref{Eq:axial_spin32}, yielding $G_A^B(0)$ and $g_1^B(0)$.


\subsection{Baryon interpolating fields}

In the lattice formulation hadron states of interest are obtained by acting on the vacuum  with interpolating fields  constructed  to have the quantum numbers of the hadron under study. For low-lying states, we usually consider interpolating fields that reduce to the quark model wave functions in the non-relativistic limit. Baryons made out of three combinations of the  $u$, $d$, $s$ and $c$ quarks belong to SU(4) multiplets, and thus we use SU(3) subgroups of the SU(4) symmetry to identify their interpolating fields. In general, the interpolating fields of baryons can be written as a sum of terms of the form $\epsilon_{abc}\left[(q_1)_a^T\Gamma^A (q_2)_b\right]\Gamma^B (q_3)_c$, apart from normalization constants. The structures $\Gamma^A$ and $\Gamma^B$ are such that they give rise to the quantum numbers of the baryon state in interest. For spin-1/2 baryons, we will use the combination $(\Gamma^A,\Gamma^B)=(C\gamma_5,\mathbb{1})$ and for spin-3/2 baryons we will use $(\Gamma^A,\Gamma^B)=(C\gamma_\mu,\mathbb{1})$, taking spatial  $\mu=1,\ldots,3$ and $C$ is the charge conjugation matrix.

\begin{figure}[!ht]
\begin{minipage}[t]{0.35\linewidth}
\includegraphics[width=\linewidth]{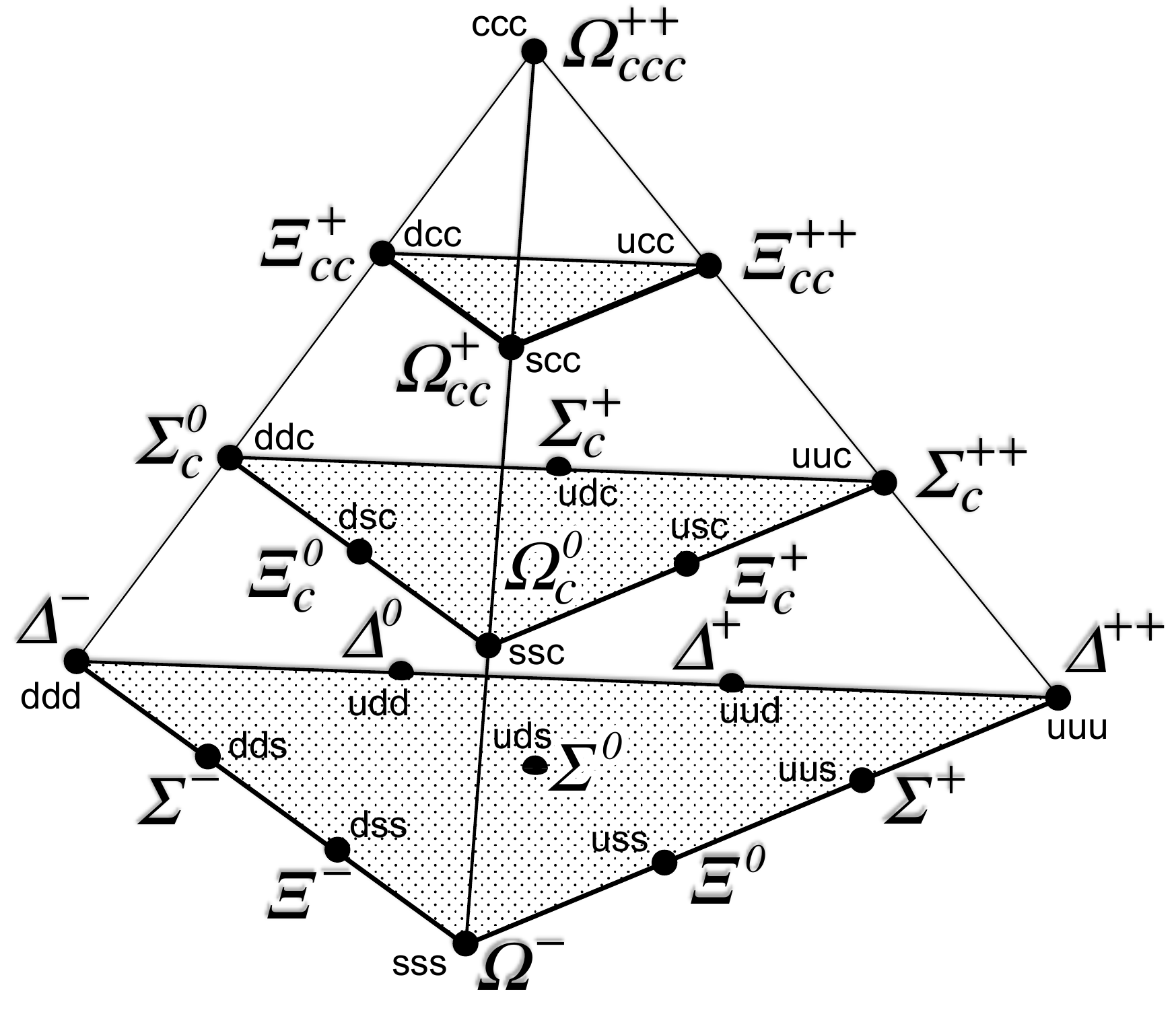}
\end{minipage} \hfill
\begin{minipage}[t]{0.35\linewidth}
\includegraphics[width=\linewidth]{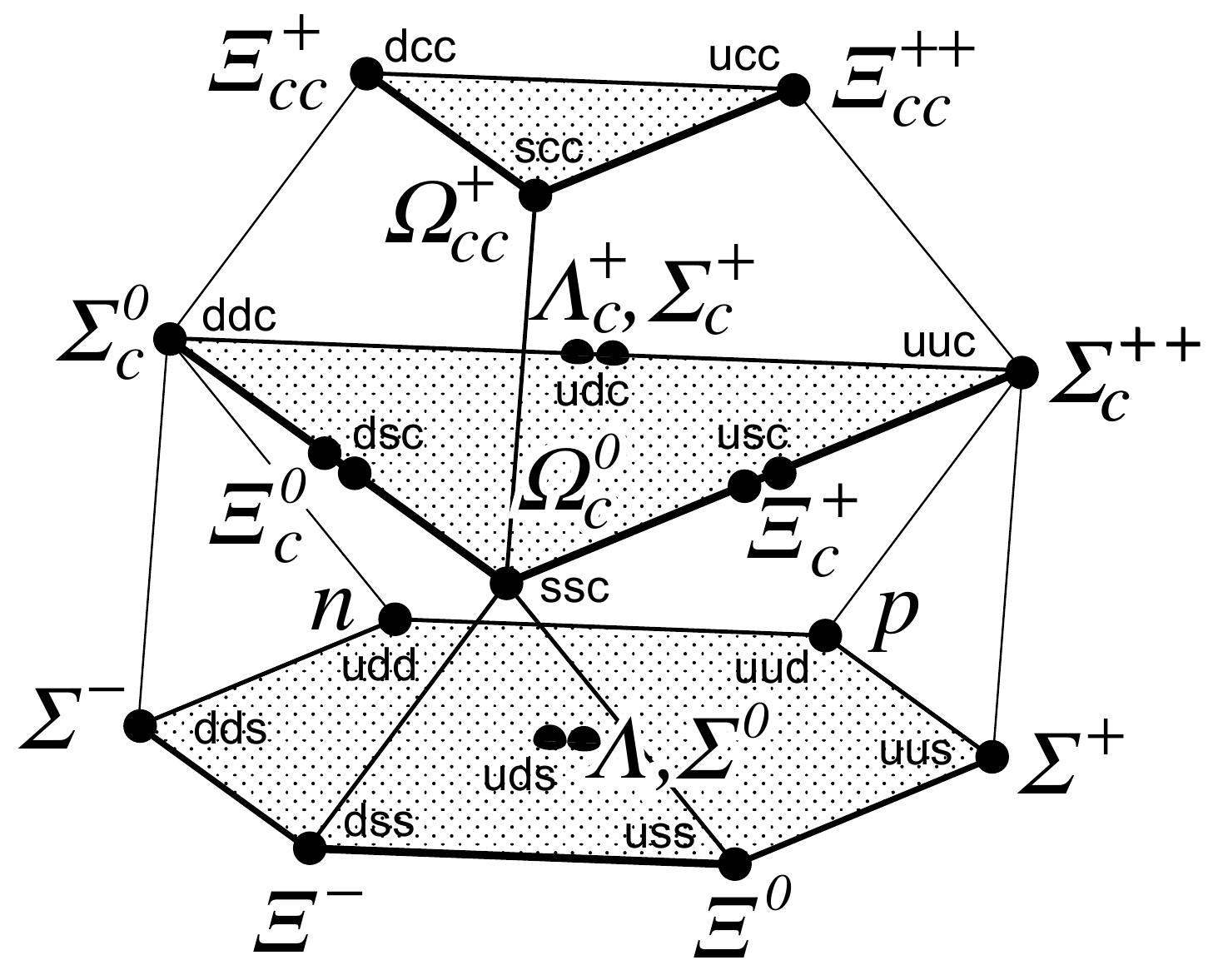}
\end{minipage} \hfill
\begin{minipage}[t]{0.28\linewidth}
\includegraphics[width=\linewidth]{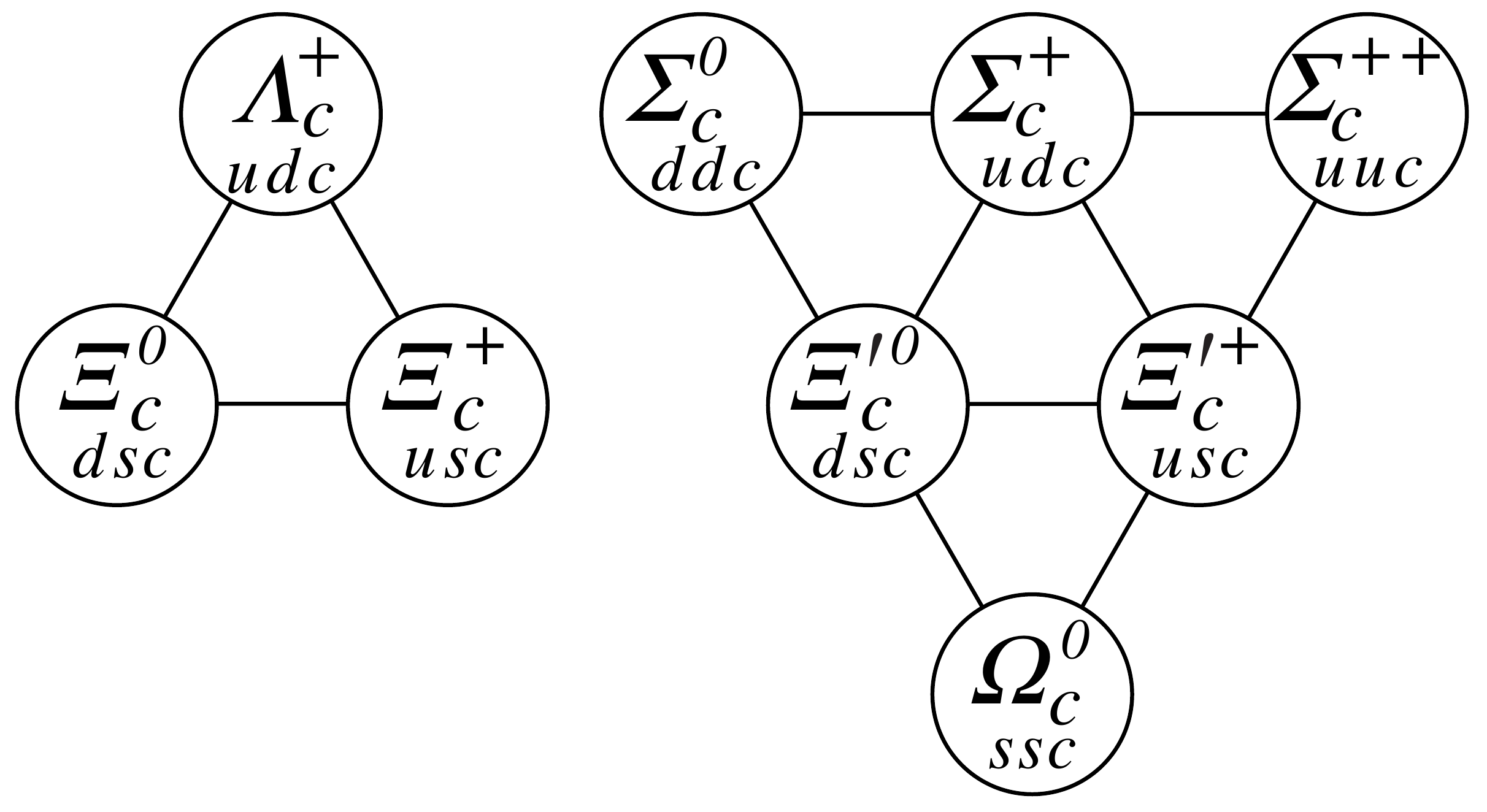}
\end{minipage} \hfill
\caption{The baryon multiplets constructed using the SU(4) group. The left diagram shows the spin-3/2 ${\bf 20}$-plet, the center diagram shows the spin-1/2 ${\bf 20^\prime}$-plet and in the right diagram we show the decomposition of the $c=1$ level of the spin-1/2 ${\bf 20^\prime}$-plet of the center diagram. All diagrams are taken from the PDG~\cite{Agashe:2014kda}.}  
\label{fig:spin12_32}
\end{figure}

The multiplet numerology is $4\otimes 4\otimes 4 = {\bf 20} \oplus {\bf 20_1^\prime} \oplus {\bf 20_2^\prime} \oplus {\bf \bar{4}}$. All the baryons in a given multiplet have the same spin and parity. The ${\bf 20}$-plet consists of the spin-3/2 baryon states and can be further decomposed according to the charm content of the baryons into ${\bf 20} = {\bf 10} \oplus {\bf 6} \oplus {\bf 3} \oplus {\bf 1}$, where the ${\bf 10}$ is the standard $c=0$ decuplet and ${\bf 1}$ is the triply charm $\Omega_{ccc}^{++}$ singlet. The singly charmed baryon states belonging to the ${\bf 6}$ multiplet are symmetric under the interchange of $u$, $d$ and $s$ quarks, following the rule that the diquark $\left[(q_1)_a^T C\gamma_\mu (q_2)_b\right]$ is symmetric under interchanging $q_1\leftrightarrow q_2$. Finally, the doubly charmed ${\bf 3}$-plet consists of the isospin partners $\Xi^*_{cc}$ and the singlet $\Omega_{cc}^{*+}$. The ${\bf 20}$-plet is shown schematically in the left panel of~\fig{fig:spin12_32}. The corresponding interpolating fields of the spin-3/2 baryons are collected in~\tbl{spin32_tab} of Appendix A.

The ${\bf 20^\prime}$-plet consists of the spin-1/2 baryons shown schematically in the center panel of~\fig{fig:spin12_32}. It can be decomposed as ${\bf 20^\prime} = {\bf 8} \oplus {\bf 6} \oplus {\bf \bar{3}} \oplus {\bf 3}$. The ground level $c=0$ comprises the well-known baryon octet, whereas the first level $c=1$ splits  into two SU(3) multiplets, a ${\bf 6}$ and a ${\bf \bar{3}}$. The states of the ${\bf 6}$ are symmetric under interchanging $u$, $d$ and $s$ where the states of the ${\bf \bar{3}}$ are anti-symmetric. We show these states explicitly in the right panel of~\fig{fig:spin12_32}. We note that the diquark $\left[(q_1)_a^T C\gamma_5 (q_2)_b\right]$ appearing the interpolating field of spin-1/2 baryons, is anti-symmetric under interchanging $q_1\leftrightarrow q_2$. The top level consists of the ${\bf 3}$-plet with $c=2$. The interpolating fields of the spin-1/2 baryons are collected in~\tbl{spin12_tab} of Appendix A. The fully antisymmetric ${\bf \bar{4}}$-plet is not considered in this work.

The interpolating fields of spin-3/2 baryons as defined in \tbl{spin32_tab} can have an overlap with spin-1/2 excited states. To remove the unwanted contributions and isolate the desired spin-3/2 ground state we project the spin-3/2 components by acting with the 3/2-projector on the interpolating fields as
\begin{equation}
J_{B_{3/2}}^\mu = P^{\mu\nu}_{3/2} J_B^\nu.
\end{equation}  
For non-zero momentum the projector is given by \citep{Benmerrouche:1989uc}
\begin{equation}\label{eq:proj32}
P^{\mu\nu}_{3/2} = \delta^{\mu\nu} - \frac{1}{3} \gamma^\mu \gamma^\nu - \frac{1}{3p^2} ( \slash{p} \gamma^\mu p^\nu + p^\mu \gamma^\nu \slash{p}).
\end{equation}
The spin-1/2 projector is obtained by $P^{\mu\nu}_{1/2} = \delta^{\mu\nu} - P^{\mu\nu}_{3/2}$ yielding
\begin{equation}\label{eq:proj12}
P^{\mu\nu}_{1/2} =  \frac{1}{3} \gamma^\mu \gamma^\nu + \frac{1}{3p^2} ( \slash{p} \gamma^\mu p^\nu + p^\mu \gamma^\nu \slash{p}).
\end{equation}
 In this work we are interested in correlation functions in the rest frame where $\vec{p}=\vec{0}$ thus the last term of Eqs.~(\ref{eq:proj32}) and~(\ref{eq:proj12}) involving momentum terms  vanishes. The form of the two-point correlation functions when the projectors to spin-1/2 and spin-3/2 are applied to the corresponding interpolating fields is given by
\bea \label{eq:correlators_32_12}
C_{\frac{3}{2}} (t) &=& \frac{1}{3}\tr [C(t)] + \frac{1}{6} \sum_{i\ne j}^3 \gamma_i \gamma_j C_{ij}(t)\;, \nonumber\\
C_{\frac{1}{2}} (t) &=& \frac{1}{3}\tr [C(t)] - \frac{1}{3} \sum_{i\ne j}^3 \gamma_i \gamma_j C_{ij}(t)\;,
\eea
where $\tr[C] = \sum_i C_{ii}$.
For some of the spin-3/2 baryons, the inclusion of the spin-3/2 projector does not have a significant effect in the correlation function, since the spin-1/2 is an excitation with a large energy splitting from the spin-3/2 ground state. This is the case, for instance, for the $\Delta$. However, for other baryons, such as the $\Xi^*$s, the projector is required to isolate the ground state. Thus, in order to ensure that we measure the desired spin-3/2 ground state, we always apply the spin-3/2 projector to the interpolating fields of \tbl{spin32_tab}. The reader interested in more details on the effects of these projectors on the baryon masses is referred to Ref.~\cite{Alexandrou:2014sha}.


\subsection{Correlation functions}

\begin{figure}[h!]
  \begin{minipage}[t]{0.4\linewidth}
    \includegraphics[width=\linewidth]{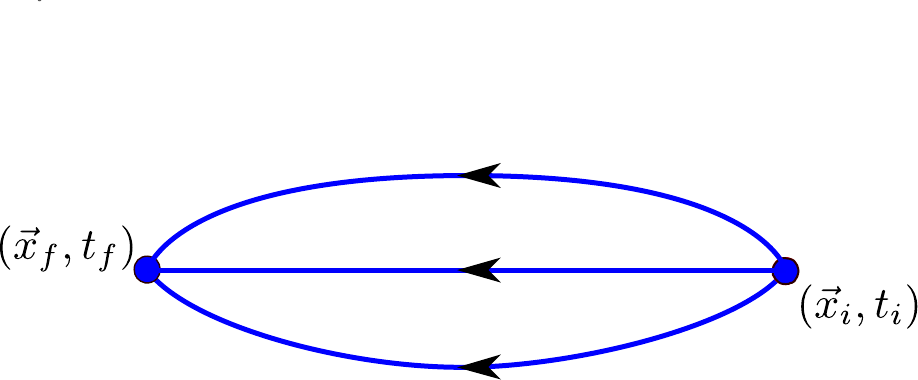}
  \end{minipage} \hfill
  \begin{minipage}[t]{0.4\linewidth}
    \includegraphics[width=\linewidth]{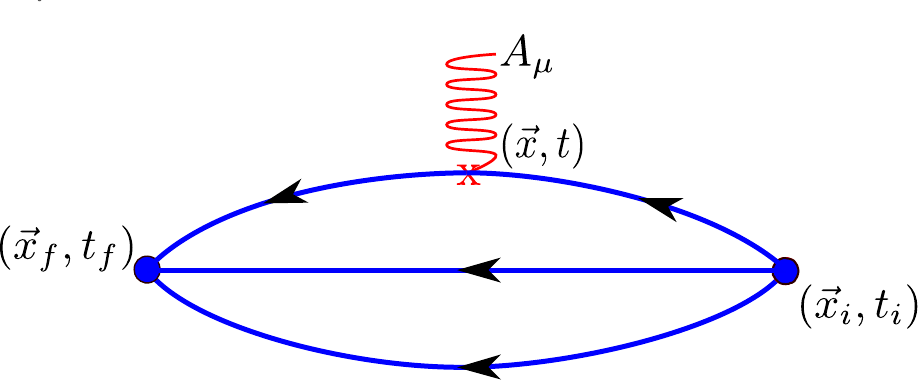}
  \end{minipage}
    \caption{\small The baryon two-point
      function (left) and the connected 
      three-point function (right) are shown diagrammatically. The solid lines represent fully dressed quark propagators.}
        \label{fig:2pt_3pt_diag}
\end{figure}

The matrix elements required for the calculation of the axial charges are extracted from dimensionless ratios involving two- and three-point correlation functions. The diagrams of the two-point function and the connected part of the three-point function involved in our calculations are depicted in \fig{fig:2pt_3pt_diag}. To extract the axial charges we consider kinematics for which the final and initial momentum are $\vec{p}_f= \vec{p}_i =0$. Since we only compute diagonal 
matrix elements, we consider three-point functions with the same baryon state at both source and sink. The time-independent ratio is obtained by dividing  the three-point function with the corresponding zero-momentum  two-point function.
  For the case of spin-1/2 baryons the two- and three-point functions
 are given by \citep{Alexandrou:2015yqa,Alexandrou:2010hf}
\be\label{eq:2pt_function}
	G_{2{\rm pt},B_{1/2}}(\vec{p}_f,t_f-t_i)=\sum_{\vec{x}_f}e^{-i(\vec{x}_f-\vec{x}_i)\cdot\vec{p}_f} \mathrm{Tr}[\Gamma^0\langle J_B(t_f, \vec{x}_f)\bar{J}_B(t_i,\vec{x}_i)\rangle] \rightarrow \frac{M_B}{E_B(\vec{p})} \vert Z_{1/2} \vert^2 e^{-E_B(\vec{p}_f) (t_f-t_i)} \mathrm{Tr}[\Gamma^0 \Lambda_{1/2}(p)] \;,
\ee
\begin{eqnarray}\label{eq:3pt_function}
G^{\mu\nu}_{3{\rm pt},B_{1/2}}(\vec{p}_f,t_f;t;\vec{p}_i,t_i)&=&\sum_{\vec{x},\vec{x}_f}e^{-i(\vec{x}_f-\vec{x}_i)\cdot\vec{p}_f} \mathrm{Tr}[ \Gamma^\nu \langle J_B (t_f,\vec{x}_f)A^\mu(t,\vec{x})\bar{J}_B(t_i,\vec{x}_i)\rangle ] e^{-i(\vec{x}-\vec{x}_i)\cdot\vec{p}_i} \nonumber \\
 &&  \rightarrow \frac{M_B}{\sqrt{E_B(\vec{p_f}) E_B(\vec{p_i})}} \vert Z_{1/2} \vert^2 e^{-E_B(\vec{p_f}) (t_f-t)} e^{-E_B(\vec{p_i})(t-t_i) } \mathrm{Tr}[\Gamma \Lambda_{1/2}(p_f) \mathcal{O}^\mu \Lambda_{1/2}(p_i) ] \;. 
\end{eqnarray}
For the case of spin-3/2 baryons the traces of the corresponding 
two- and three-point functions are given by~\citep{Alexandrou:2013opa,Alexandrou:2011py}
\be \label{Eq:2pt_function_32}
G_{2{\rm pt},B_{3/2}}(\vec{p}_f,t_f-t_i) \rightarrow \frac{M_B}{E_B(\vec{p}_f)} \vert Z_{3/2} \vert^2 e^{-E_B(\vec{p}_f) (t_f-t_i)} \mathrm{Tr}[\Gamma^0 P_{3/2}^{\sigma\tau}(\vec{p}_f) \Lambda_{3/2}^{ \tau\rho}(\vec{p}_f) P_{3/2}^{\rho\sigma}(\vec{p}_f) ] \;,
\ee

\bea\label{Eq:3pt_function_32}
	G^{\mu\nu}_{3{\rm pt},B_{3/2}}(\vec{p}_f,t_f;t;\vec{p}_i,t_i)&\rightarrow& \frac{M_B}{\sqrt{E_B(\vec{p_f}) E_B(\vec{p_i})}} \vert Z_{3/2} \vert^2 e^{-E_B(\vec{p_f}) (t_f-t)} e^{-E_B(\vec{p_i})(t-t_i) }\times\nonumber\\
	 &\times& \mathrm{Tr}[\Gamma^\nu P^{\sigma\tau}_{3/2}(\vec{p}_f) \Lambda_{3/2}^{\tau\rho}(p_f) \mathcal{O}^{\rho\pi;\mu} \Lambda_{3/2}^{\pi\kappa}(p_i) P_{3/2}^{\kappa\sigma}(\vec{p}_i)] \;.
\eea

 The projection matrices $\Gamma^0$ and $\Gamma^\nu$ are given by
\be\label{eq:proj_matrices}
\Gamma^0 = \reci{4}(\idnty + \gamma^0)\;,\quad \Gamma^\nu=\Gamma^0 i \gamma^5 \gamma^\nu\;.
\ee


\subsection{Smearing techniques}

In order to increase the overlap with the baryon ground state, we apply Gaussian smearing at the source and the sink~\cite{Alexandrou:1992ti,Gusken:1989qx}. The smeared interpolating fields are given by
\be \label{eq:smeared_fields}
q_{\rm smear}^a(t,\vec{x})=\sum_{\vec{y}}F^{ab}(\vec{x},\vec{y};U(t))q^b(t,\vec{y})\;,
\ee
where
\be
F=(\idnty +a_G H)^{N_G}
\ee
and $H$ is the hopping matrix given by
\be\label{eq:hop_mat}
H(\vec{x},\vec{y};U(t))=\sum_{i=1}^3\left[U_i(x)\delta_{x,y-\hat{i}}+U_i^\dag(x-\hat{i})
\delta_{x,y+\hat{i}}\right]\;.
\ee
We also apply APE-smearing to the gauge fields $U_\mu$ entering the hopping matrix. The parameters for the Gaussian smearing $a_G$ and $N_G$ are optimized using the nucleon ground state~\cite{Alexandrou:2008tn}. Various combinations of Gaussian smearing parameters, $N_G$ and $a_G$ have been tested and it was found that combinations giving a root mean square radius of about $0.5$ fm are optimal for suppressing excited states in the case of the nucleon. We adopt the same parameters
 here, which have the following values
\begin{table}[!ht]
\begin{center}
\begin{tabular}{c c c c c}
$\beta=1.95$ : & $N_G=50$  &, $a_G=4$ &, $N_{APE}=20$ &, $a_{APE}=0.5$, \\
$\beta=2.10$ : & $N_G=110$ &, $a_G=4$ &, $N_{APE}=50$ &, $a_{APE}=0.5$.\\
\end{tabular}
\end{center}
\end{table}


\subsection{Plateau method to extract axial charge}\label{sec:meth}

The computation of the axial charges proceeds through the evaluation of the diagrams shown in \fig{fig:2pt_3pt_diag}. As already mentioned, when taking the isovector combination of the axial current  the disconnected diagrams are zero up to lattice artifacts and can be safely neglected when close to the continuum limit. In such cases the connected contribution depicted in  \fig{fig:2pt_3pt_diag}
yields the whole contribution. For the rest of the cases the disconnected contributions  are neglected in this first computation. The creation operator of the baryon of interest is taken at a fixed source position $\vec{x}_i=\vec{0}$ with zero-momentum. Since, as discussed above, the axial charges are extracted directly from the matrix elements at $Q^2=0$, the annihilation operator at a later time $t_f$ also carries momentum $\vec{p}_f=0$. The current couples to a quark at an intermediate time $t$ and carries zero momentum ($\vec{q}=0$).  To compute the connected three-point function we use the so-called fixed-current method~\cite{Martinelli:1988rr} where the current-type and the time separation between the source and the current insertion, $t-t_i$ are fixed. The advantage of the sequential inversion through the current is that with one set of sequential inversions per choice of momentum and insertion time we obtain results for all possible sink times, any particle state and any choice of the projectors given in \eq{eq:proj_matrices}. An alternative approach that computes the spatial all-to-all propagator using stochastic methods was shown to be suitable for the evaluation of baryon three-point functions~\cite{Alexandrou:2013xon}. With this method one can include any current at the insertion point for any particle state and any projector at the sink without needing additional inversions. However, the disadvantage is that one introduces stochastic noise, so one has to check convergence as a function of the number of noise vectors. In this work, since we are only interested in the axial charges,  we instead adopt the sequential method through the current also referred to as fixed current method.

A standard way of isolating the matrix elements,  is to form appropriate ratios with the use of the two- and three-point functions of Eqs.~\ref{eq:2pt_function} and~\ref{eq:3pt_function} for the spin-1/2 baryons and Eqs.~\ref{Eq:2pt_function_32} and~\ref{Eq:3pt_function_32} for the spin-3/2 baryons. 

In the limit $t_f-t\gg 1$ and $t-t_i\gg 1$  the unknown overlap terms and Euclidean time dependence cancel thus yielding  a time-independent result as a function of the sink time, referred to as plateau region. A constant fit is then performed to extract the axial charge. The traces involved in the two- and three-point functions can be calculated using Dirac trace algebra. Since zero-momentum kinematics are employed the final relations acquire simple forms. Specifically, for the spin-1/2 baryons one obtains
\be \label{Eq:ratio_12}
R^{ij}_{1/2}(t_f-t,t-t_i) = \frac{ G^{ij}_{1/2}(t_f-t,t-t_i) }{G_{1/2}(t_f-t_i) } \Longrightarrow \lim_{t_f-t\rightarrow\infty}\lim_{t-t_i\rightarrow\infty} R^{ij}_{1/2}(t_f-t,t-t_i) = \Pi^{ij}_{1/2} = G_A(0) \delta^{ij}\, ,
\ee
while the corresponding expression for spin-3/2 baryons yields \citep{Alexandrou:2013opa}
\be \label{Eq:ratio_32}
R^{ij}_{3/2}(t_f-t,t-t_i) = \frac{G^{ij}_{3/2}(t_f-t,t-t_i) }{G_{3/2}(t_f-t_i) } \Longrightarrow \lim_{t_f-t\rightarrow\infty}\lim_{t-t_i\rightarrow\infty} R^{ij}_{3/2}(t_f-t,t-t_i) = \Pi^{ij}_{3/2} = \frac{5}{9} g_1(0) \delta^{ij}\,.
\ee
The  plateau value thus yields the unrenormalised  charges, which after
renormalization with $Z_A$, gives directly the axial charge of the baryon. The renormalization constants used in this work are $Z_A^{\beta=1.95}=0.7556(5)(85)$ and $Z_A^{\beta=2.10}=0.7744(7)(31)$, taken from Ref.~\cite{Abdel-Rehim:2015owa}.

\subsection{Flavor structure of the axial-vector current}

The  baryon axial charges govern processes like $n\rightarrow p e^{-}\bar{\nu}_e$ and $\Sigma^-\rightarrow \Sigma^0 e^{-}\bar{\nu}_e$. They can be extracted by considering the matrix elements $\langle B|A_\mu^3|B\rangle_{Q^2=0}$ where $B=N,\Delta,\Sigma,\ldots$~\cite{Choi:2010ty} and $A_\mu^3$ is the isovector combination for the axial-vector current.  Given that we have four quark flavours, we can construct for the axial-vector current combinations corresponding to the generators of the $SU(4)$ gauge group. In this study besides the isovector that corresponds to the  $\frac{1}{2}\lambda_3$ generator we consider combinations of the other two diagonal SU(4) generators, namely $\frac{1}{2}\lambda_8$ and $\frac{1}{2}\lambda_{15}$.

%

The generator $\frac{1}{2}\lambda_3$ gives the well-known isovector combination, which produces the  axial coupling between the pion and the baryon effective fields. In the SU(4) limit disconnected contributions will cancel for all three  combinations  given by the currents
\bea\label{isov_currents}
A_\mu^3&=&\frac{1}{2}\left(\bar{q}_{f_1} \gamma_\mu\gamma_5 q_{f_1}- \bar{q}_{f_2} \gamma_\mu\gamma_5 q_{f_2}\right)\nonumber\\
A_\mu^8&=&\frac{1}{2}\left(\bar{q}_{f_1} \gamma_\mu\gamma_5 q_{f_1}+\bar{q}_{f_2} \gamma_\mu\gamma_5 q_{f_2}-2\bar{q}_{f_3} \gamma_\mu\gamma_5 q_{f_3}\right)\\
A_\mu^{15}&=&\frac{1}{2}\left(\bar{q}_{f_1} \gamma_\mu\gamma_5 q_{f_1}+ \bar{q}_{f_2} \gamma_\mu\gamma_5 q_{f_2}+ \bar{q}_{f_3} \gamma_\mu\gamma_5 q_{f_3}-3\bar{q}_{f_4} \gamma_\mu\gamma_5 q_{f_4}\right)\quad \nonumber .
\eea
In what follows, for a given baryon, we denote the  flavor combination  of the current corresponding to $\lambda_3$  as $g_A^B$, to $\lambda_8$  as $g_{8}^B$ and to $\lambda_{15}$ as $g_{15}^B$.
In the case of $A^\mu_{15}$ at least one term in the current will yield a purely disconnected contribution, which will be neglected here.
In addition, we consider the  isoscalar combination 
\be
A^\mu_{0} = \sum_{i=1,\cdots,4}\bar{q}_{f_i} \gamma_\mu\gamma_5 q_{f_i}\;.
\label{isos_current} 
\ee
Having these combinations one can extract the axial charge corresponding to each quark flavor $g_A^q$. In Eqs.~\ref{isov_currents} and~\ref{isos_current} $f_1=u,\,f_2=d,\,f_3=s\,$ and $f_4=c$. Depending on the quark flavor content of the baryon some terms will give purely disconnected contributions and will be neglected.

We note that $g_A^q$ determines the intrinsic spin carried by the quark $q$ inside the given baryon.


\subsection{Fixing the insertion time}

In the fixed current method that involves sequential inversion through the current, the time separation between the source and the current insertion, $t-t_i$ is fixed. Optimally, one would choose a source-insertion separation small enough to keep the statistical errors as small as possible and still large enough to ensure that excited state contributions are sufficiently suppressed.
\begin{figure}[!ht]
\includegraphics[width=0.47\linewidth]{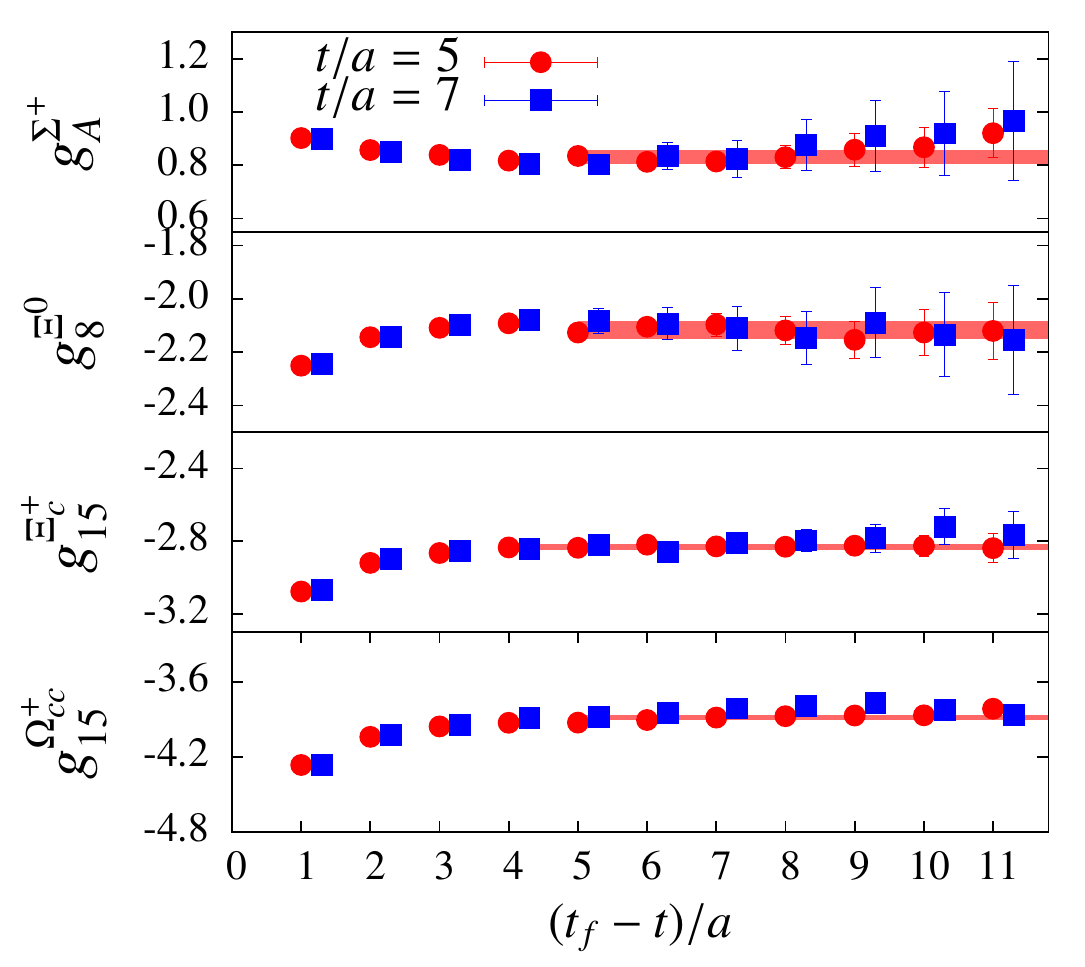}
\caption{\small Results for the axial charges of hyperons and charmed baryons for two choices of the current insertion time are shown, namely $t/a=5$ with red circles and $t/a=7$ with blue squares as a function of  $(t_f-t)/a$.}
\label{fig:time_comparison}
\end{figure}

While recent studies have shown that the optimal source-sink time separation is operator dependent~\cite{Alexandrou:2011aa,Dinter:2011sg}, for the axial charge the excited state contamination was generally found to be small at least for pion masses larger than physical~\cite{Alexandrou:2013joa}.  Still we need to ensure that the insertion-source time separation is sufficiently large to be free of large excited state contaminations. We examine  two values of the insertion time, namely $t/a=5$ and $t/a=7$ for our B-ensemble with $a\mu=0.0055$ or $m_\pi=373$~MeV, with results shown in \fig{fig:time_comparison}. As can be seen, the results are compatible for these two values of the current time insertion. Thus, we  fix $t/a=5$ and seek a plateau as a function of $t_f-t$.

A plateau region starting at $(t_f-t)/a=5$ is obtained confirming ground state dominance at a time separation of $10a$ from the source and the sink. For the D-ensemble we take $t/a=7$ to keep the time separation in physical units about the same.


\section{Effective Lagrangian}

Before we present our lattice QCD results, we discuss briefly the effective meson-baryon Lagrangians where these axial couplings are defined.
Heavy baryon chiral perturbation theory (HB$\chi$PT) is most commonly applied to the octet and decuplet baryons. The lowest order (tree-level)  meson-baryon effective interaction for the octet can be written in terms of two SU(3) scalars. Arranging these two scalars into symmetric and antisymmetric combinations we have~\citep{Jenkins:1990jv}

\be
{\cal L}^{(1)}_{1/2}=2D {\rm Tr}\bar{B} S^\mu\{A_\mu,B\}+ 2F{\rm Tr}\bar{B}S^\mu[A_\mu,B],
\label{chiral octet}
\ee
where $B$ is the traceless $3\times 3$ octet field
\be
 B=\sum_{a=1}^8 \frac{B_a\lambda_a}{\sqrt{2}} = \left (\begin{array}{ccc}
\frac{1}{\sqrt{2}}\Sigma^0+\frac{1}{\sqrt{6}}\Lambda & \Sigma^{+} & p \\
\Sigma^{-} & -\frac{1}{\sqrt{2}}\Sigma^0+\frac{1}{\sqrt{6}}\Lambda & n\\
\Xi^{-} & \Xi^0 & -\frac{2}{\sqrt{6}}\Lambda
\end{array} \right )\;.
\label{eq:eff_lagr_B}
\ee
$A_\mu$ is written in terms of $\xi={\rm exp}(-i\pi/f_\pi)$   and  it is the combination of meson fields that transform like an axial-vector current. Here we follow standard notation and take  $\pi$ the $3\times 3$ matrix of the pseudoscalar mesons, $S^\mu$ the spin operator acting on the baryon fields, while we suppress the velocity index on $B$ and $S^\mu$.

In the limit of $SU(3)$ flavour symmetry, the axial couplings are thus given in terms of the two low-energy constants $D$ and $F$ appearing in the Lagrangian.  For the pion-baryon axial couplings we  thus  have
\begin{equation}\label{eq:ax_param}
g_{\pi NN} = F + D\equiv g_A^N, \;\;\; g_{\pi \Xi\Xi} = F-D\equiv g_A^\Xi, \;\;\; g_{\pi \Sigma\Sigma} = 2F\equiv g_A^\Sigma,
\end{equation}  
 while for the octet $\eta_8$-baryon couplings
\be
g_{\eta_8 NN}=-\frac{1}{\sqrt{3}}(D-3F)\equiv g_8^N,\,\, g_{\eta_8 \Lambda \Lambda}=-\frac{2}{\sqrt{3}}D\equiv g_8^\Lambda,\,\, g_{\eta_8 \Sigma \Sigma}=\frac{2}{\sqrt{3}}D\equiv g_8^\Sigma,\,\,  g_{\eta_8 \Xi \Xi}=-\frac{1}{\sqrt{3}}(D+3F)\equiv g_8^\Xi.
\label{eta-baryon}
\ee
There are five transition coupling constants in addition to the above, namely $g_{\pi \Lambda \Sigma}$, $g_{KN\Lambda}$, $g_{KN\Sigma}$, $g_{KN\Xi}$ and $g_{K\Sigma\Xi}$, which are also written in terms of $D$ and $F$. These  require  the computation of
non-diagonal matrix elements and are not considered in this work.  

For decuplet baryons one can only construct one SU(3) scalar and thus the axial coupling constants are given in terms of one constant ${\cal H}$. The lowest order interaction Lagrangian involving diagonal terms is given by~\citep{Jenkins:1991es}
\be
{\cal L}^{(1)}_{3/2}={\cal H} \bar{T}^\mu \gamma_\nu \gamma_5 A^\nu T_\mu=2{\cal H} \bar{T}^\mu S_\nu  A^\nu T_\mu \,,
\label{chiral decuplet}
\ee
where we suppress the velocity index on the tensor $T_\mu$. Suppressing the Lorenz index $\mu$, $T$ is given by~\cite{Butler:1992pn}
\bea
& & T^{111}=\Delta^{++},\,\, T^{112}=\frac{1}{\sqrt{3}}\Delta^+,\,\, T^{122}=\frac{1}{\sqrt{3}}\Delta^0,\,\, T^{222}=\Delta^{-},\,\, T^{333}=\Omega^-\nonumber \\
& & T^{113}=\frac{1}{\sqrt{3}}\Sigma^{*+},\,\, T^{123}=\frac{1}{\sqrt{6}}\Sigma^{*0},\,\, T^{223}=\frac{1}{\sqrt{3}}\Sigma^{*-},\,\, T^{133}=\frac{1}{\sqrt{3}}\Xi^{*0},\,\,T^{233}=\frac{1}{\sqrt{3}}\Xi^{*-}\quad.
\eea
In Eq.~\ref{chiral decuplet} we have not written the coupling of the decuplet to the octet baryons, which introduces another axial transition coupling constant since this will also involved non-diagonal matrix elements which are not computed here.
In the  SU(3) limit the decuplet axial couplings are given by~\cite{Tiburzi:2008bk}
\be\label{eq:decuplet_H_dep}
g_{\pi\Delta\Delta}={\cal H}\equiv g_A^\Delta,\,\, g_{\pi\Sigma^*\Sigma^*}=\frac{2}{3}{\cal H}\equiv g_A^{\Sigma^*},\,\, g_{\pi\Xi^*\Xi^*}=\frac{1}{3}{\cal H}\equiv g_A^{\Xi^*}.
\ee 

Only a few groups have considered charmed baryons within HB$\chi$PT, see e.g. Refs~\cite{Jiang:2014ena,Li:2012bt,Liu:2012uw,Liu:2012sw}, and these studies focus only on the singly charmed baryons. For completeness we give here the Lagrangian for singly charmed baryons. The baryon fields for the symmetric $\bf 6$-tet and the antisymmetric $\bf\bar{3}$-plet of spin-1/2 charmed baryons are defined as follows

\be
 B_{\bar{3}} =  \left (\begin{array}{ccc}
 0             & \Lambda_c^+ & \Xi_c^+ \\
 -\Lambda_c^+  &       0     & \Xi_c^0 \\
 -\Xi_c^+      &  -\Xi_c^0   &    0    
\end{array} \right )\;,\;\;\;  B_{6} =  \left (\begin{array}{ccc}
         \Sigma_c^{++}            & \frac{1}{\sqrt{2}}\Sigma_c^+     &\frac{1}{\sqrt{2}}\Xi_c^{\prime +}\\
\frac{1}{\sqrt{2}}\Sigma_c^{+}    &         \Sigma_c^0               &\frac{1}{\sqrt{2}}\Xi_c^{\prime 0}\\
\frac{1}{\sqrt{2}}\Xi_c^{\prime +}&\frac{1}{\sqrt{2}}\Xi_c^{\prime 0}&  \Omega_c^0 
\end{array} \right )\;.
\label{eq:eff_lagr_heavyb}
\ee
The definition of $B_6^*$ for the spin-3/2 $\bf 6$-tet is similar to that of $B_6$. The effective Lagrangian at tree-level can be written in terms of the couplings $g_i\;,i=1\ldots 6$ and reads~\cite{Liu:2012uw,Liu:2012sw}
\bea\label{eq:eff_lagr_charm}
\mathcal{L}_{\rm ch.b.}^{(1)} &=& 2g_1{\rm Tr}\left(\bar{B}_6S\cdot u B_6\right)+ 2g_2{\rm Tr}\left(\bar{B}_6S\cdot u B_{\bar{3}}+{\rm H.c.}\right)+g_3{\rm Tr}\left(\bar{B}_{6\mu}^* u^\mu B_6+{\rm H.c.}\right) +\nonumber\\
&+& g_4{\rm Tr}\left(\bar{B}_{6\mu}^* u^\mu B_{\bar{3}}+{\rm H.c.}\right)+2g_5{\rm Tr}\left(\bar{B}_{6}^*S\cdot u B_6^*\right)+2g_6{\rm Tr}\left(\bar{B}_{\bar{3}}S\cdot u B_{\bar{3}}\right)\;.
\eea
Similarly to the octet case, $u_\mu$ is written in terms of $\exp(i\pi/f_\pi)$, where $\pi$ is the $3\times 3$ pseudoscalar meson field and $S_\mu$ is the spin matrix acting on the baryon fields.

For hyperons $SU(3)$ breaking arises  as a result of the larger strange quark mass and lattice QCD provides a framework to study the $SU(3)$ breaking as a function of the quark mass.  Although a similar approach can be used for charmed baryons, the much larger mass of the charm-quark can make symmetry patterns more difficult or even impossible to disentangle.


\section{Lattice results}

In this section we present our results on the axial charges for the four B- and the one D-ensembles.  Comparisons with other lattice calculations are shown for the axial charge of spin-1/2 hyperons results wherever available.  A study of the SU(3) flavour breaking  for the octet and decuplet baryons is also presented. A similar analysis is carried out for charmed baryons where corresponding relations hold when replacing the strange with the charm quark although the breaking is expected to be larger. All the lattice data on the axial couplings considered in this work are collected in Tables~\ref{Table:axial_IV}-\ref{Table:axial_charm} of Appendix B.

\subsection{Axial charges of octet and decuplet baryons}

\subsubsection{Octet baryons}
\begin{figure}[!ht]
  \begin{minipage}[t]{0.49\linewidth}
    \includegraphics[width=\linewidth]{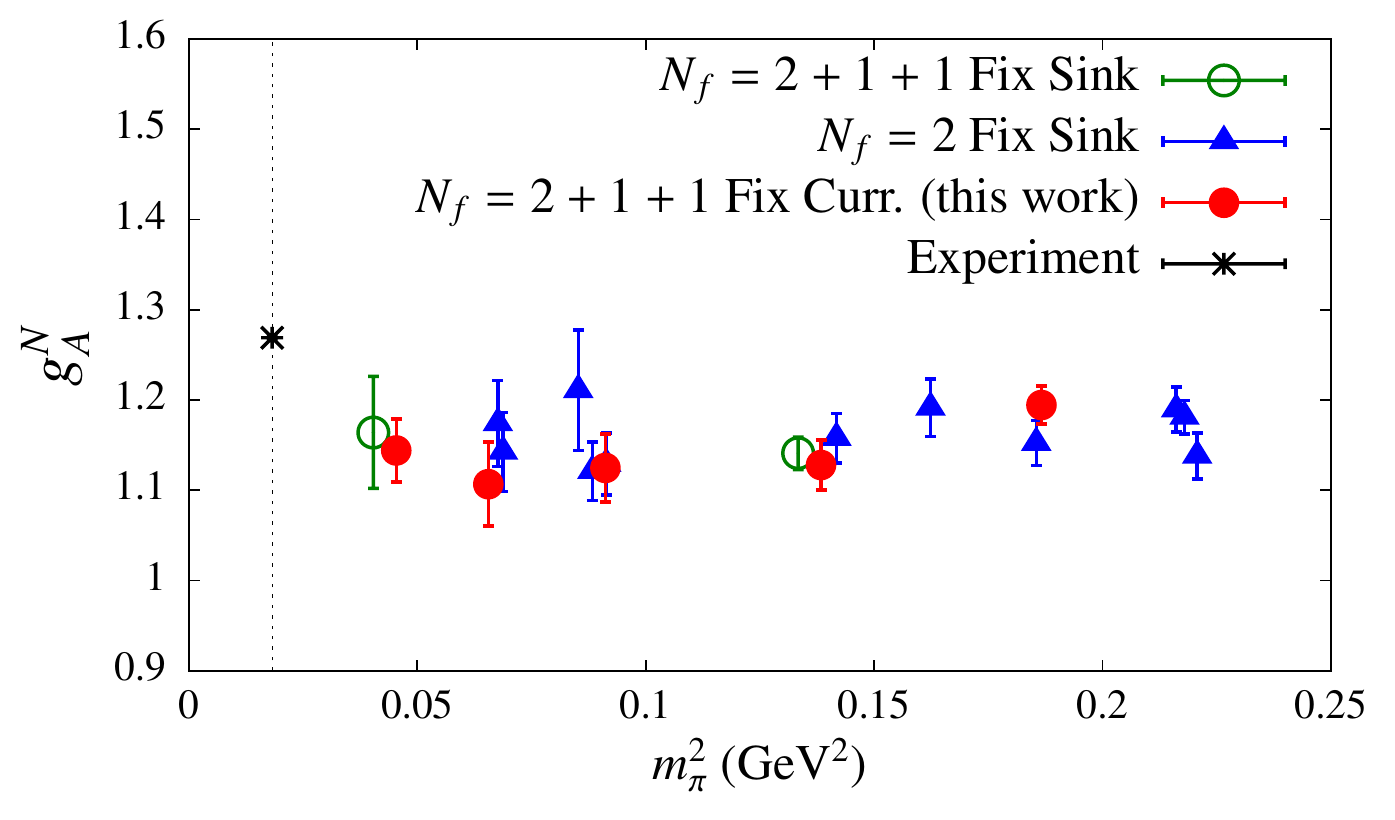}
  \end{minipage}
  \hfill
  \begin{minipage}[t]{0.49\linewidth}
    \includegraphics[width=\linewidth]{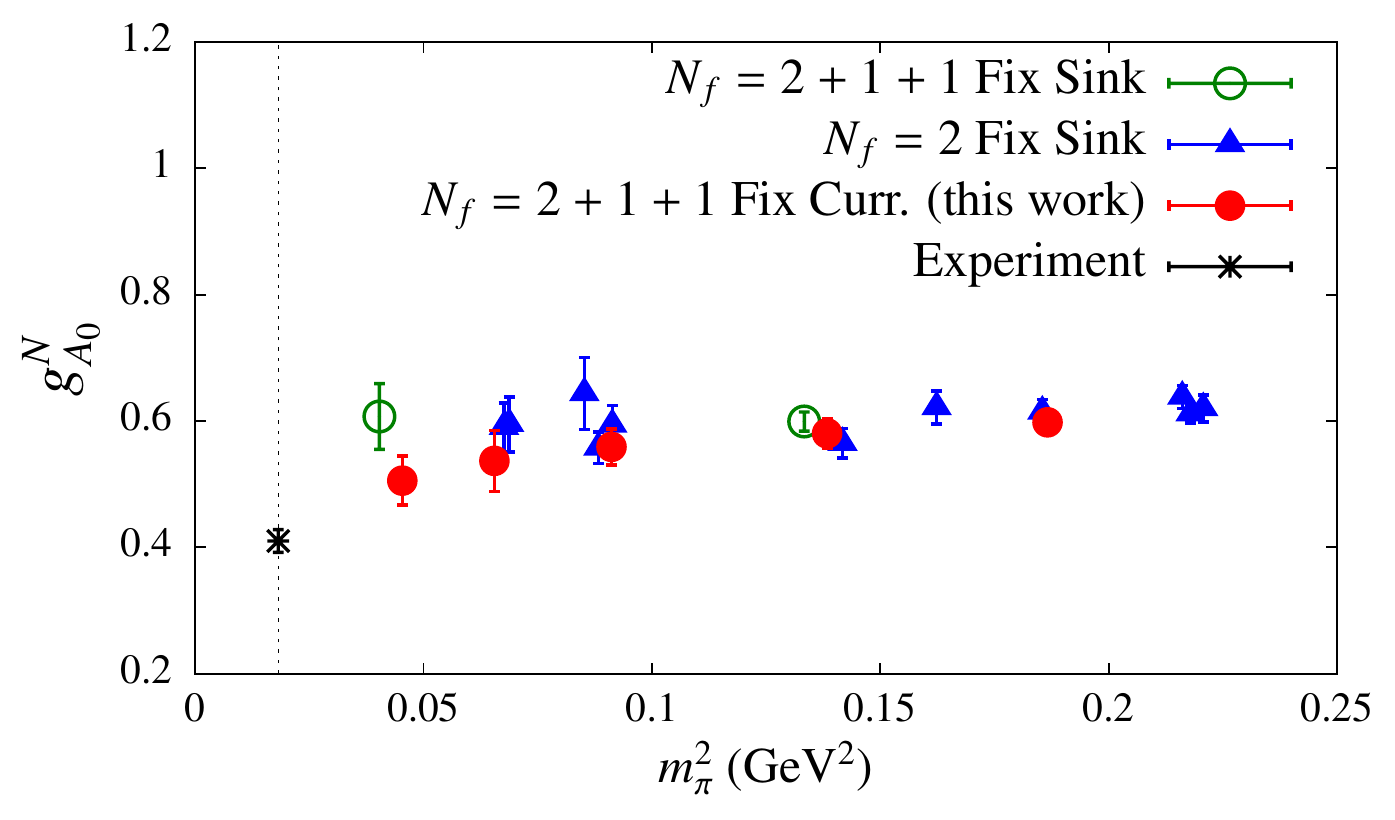}
  \end{minipage}
\caption{Axial charge for the nucleon as a function of the pion mass for the isovector (left) and isoscalar (right) combinations. With red circles we show the results of this work, in blue triangles we show results from $N_f=2$ twisted mass fermions (TMF) from Ref.~\cite{Alexandrou:2010hf} and with open green circles we show results using $N_f=2+1+1$ TMF ensembles (B55.32 and D15.48) but obtained with  using the fixed sink method from Ref.~\cite{Alexandrou:2013joa} (shifted to the left for clarity). The experimental values shown with the black asterisk are taken from PDG~\cite{Agashe:2014kda}.}
\label{fig:gA_Nucleon}
\end{figure}

As already pointed out, the axial charge of the nucleon is well measured and it
is  thus considered as a benchmark quantity within lattice QCD. Before discussing results on the axial charges of other baryons, we first compared the nucleon axial charge, $g_A^N$, using the fixed current approach adopted here with the results obtained with the fixed sink method. The latter approach is the one routinely used
to extract the nucleon axial form factors. In a
 previous work we calculated $g_A^N$ for  the B55.32  and the D15.48
ensembles using the fixed sink method.  In \fig{fig:gA_Nucleon} we show results  as a function of the pion mass for both the isovector  and the isoscalar  axial charges using the fixed current approach, as well as  using the fixed sink method~\cite{Alexandrou:2010hf,Alexandrou:2013joa}.
One can see that the two methods give compatible results. A general observation is the underestimation of the nucleon isovector axial charge for larger than physical pion masses. A recent computation using $N_f=2$ twisted mass clover-improved fermions at a physical value of the pion mass yields a value consistent with the experimental value albeit with large statistical uncertainty~\cite{Abdel-Rehim:2015owa}.  Similarly, the connected part of the isoscalar charge $g_{A_0}^N$ is overestimated for larger pion masses.  Disconnected contributions are found to be negative and will thus  decrease this value. As expected, as we approach the physical pion mass larger statistics are required in order to obtain a more robust result. 

\begin{figure}[!ht]
\includegraphics[width=0.5\linewidth]{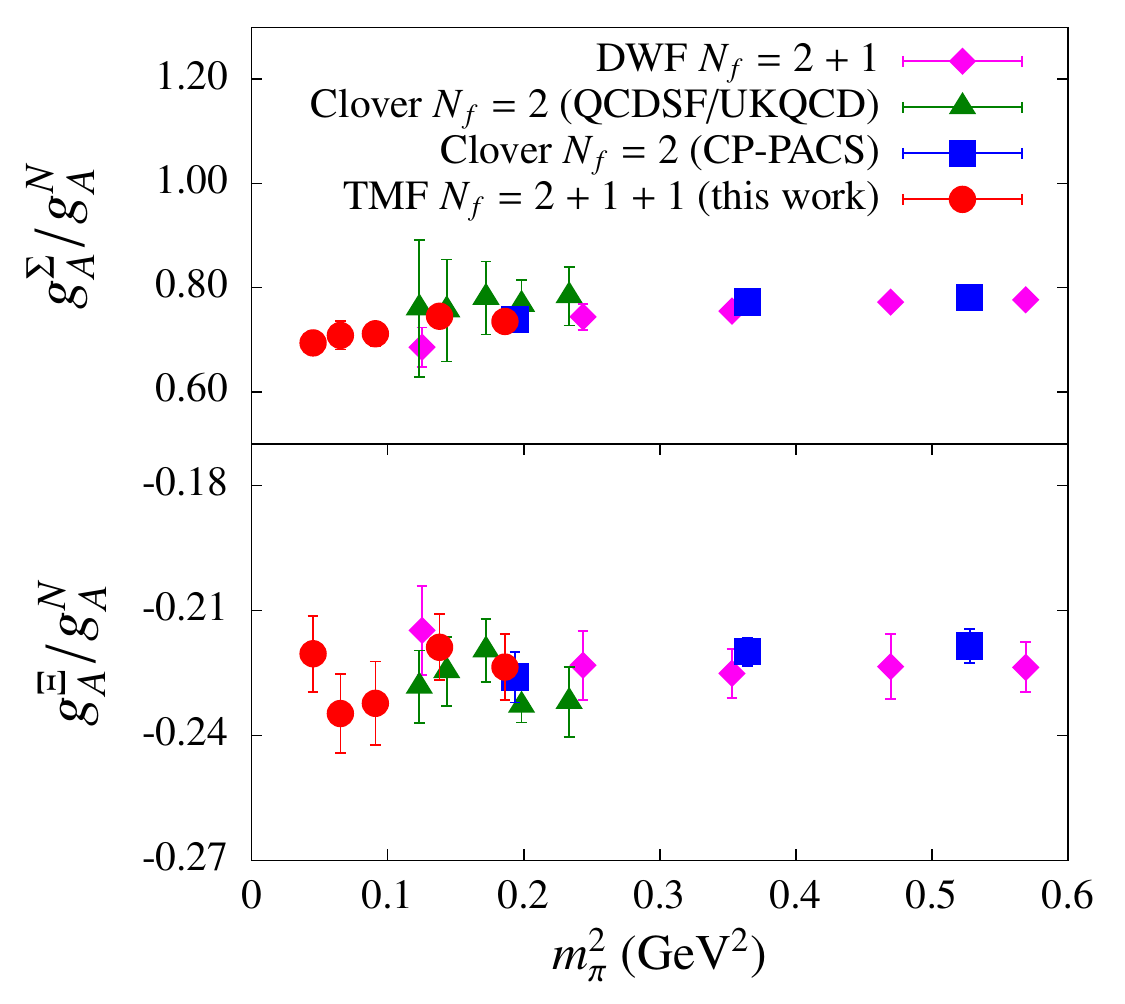}
\caption{\small Comparison of the ratios $g_A^\Sigma/g_A^N$ (upper) and $g_A^\Xi/g_A^N$ (lower) of this work (red circles), $N_f=2+1$ domain wall fermions (DWF) from Ref.~\cite{Lin:2007ap} with $a=0.123$~fm, $L=2.6$~fm (pink diamonds), $N_f=2$ Clover fermions from QCDSF/UKQCD~\cite{Gockeler:2011ze} with $a=0.078$~fm, $L=1.9$~fm (green triangles) and $N_f=2$ Clover fermions from CP-PACS~\cite{Erkol:2009ev} with $a=0.156$~fm, $L=2.5$~fm (blue squares).}
\label{fig:gA_sigma_xi_comp}
\end{figure}

The other particles within the octet are the $\Sigma$ and $\Xi$ isospin multiplets and the $\Lambda^0$ singlet. Contrary to the nucleon, the short lifetime of these baryons makes the experimental determination of their axial couplings difficult, and therefore very limited experimental data are available. Additionally, theoretical estimates are rather imprecise. On the lattice, only a handful of other calculations have considered the octet axial charges~\cite{Lin:2007ap,Gockeler:2011ze,Erkol:2009ev}. We compare previous lattice QCD results with our values in \fig{fig:gA_sigma_xi_comp}, where we show the renormalization independent ratios $g_A^\Sigma/g_A^N$ and $g_A^\Xi/g_A^N$. There is an agreement among all the data within the whole pion mass range despite the different discretizations, lattice spacings and volumes, indicating that lattice artefacts are small for the parameters used in these simulations. An estimate of $g_A^\Sigma$ and $g_A^\Xi$ from $\chi$PT is found in Ref.~\cite{Jiang:2009sf}, giving values of $g_A^\Sigma = 0.73$ and $g_A^\Xi = -0.23$, not far from our lattice values. These couplings have been also obtained from relativistic constituent quark model (RCQM) calculations in Ref.~\cite{Choi:2010ty}, yielding values  $g_A^\Sigma = 0.919$ and $g_A^\Xi = -0.22$. The value of $g_A^\Sigma$ from the latter calculation notably overestimates our lattice result \footnote{ We note here that due to a different definition of $g_A^\Xi$, the value quoted here has a sign opposite to the one in Ref.~\cite{Jiang:2009sf}. Similarly, the value of $g_A^\Sigma$ in Ref.~\cite{Choi:2010t} is smaller by $\sqrt{2}$ to the one quoted here.}.

\begin{figure}[!ht]
  \begin{minipage}[t]{0.49\linewidth}
    \includegraphics[width=\linewidth]{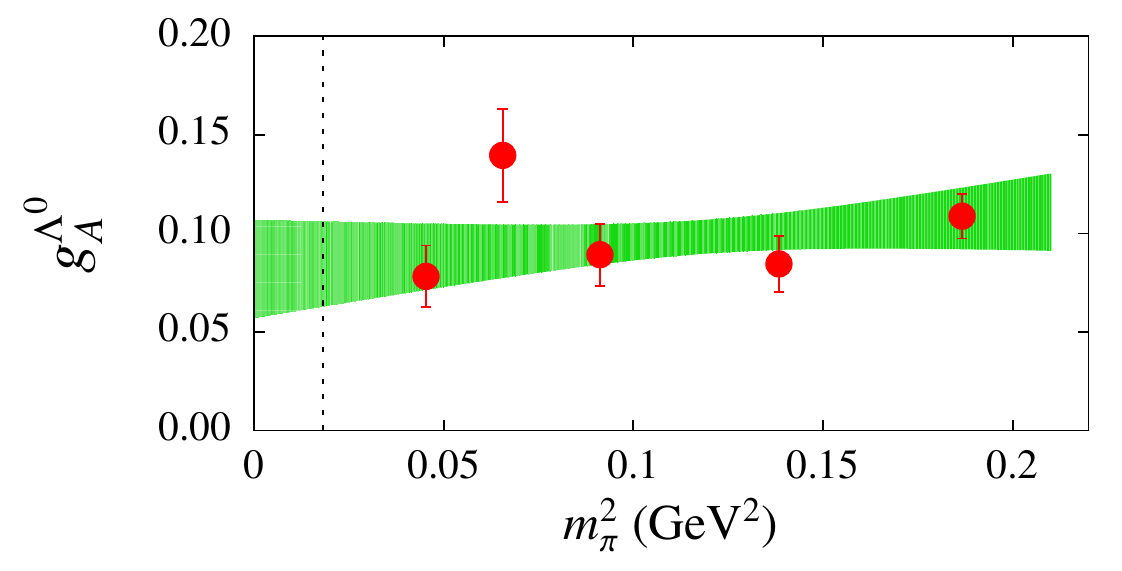}
  \end{minipage}
  \hfill
  \begin{minipage}[t]{0.49\linewidth}
    \includegraphics[width=\linewidth]{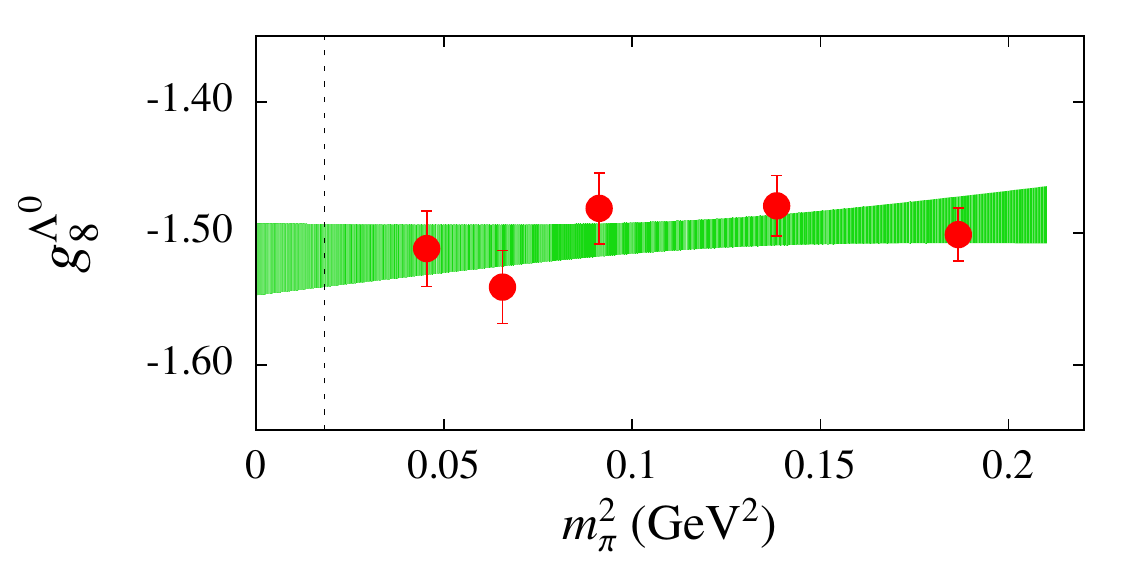}
  \end{minipage}
\caption{\small Results for the  axial charges for the $\Lambda^0$ baryon. Left: $\lambda_3$ (isovector) combination. Right: $\lambda_8$ (octet) combination.}
    \label{fig:gA_lambda}
\end{figure}

\begin{figure}[!ht]
  \begin{minipage}[t]{0.49\linewidth}
    \includegraphics[width=1.02\linewidth]{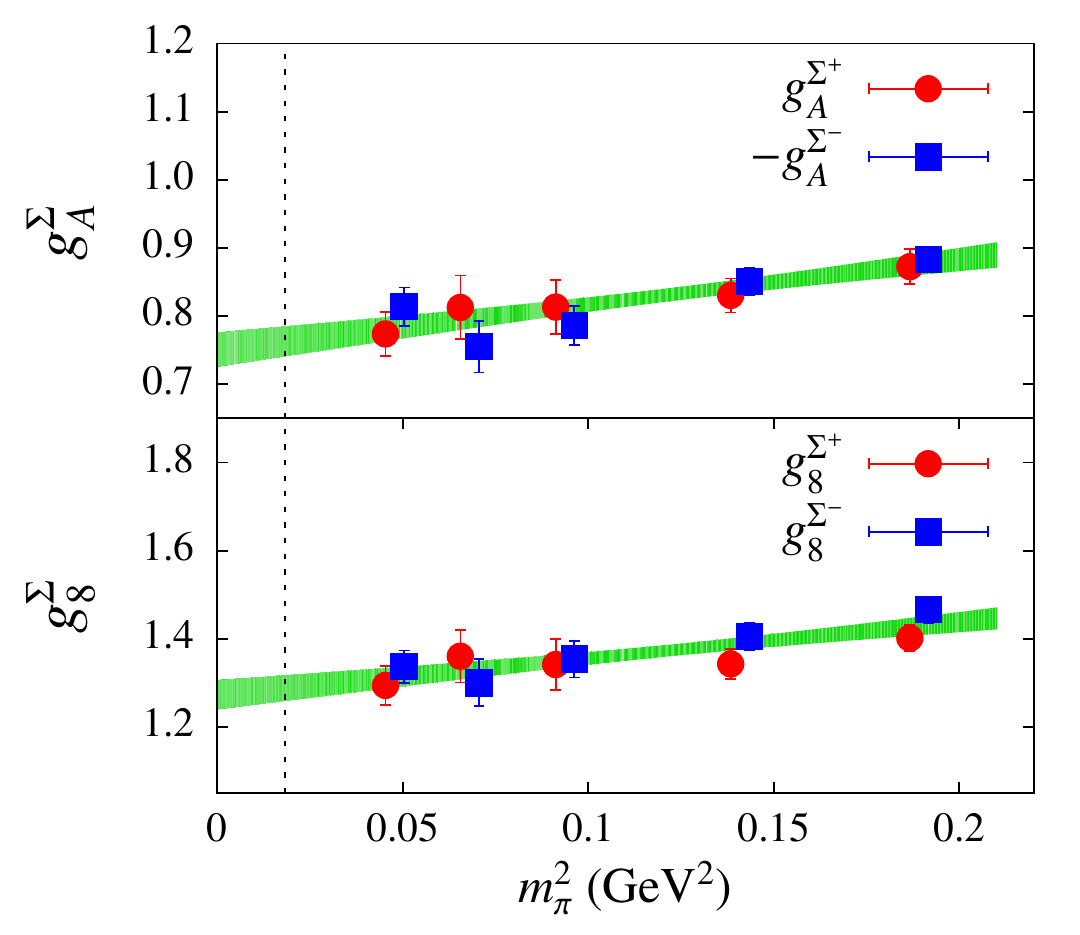}
  \end{minipage}
  \hfill
  \begin{minipage}[t]{0.49\linewidth}
    \includegraphics[width=1.02\linewidth]{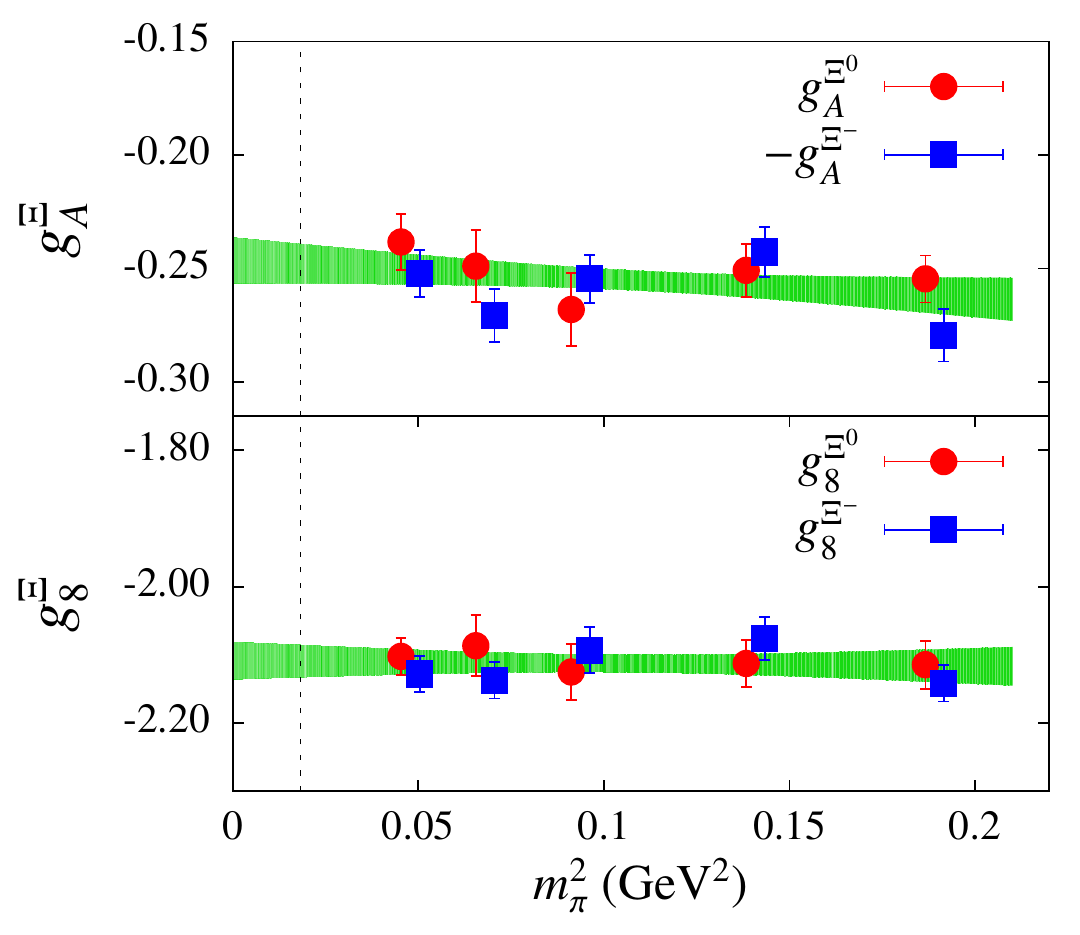}
  \end{minipage}
\caption{\small The $\lambda_3$ (isovector) and $\lambda_8$ (octet) combinations for the $\Sigma$ (left) and $\Xi$ (right) isospin multiplets.}
    \label{fig:gA_sigma_xi}
\end{figure}

In Figs.~\ref{fig:gA_lambda} and~\ref{fig:gA_sigma_xi} we show the pion mass dependence for the $\Lambda^0$, the $\Sigma$ and $\Xi$ multiplets for the flavor combination of the two diagonal generators $\lambda_3$ and $\lambda_8$. As can be seen, results are fully compatible between isospin partners, indicating that the isospin symmetry breaking effects, due to the finite lattice spacing, are small. All data exhibit weak dependence on the pion mass over the range of pion masses studied in this work. Comparing the results using the B25.32 and D15.48, which have similar pion mass, we observe consistent values for the axial charges, an indication that cut-off effects are small.

In order to obtain an estimate of the axial charges of the hyperons at the physical pion mass, we perform a chiral extrapolation according to the Ansatz $a + b m_\pi^2$, where $a$ and $b$ are fit parameters. This Ansatz proves to be preferable by our results as a leading-order expression, as the $\chi^2/$d.o.f from these linear fits ranges from $0.21\sim1.55$. For the $\Sigma$ and $\Xi$ states, the fit is performed on the average of the isospin partners since the isospin symmetry breaking effects are found to be negligible within our statistical accuracy. Although NLO expressions for $\Sigma$ and $\Xi$ exist in Ref.~\cite{Jiang:2009sf},  we refrain from using these expressions to avoid introducing new low-energy constants. The linear fits are shown with the green error bands on the plots. The extrapolated values at the physical point are collected in~\tbl{Table:extrap_values} of Appendix C. In order to correctly estimate the error band, we apply an extended version of the standard jackknife error procedure known as super-jackknife analysis~\cite{Bratt:2010jn}. Briefly, this generalized method is applicable for analyzing data computed on several gauge ensembles. Despite the fact that data sets from different gauge ensembles are uncorrelated, there is correlation among the data within each ensemble. This analysis method allows us to consider a different number of lattice QCD measurements for each ensemble while correlations within each ensemble are appropriately taken into account.


\subsubsection{Decuplet baryons}

\begin{figure}[!ht]
\begin{minipage}[t]{0.49\linewidth}
\includegraphics[width=\linewidth]{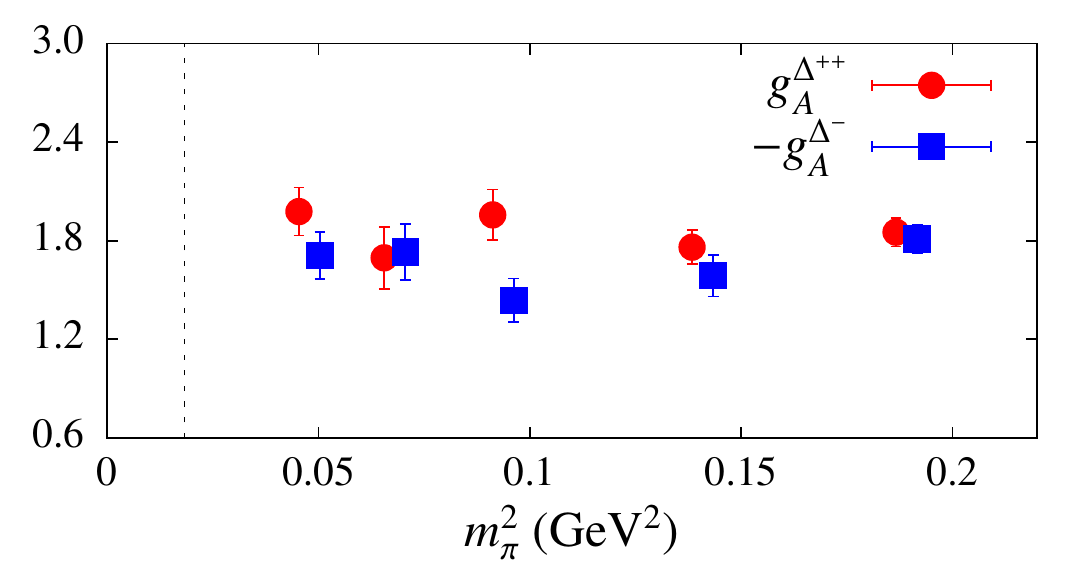}
\end{minipage}
\hfill
\begin{minipage}[t]{0.49\linewidth}
\includegraphics[width=\linewidth]{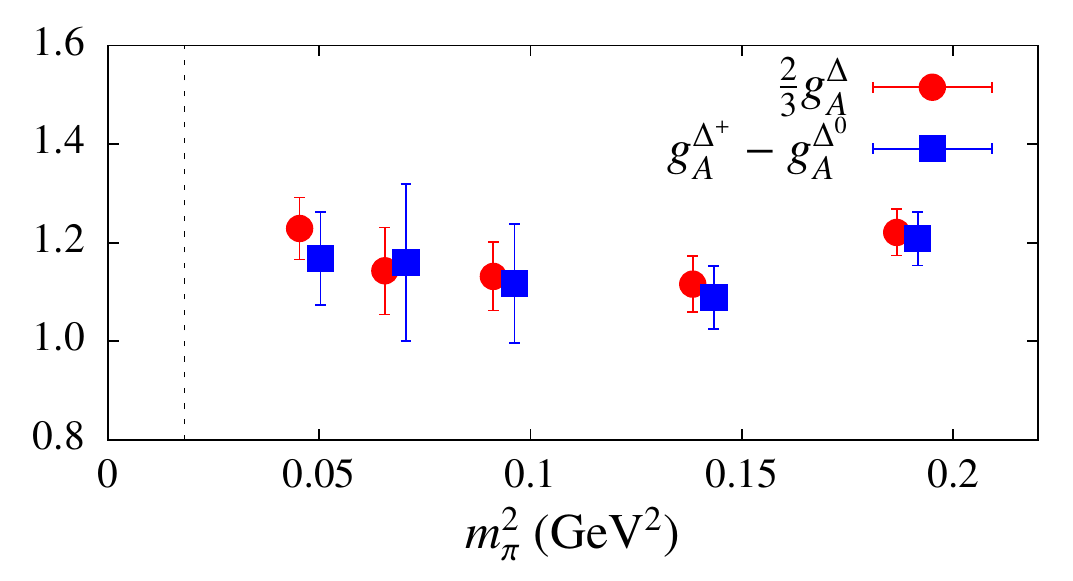}
\end{minipage}
\caption{Results for $g_A^{\Delta^{++}}$ and $g_A^{\Delta^{-}}$ (left) and  $g_A^{\Delta^{+}}-g_A^{\Delta^{0}}$ and $\frac{2}{3}g_A^\Delta$ (right), according to the definition given in Eq.~\ref{eq:gA_delta_plus_plus}.}
\label{fig:gA_delta}
\end{figure}

The axial coupling of the  $\Delta$ $g_A^\Delta$ enters chiral Lagrangians that explicitly contain $\Delta$ degrees of freedom and thus its value is needed as an input in chiral perturbation expressions for many important quantities. Due to the fact that its value is not known, it is usually treated as a fit parameter. Lattice QCD can provide a determination of $g_A^\Delta$ with the formalism described in  Refs.\citep{Alexandrou:2011py,Alexandrou:2013opa} where first results were given using domain wall fermions.  According to Ref.~\cite{Jiang:2008we} and using our notation, the axial charge of $\Delta$ can be defined as 
\begin{equation}
 g_A^{\Delta^{++}} - g_A^{\Delta^{-}} \equiv 2 g_A^{\Delta}\;,\quad
g_A^{\Delta^+} - g_A^{\Delta^0} = \frac{2}{3} g_A^{\Delta}\;,
\label{eq:gA_delta_plus_plus}
\end{equation}
where the second relation results from the isospin Clebsch-Gordan coefficients.
Due to isospin symmetry we expect the axial charge of the $\Delta^{++}$ to be the same as that of $\Delta^{-}$, apart from a minus sign. Indeed, our results on $g_A^{\Delta^{++}}$ and $g_A^{\Delta^-}$, $g_A^{\Delta^+}$ and $g_A^{\Delta^0}$ shown in~\fig{fig:gA_delta} are in agreement confirming again that cut-off effects are small.
%
%
%
%

\begin{figure}[!ht]
\begin{minipage}[t]{0.49\linewidth}
\includegraphics[width=\linewidth]{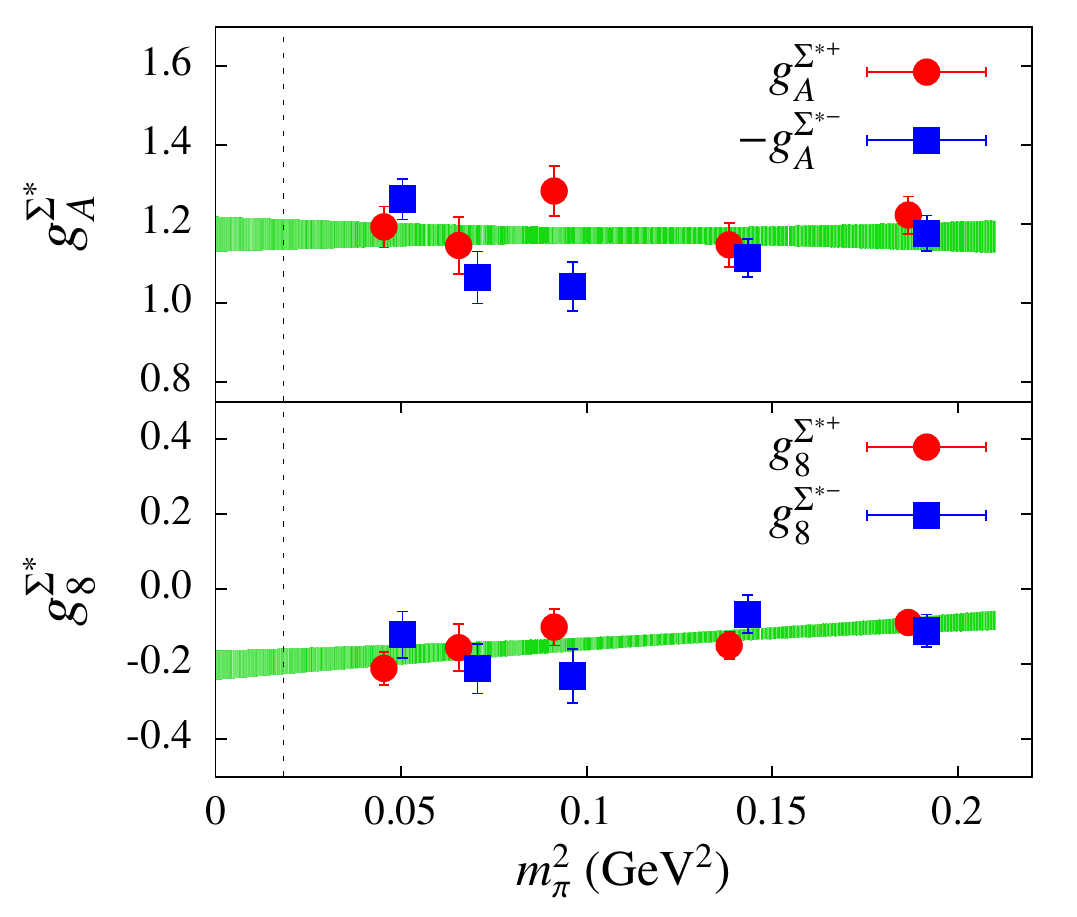}
\end{minipage}
\hfill
\begin{minipage}[t]{0.49\linewidth}
\includegraphics[width=1.01\linewidth]{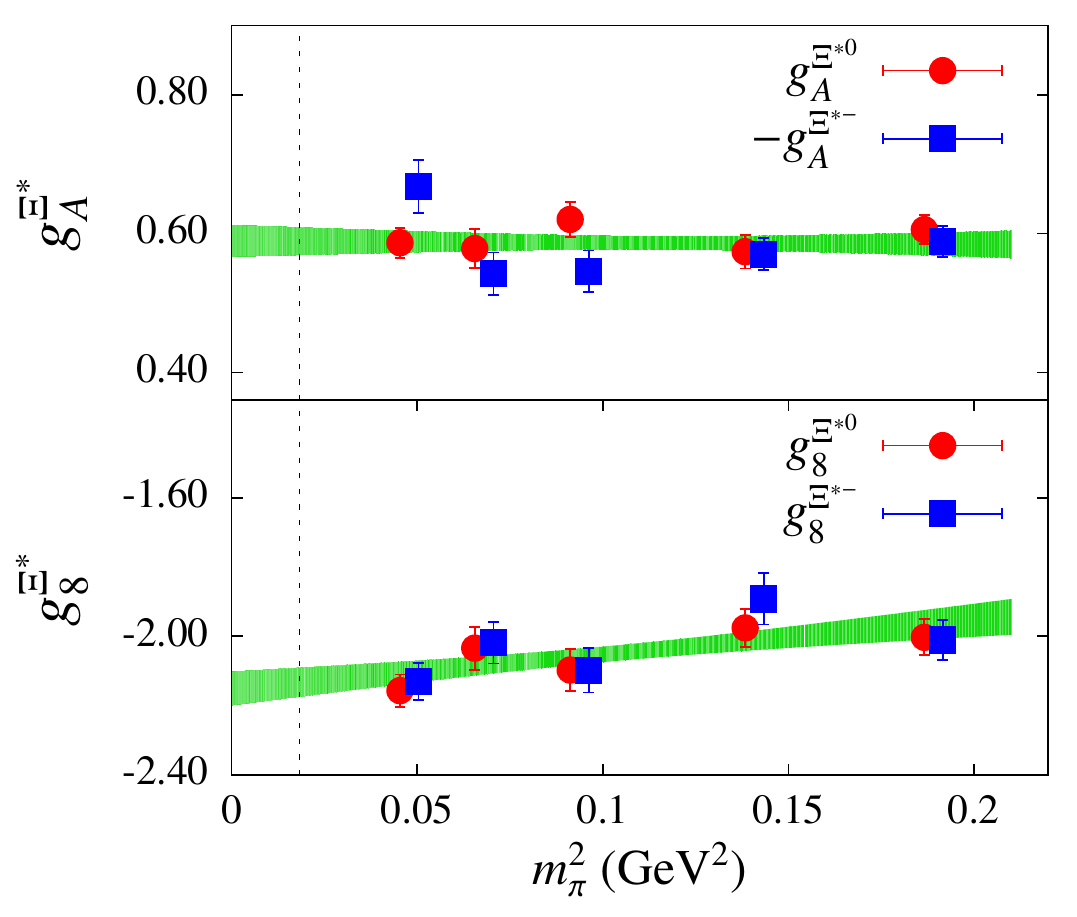}
\end{minipage}
\caption{\small The $\lambda_3$ and $\lambda_8$ combinations for the decuplet $\Sigma^*$ (left) and $\Xi^*$ (right) isospin multiplets. }
\label{fig:gA_sigmastar_xistar}
\end{figure}
\begin{figure}[!h]
\includegraphics[width=0.5\linewidth]{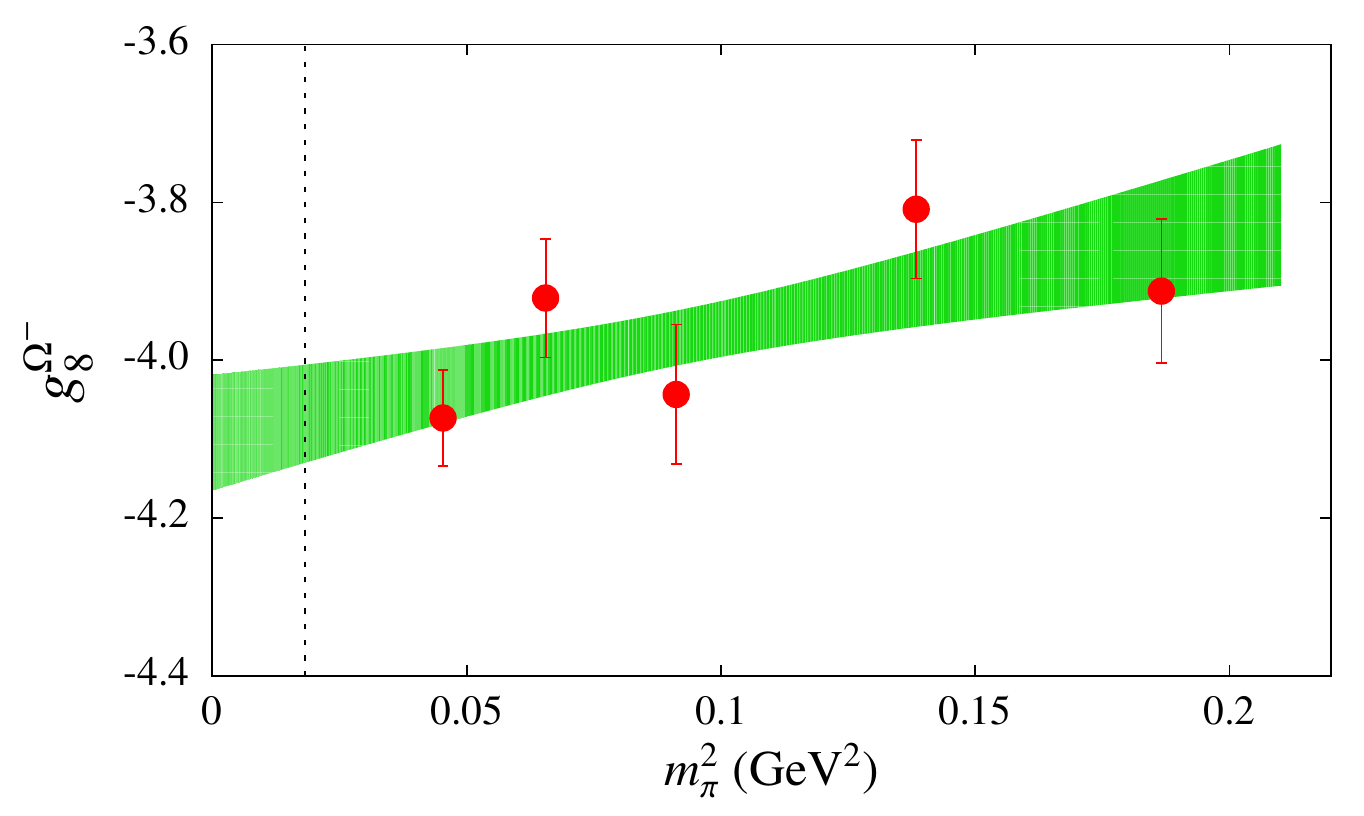}
\caption{The axial coupling $\lambda_8$ for the triply strange $\Omega^-$ baryon.}
\label{fig:gA_omega}
\end{figure}

In \fig{fig:gA_sigmastar_xistar} and~\fig{fig:gA_omega}  we show representative results for the rest of the decuplet baryons, namely the $\Sigma^*$ and $\Xi^*$ multiplets as well as the triply strange $\Omega^-$ baryon. As expected, the $\lambda_8$ axial coupling increases with the strangeness of the baryon being largest for the $\Omega^-$. We note here that the experimental measurement of the $\Omega^-$ axial charge is feasible, since it decays only via weak interactions and it has a relatively long lifetime compared to the other hyperons. The axial charges of the decuplet baryons feature weak pion mass dependence and no isospin symmetry breaking effects within our statistical accuracy.

As in the octet case, we perform a chiral extrapolation keeping the leading order  $m_\pi^2$-term. The extrapolated values at the physical point are collected in~\tbl{Table:extrap_values} of Appendix C.

In the absence of experimental or lattice QCD data for the decuplet axial couplings, we can only compare with estimates from effective field theories (EFT). As already mentioned, the isovector axial couplings of $\Delta$, $\Sigma^*$ and $\Xi^*$ can be expressed at tree-level in terms of a single low energy constant (LEC) as given in~\eq{eq:decuplet_H_dep}. Results from HB$\chi$PT~\cite{Jenkins:1991es} and Ref.~\cite{Butler:1992pn} quote values $|\mathcal{H}| = 1.9\pm 0.7$ and $|\mathcal{H}| = 2.2\pm 0.6$ respectively. The large errors on $\mathcal{H}$ make our results for $g_A^\Delta$, $g_A^{\Sigma^*}$ and $g_A^{\Xi^*}$ compatible with these calculations. However, a calculation within
relativistic constituent quark models (RCQM)~\cite{Choi:2010ty}  yields a larger value as compared to our result at the physical point, especially for $\Sigma^*$ and $\Xi^*$. The RCQM model yields values $g_A^{\Delta} = 2.20$, $g_A^{\Sigma^*} = 1.49$ and $g_A^{\Xi^*} = 0.75$, which are clearly higher than our values. As already mentioned, this is also the case with the octet baryons. We note here that due to different definitions of the axial-vector matrix elements, the values quoted here for $\Delta$ and $\Sigma^*$ are different by a factor of -2 and $-1/\sqrt{2}$, respectively, from the original paper.


\subsection{Charmed baryons axial charges}
\begin{figure}[!ht]
\begin{minipage}[t]{0.49\linewidth}
\includegraphics[width=\linewidth]{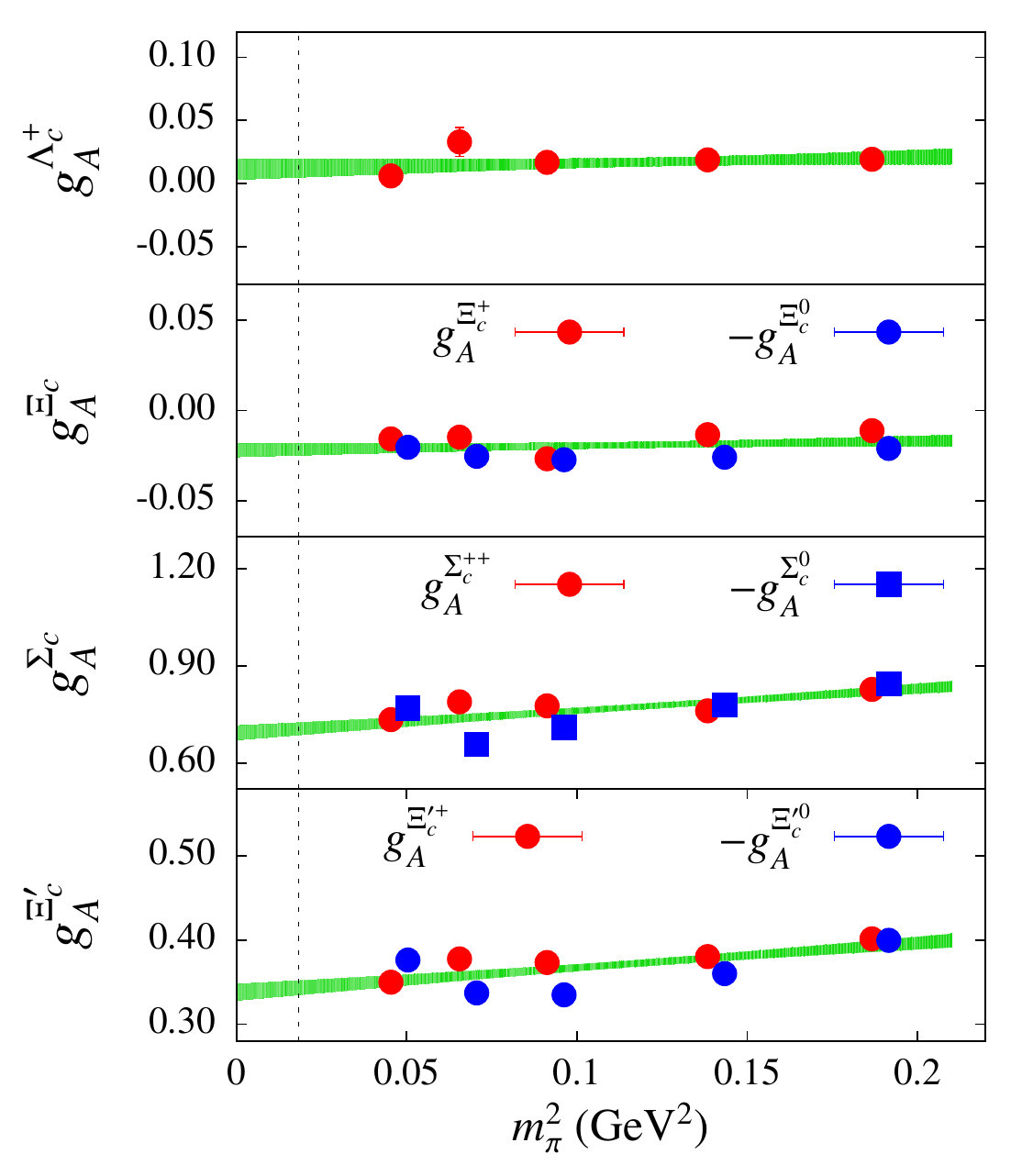}
\end{minipage}
\hfill
\begin{minipage}[t]{0.49\linewidth}
\includegraphics[width=\linewidth]{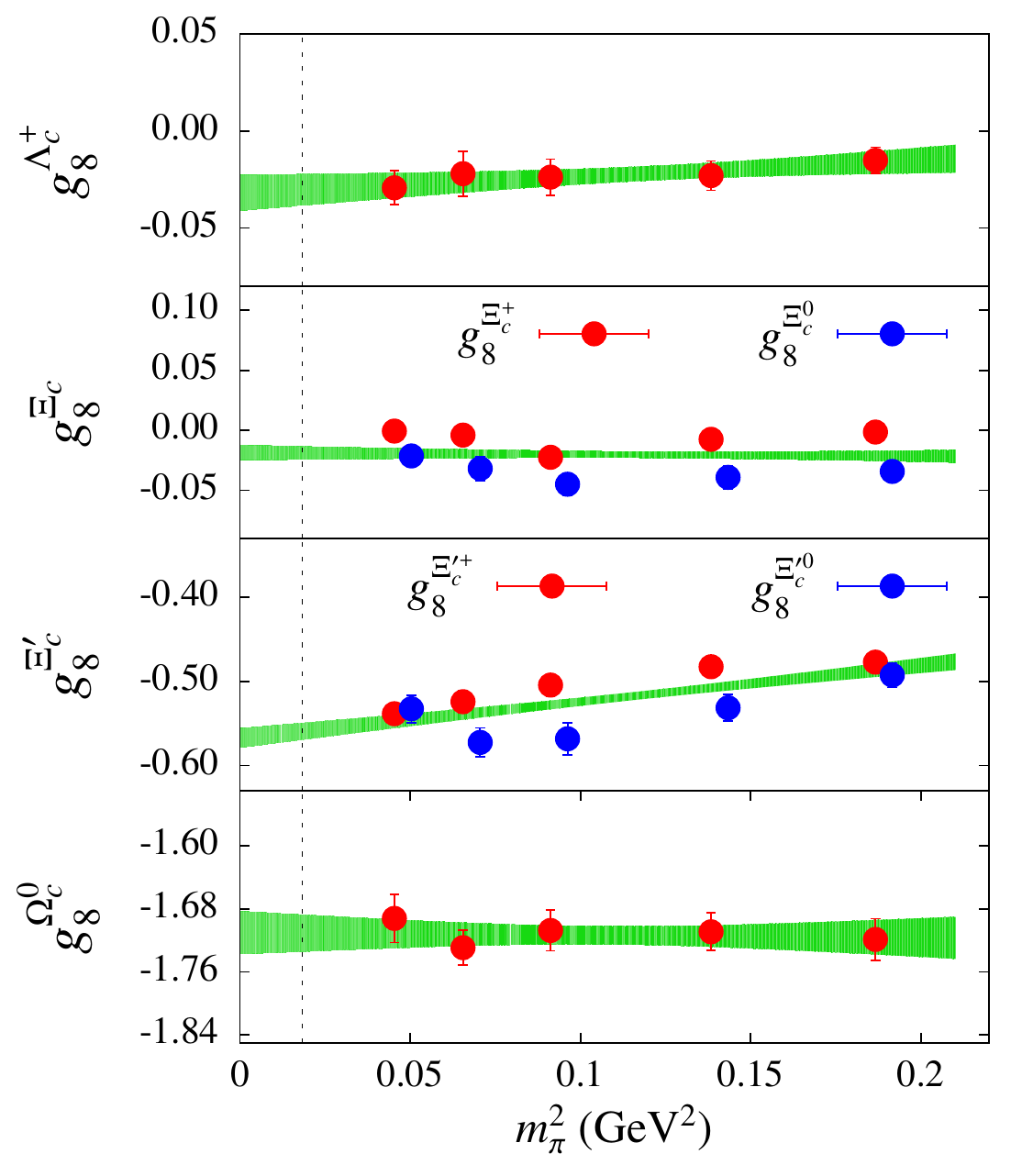}
\end{minipage}
\hfill
\includegraphics[width=0.49\linewidth]{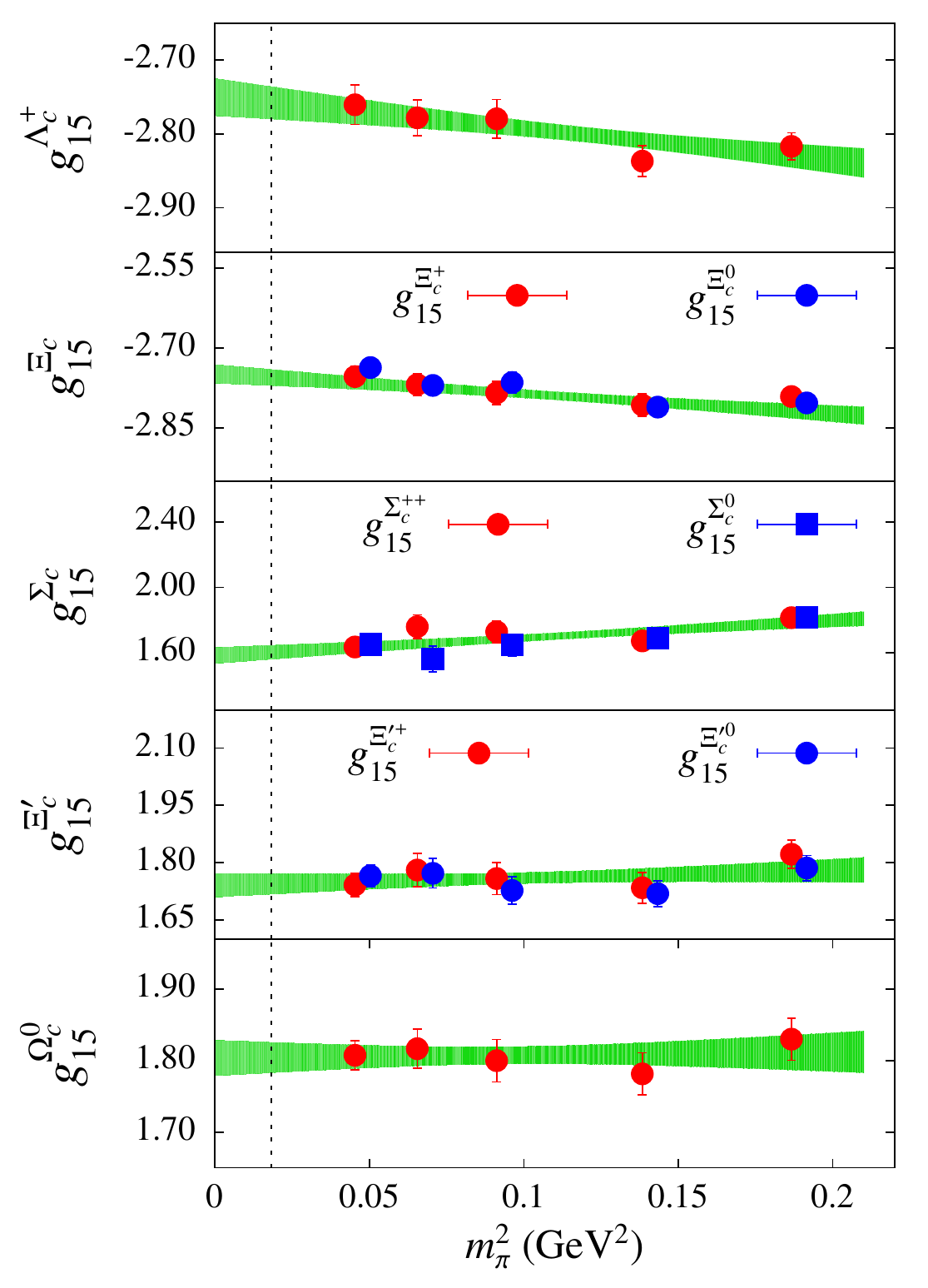}
\caption{\small Representative results on the axial charges of the singly charmed spin-1/2 baryons, for the $\lambda_3$ (top left), $\lambda_8$ (top right) and $\lambda_{15}$ (bottom) flavor combinations. For the $\Sigma_c^{++}$ and $\Sigma_c^0$ states we do not include the $\lambda_8$ combination, because it is the same as the $\lambda_3$, up to disconnected contributions. Similarly, the $\lambda_3$ combination for $\Omega_c^0$ is purely disconnected.}
\label{fig:charm_12_1}
\end{figure}  

\begin{figure}[!ht]
\begin{minipage}[t]{0.49\linewidth}
\includegraphics[width=1.005\linewidth]{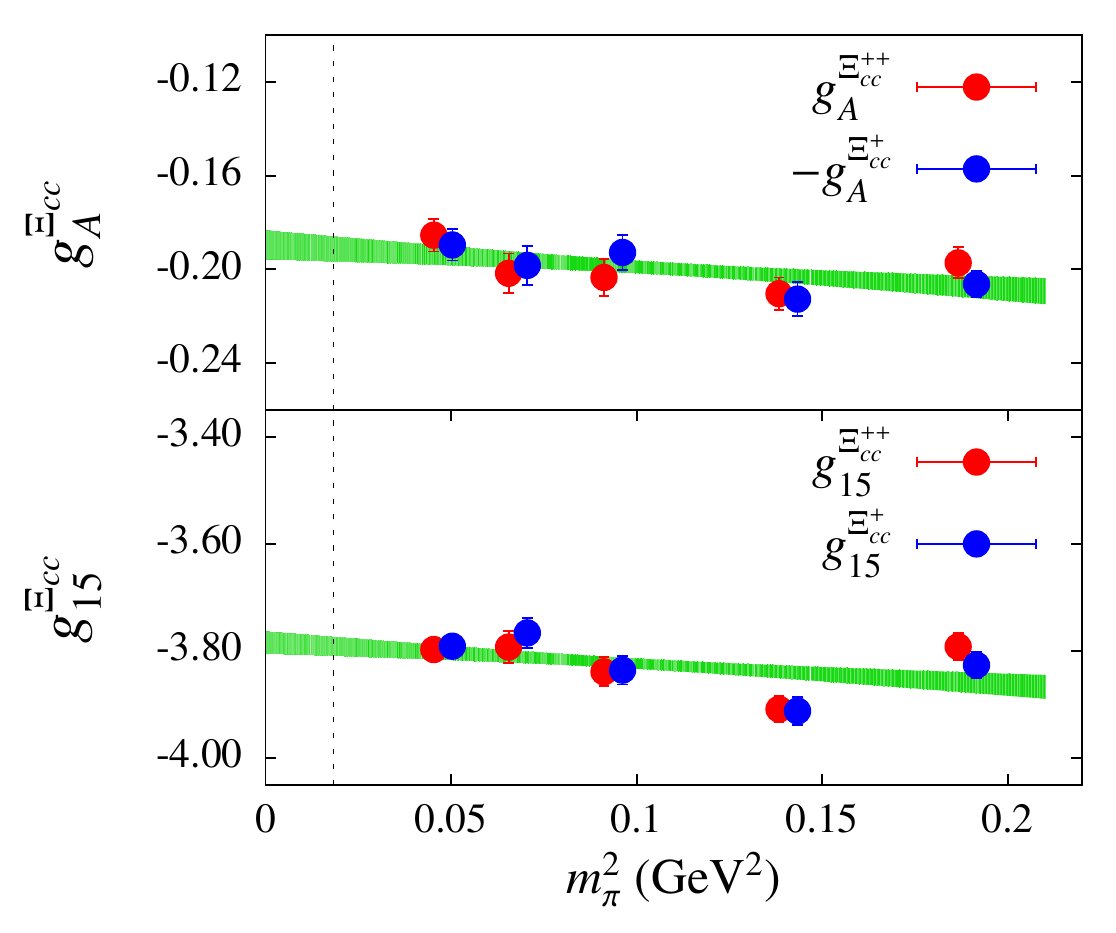}
\end{minipage}
\hfill
\begin{minipage}[t]{0.49\linewidth}
\includegraphics[width=\linewidth]{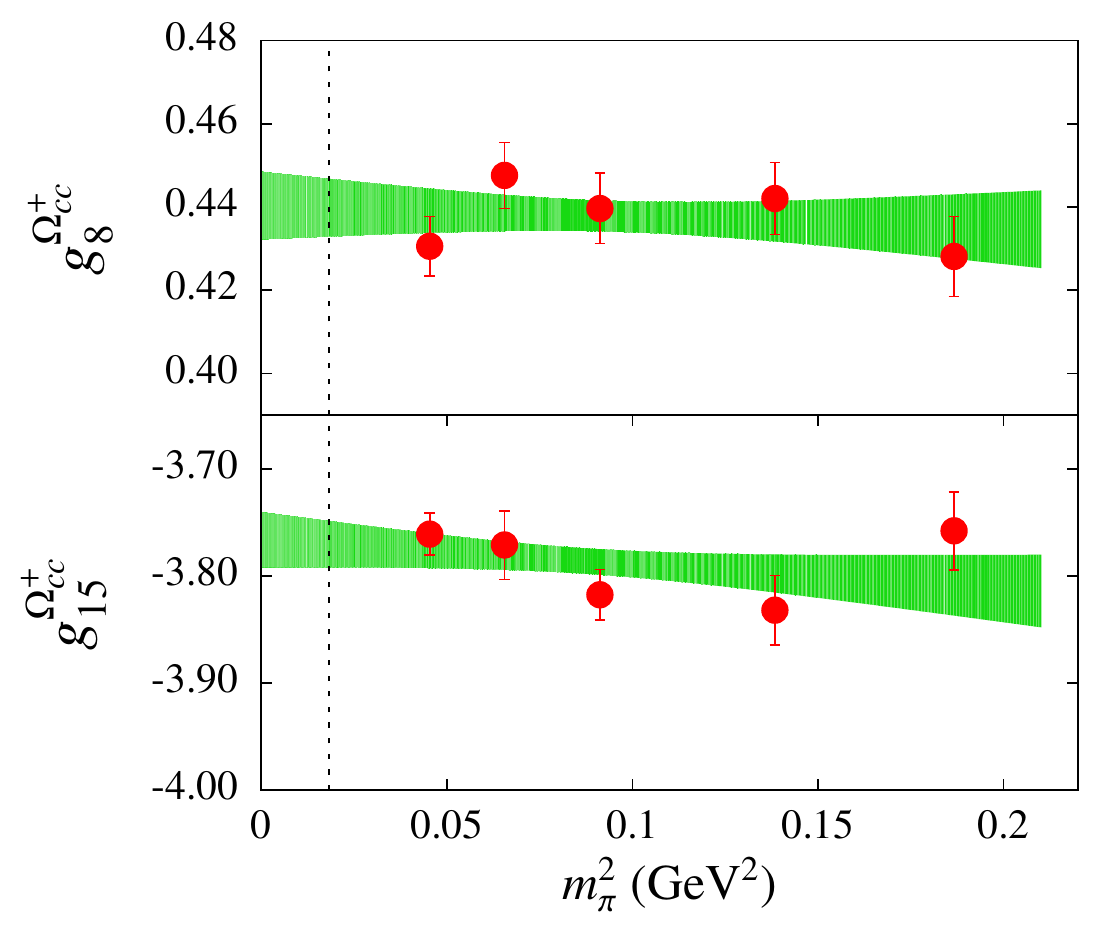}
\end{minipage}
\caption{\small Representative results on the axial charges of the doubly charmed spin-1/2 $\Xi_{cc}$ (left) and $\Omega_{cc}^+$ (right) baryons. The $\lambda_8$ flavour combination for the $\Xi_{cc}$ states is the same as the $\lambda_3$, up to disconnected contributions. The $\lambda_3$ combination for $\Omega_{cc}^+$ is purely disconnected.}
\label{fig:charm_12_2}
\end{figure}  
\begin{figure}[!ht]
\begin{minipage}[t]{0.49\linewidth}
\includegraphics[width=\linewidth]{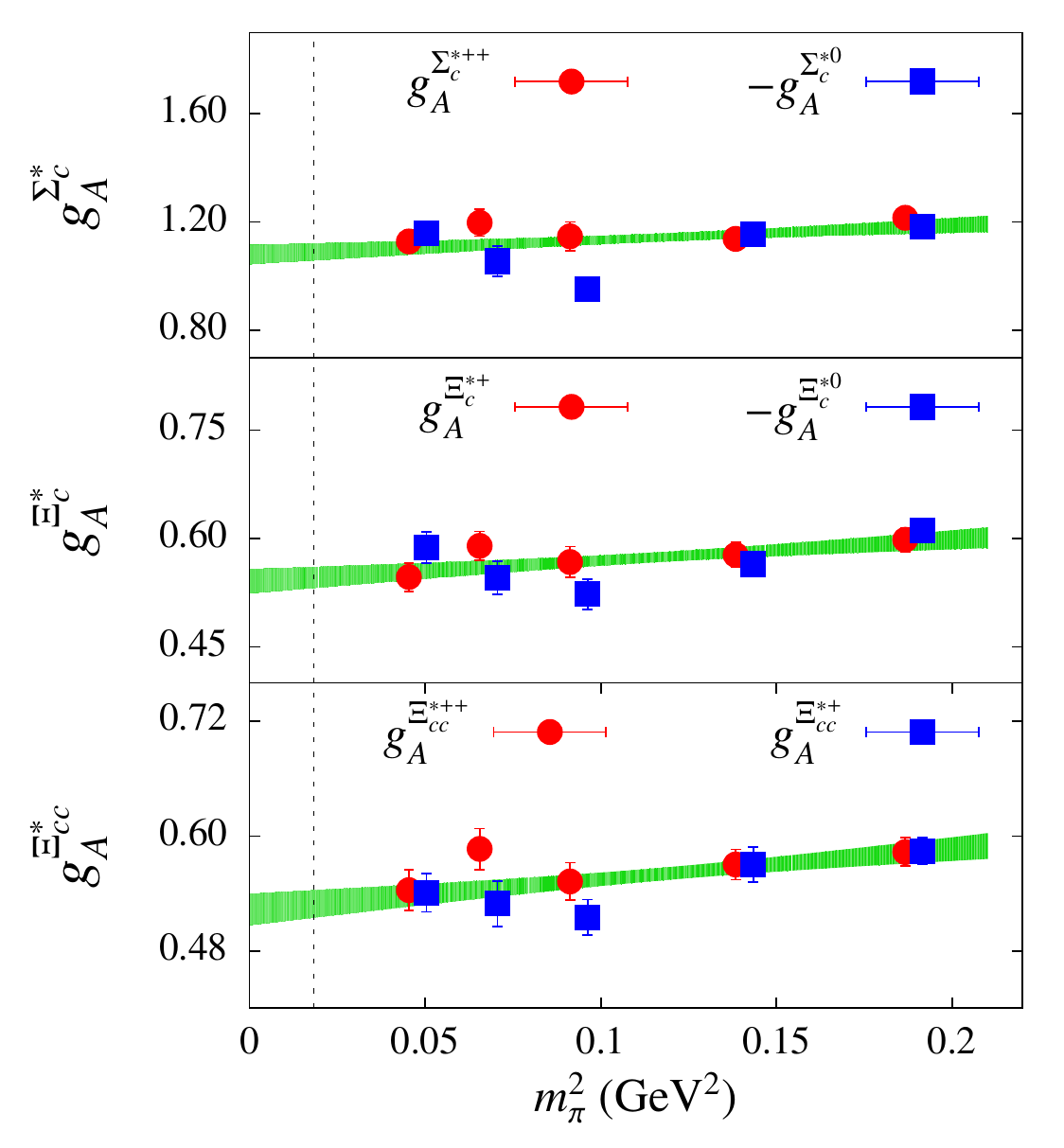}
\end{minipage}
\hfill
\begin{minipage}[t]{0.49\linewidth}
\includegraphics[width=\linewidth]{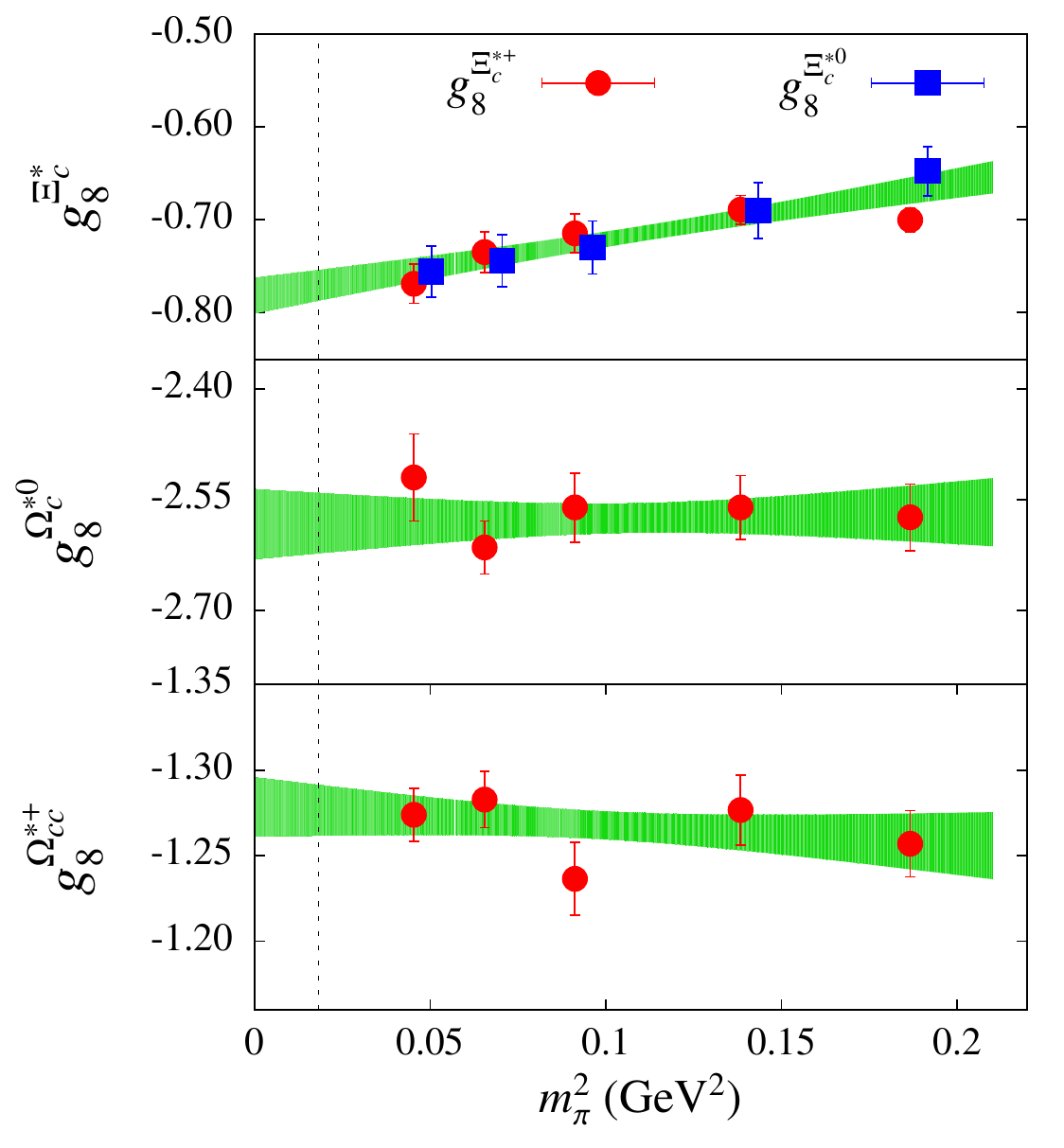}
\end{minipage}
\hfill
\begin{minipage}[t]{0.49\linewidth}
\includegraphics[width=\linewidth]{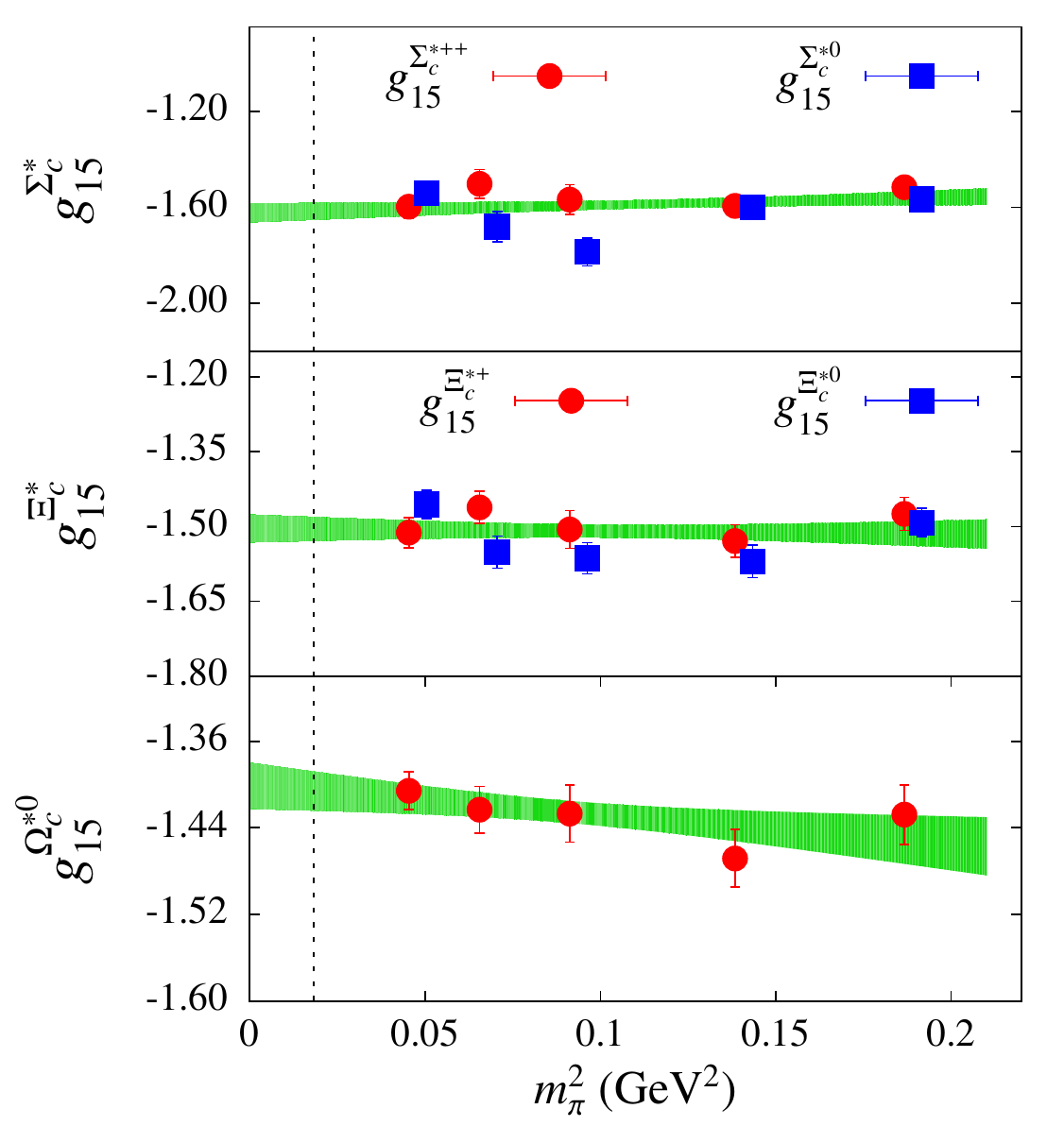}
\end{minipage}
\hfill
\begin{minipage}[t]{0.49\linewidth}
\includegraphics[width=\linewidth]{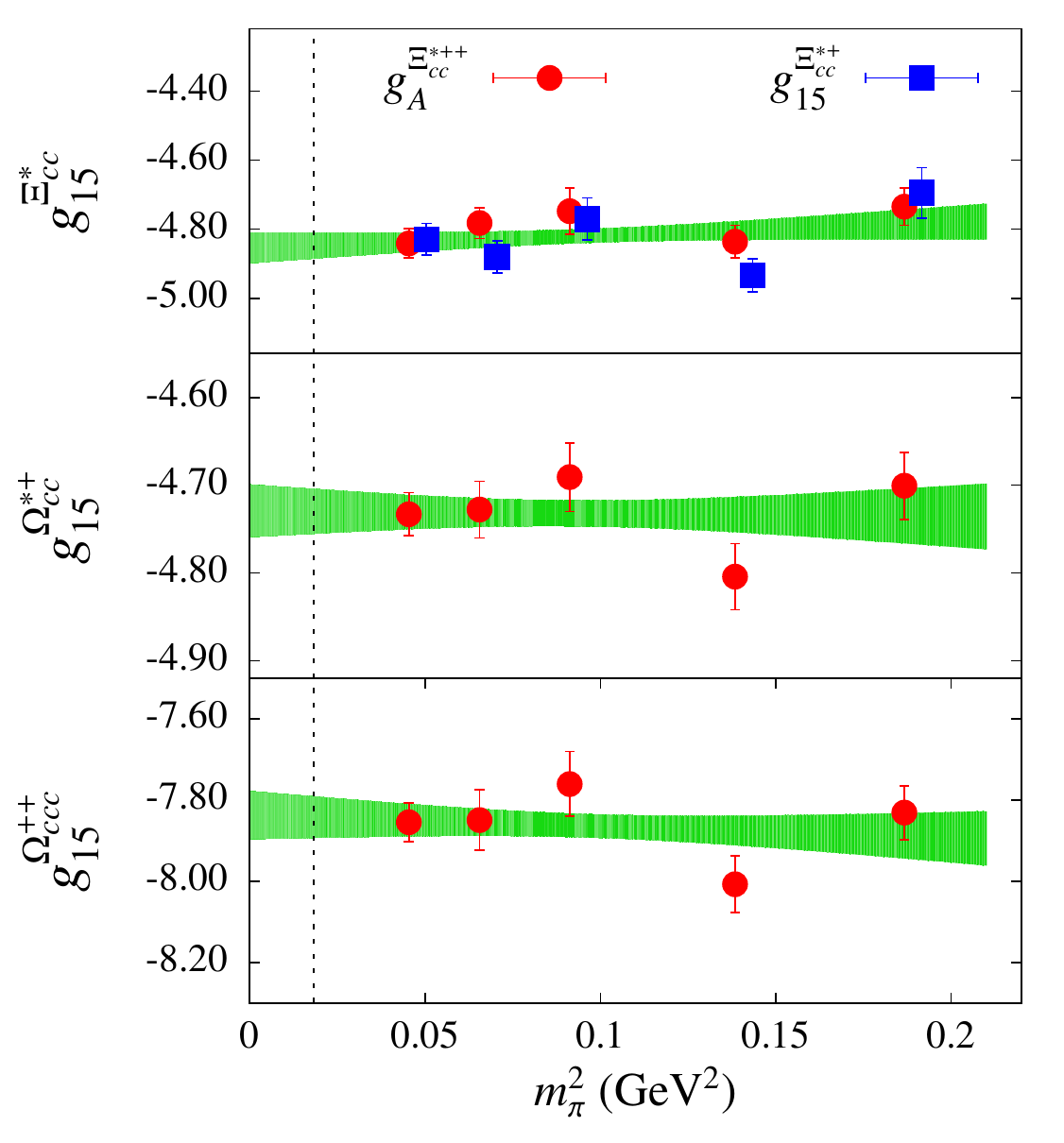}
\end{minipage}
\hfill
\caption{\small Representative results on the axial charges of the spin-3/2 charm baryons, for the $\lambda_3$ (top left), $\lambda_8$ (top right) and $\lambda_{15}$ (bottom left and right) flavour combinations. As in the spin-1/2 case, we do not include the $\lambda_8$ combination for the $\Sigma_c^{*++}$ and $\Sigma_c^{*0}$ states, as it is the same as the $\lambda_3$, up to disconnected contributions. The same holds for the $\Xi_{cc}^*$ states. The $\lambda_3$ flavour combination for the $\Omega_c^{*0}$ and $\Omega_{cc}^{*+}$ baryons is purely disconnected.}
\label{fig:charm_32}
\end{figure}   

Charmed baryons have first been observed in experiments in the 1970s, however, four of them namely the $\Omega_{cc}$, $\Xi_{cc}^*$, $\Omega_{cc}^*$ and $\Omega_{ccc}^{++}$, predicted by the quark model, have not yet been observed. There are recent results on the  charmed baryon spectrum within the lattice QCD framework showing agreement among different fermion discretization schemes~\cite{Alexandrou:2014sha,Padmanath:2015jea,Bali:2015lka}. There was also a recent lattice QCD study for the electromagnetic form factors of charmed baryons~\cite{Can:2015exa}. However, there is to date  no computation of the axial form factors. Results on the axial couplings for singly charmed baryons have  been obtained within heavy baryon chiral perturbation theory~\cite{Jiang:2014ena}, which  can thus provide a comparison to the results of this work.

We show representative results for spin-1/2 charmed baryons  in Figs.~\ref{fig:charm_12_1} and~\ref{fig:charm_12_2} and for spin-3/2 charmed baryons in~\fig{fig:charm_32}. As can be seen, the charm baryon axial charges do not show a strong pion mass dependence and there is no  breaking of the isospin symmetry due to cut-off effects between isospin partners. As for the strange sector 
  we perform linear fits using the Ansatz $a+bm_\pi^2$, which give rise to the green bands in Figs.~\ref{fig:charm_12_1} and~\ref{fig:charm_12_2}. The values
extracted from the fits yield  the axial couplings at the physical pion mass. 
We collect these values  in~\tbl{Table:extrap_values} of Appendix C.

\begin{figure}[!ht]
\includegraphics[width=0.49\linewidth]{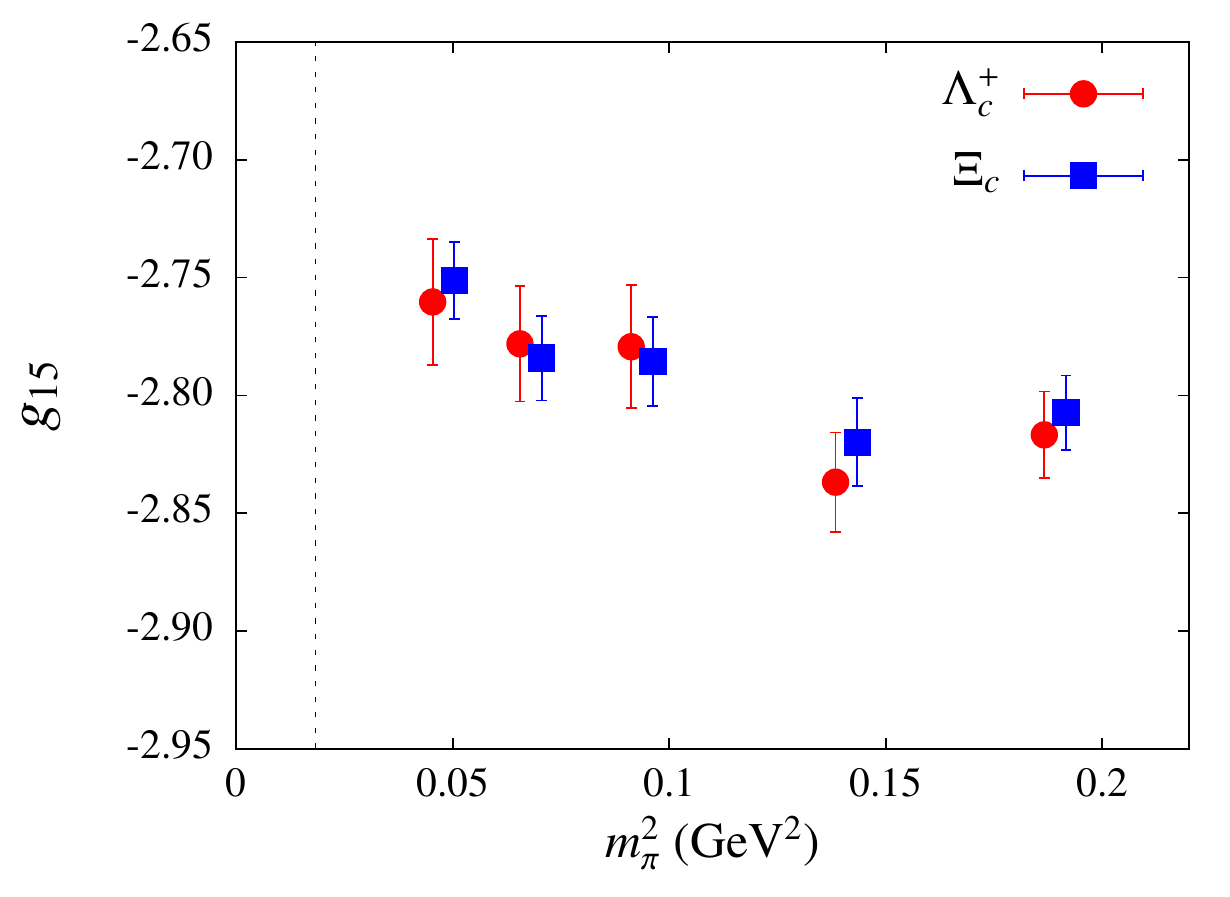}\\
\begin{minipage}[t]{0.49\linewidth}
\includegraphics[width=\linewidth]{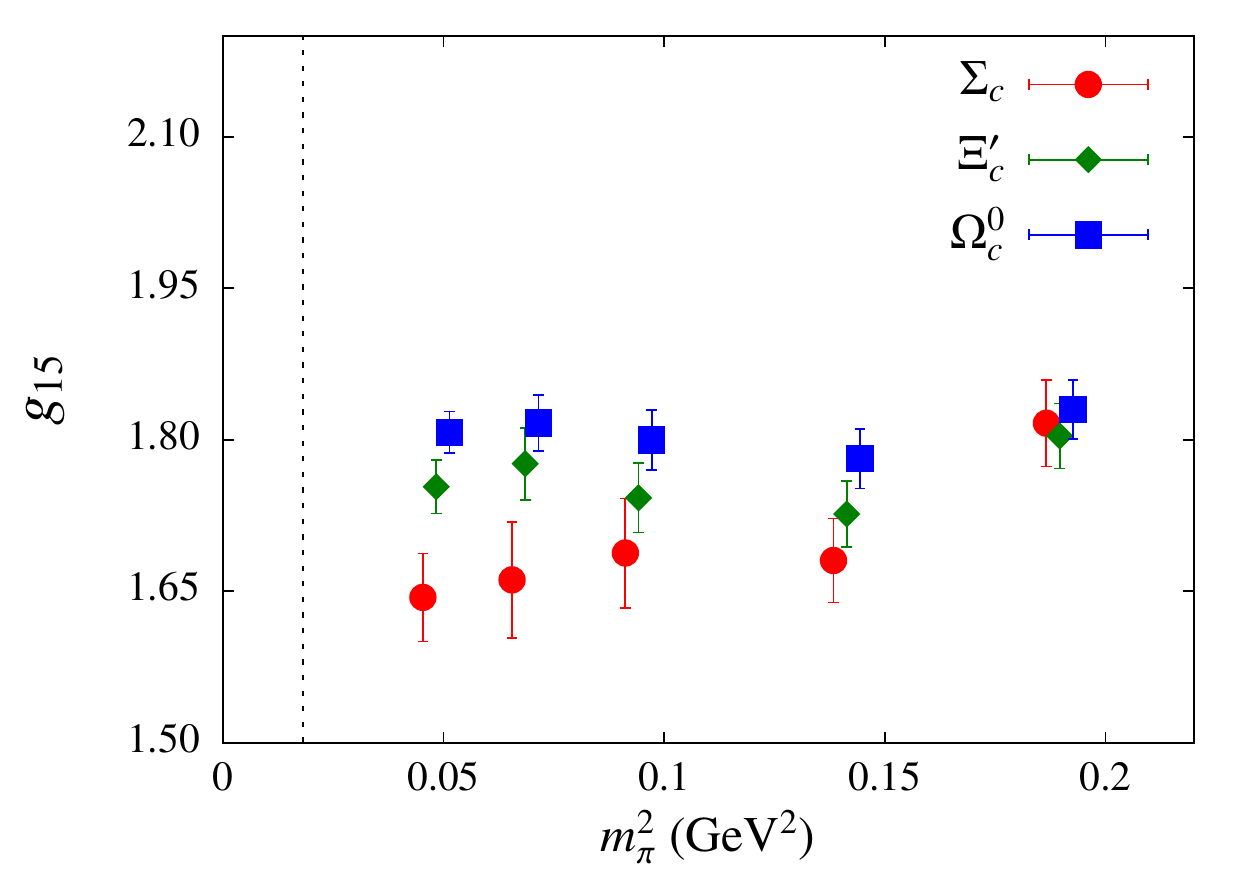}
\end{minipage}
\hfill
\begin{minipage}[t]{0.49\linewidth}
\includegraphics[width=\linewidth]{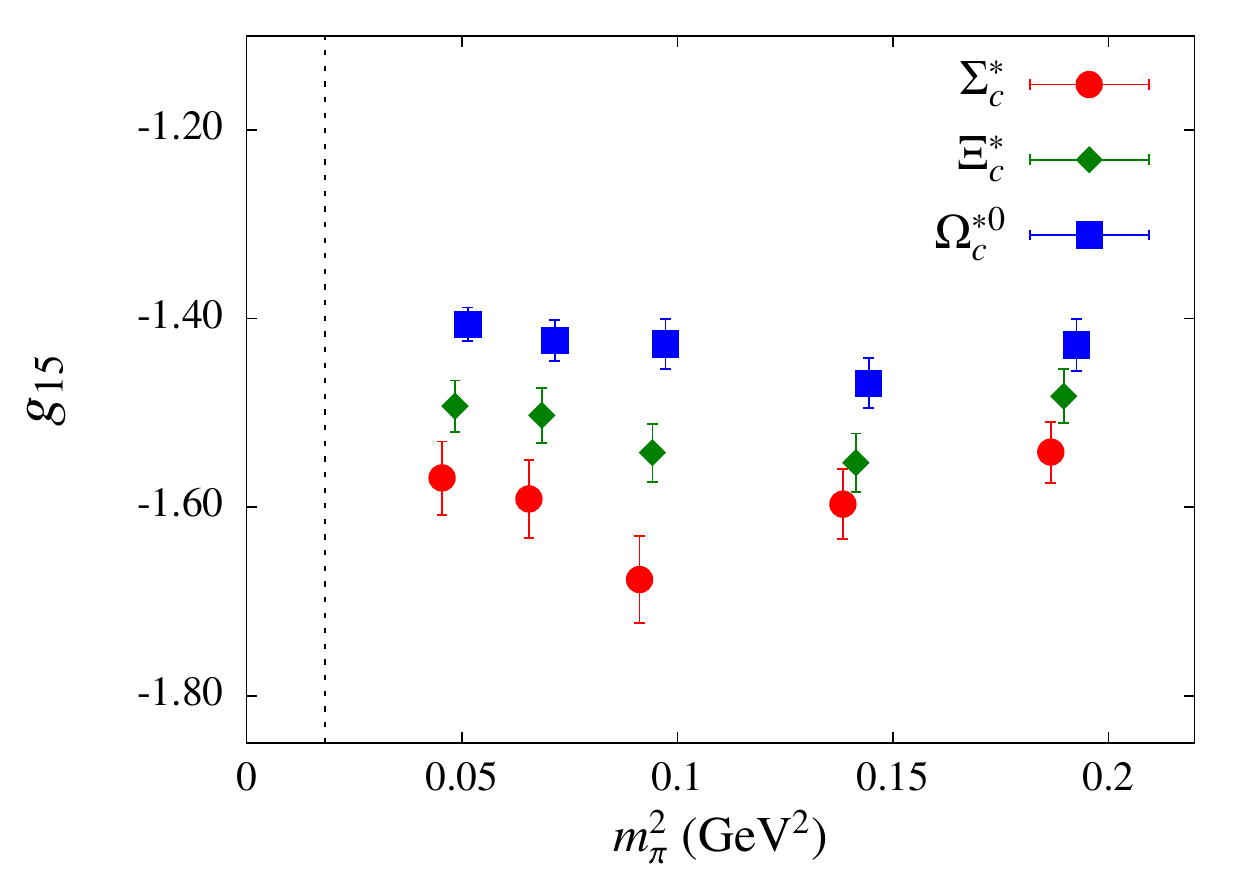}
\end{minipage}
\caption{\small Top: Comparison of the $g_{15}$ couplings of the antisymmetric $\bf\bar{3}$-plet. Bottom: Comparison of the $g_{15}$ couplings for the symmetric $\bf 6$-tets of the singly charmed spin-1/2 (left) and spin-3/2 (right) baryons. In all plots the average $g_{15}$ coupling over the various isospin partners is shown.}
\label{fig:charm1_comps}
\end{figure}

Since we only consider diagonal matrix elements, only the couplings $g_1$, $g_5$ and $g_6$ appearing in \eq{eq:eff_lagr_charm} can be probed. Conservation of  angular moment and parity forbids the coupling of pseudoscalar mesons with the $\bf\bar{3}$-plet charmed baryons. This implies that $g_6=0$~\cite{Jiang:2014ena}, therefore $g_A^{\Lambda_c^+} = g_A^{\Xi_c} = 0$. Our lattice results for $\Lambda_c^+$ and $\Xi_c$ show very small non-zero values and a tendency towards zero at the physical pion mass, consistent with the HB$\chi$PT prediction. Another interesting observation is that the $g_{15}$ couplings of $\Lambda_c^+$ and $\Xi_c$ have similar values for all our pion masses. On the other hand, the $g_{15}$ couplings of the symmetric spin-1/2 $\bf 6$-tet exhibit small splittings as the physical pion mass is approached. This effect is also present in the spin-3/2 $\bf 6$-tet, where the $g_{15}$ couplings of $\Sigma_c^*$, $\Xi_c^*$ and $\Omega_c^{*0}$ exhibit similar splitting patterns. The aforementioned comparisons are explicitly shown in~\fig{fig:charm1_comps}.

\begin{figure}[!ht]
\begin{minipage}[t]{0.49\linewidth}
\includegraphics[width=\linewidth]{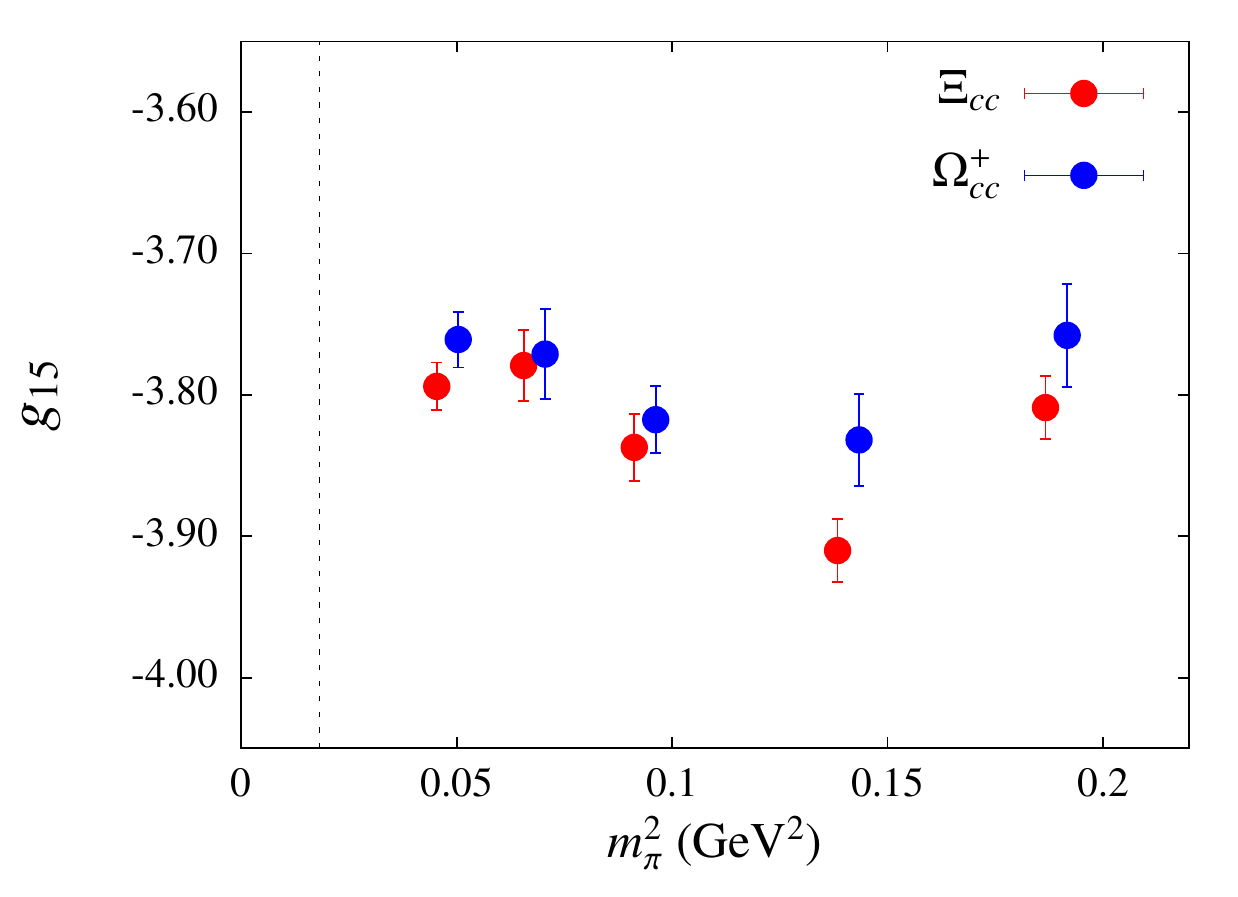}
\end{minipage}
\hfill
\begin{minipage}[t]{0.49\linewidth}
\includegraphics[width=\linewidth]{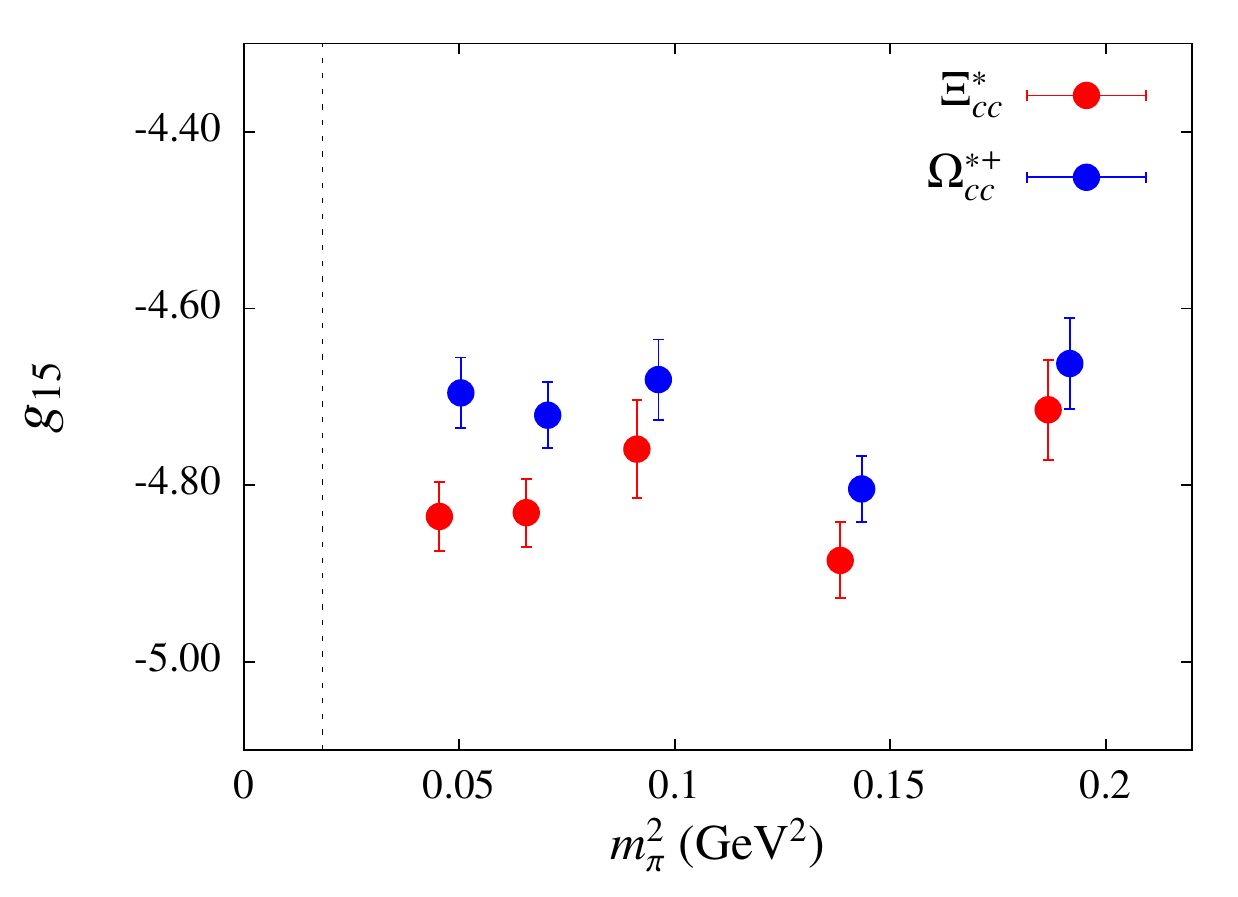}
\end{minipage}
\caption{\small Comparison of the $g_{15}$ couplings of the doubly charmed spin-1/2 (left) and spin-3/2 (right) baryons. The average value of $g_{15}$ for the $\Xi_{cc}$ and $\Xi_{cc}^*$ states is shown.}
\label{fig:charm2_comps}
\end{figure}

In the case of the symmetric $\bf 3$-plets of the doubly charmed baryons, we show the $g_{15}$ couplings in~\fig{fig:charm2_comps}. As can be seen,  the couplings for the  spin-1/2 $\Xi_{cc}$ and $\Omega_{cc}^+$ have similar values as it is also approximately the case for the corresponding spin-3/2 states.


\subsection{$SU(3)$ flavour symmetry breaking} 

\begin{figure}[!ht]
\begin{minipage}[t]{0.49\linewidth}
\includegraphics[width=\linewidth]{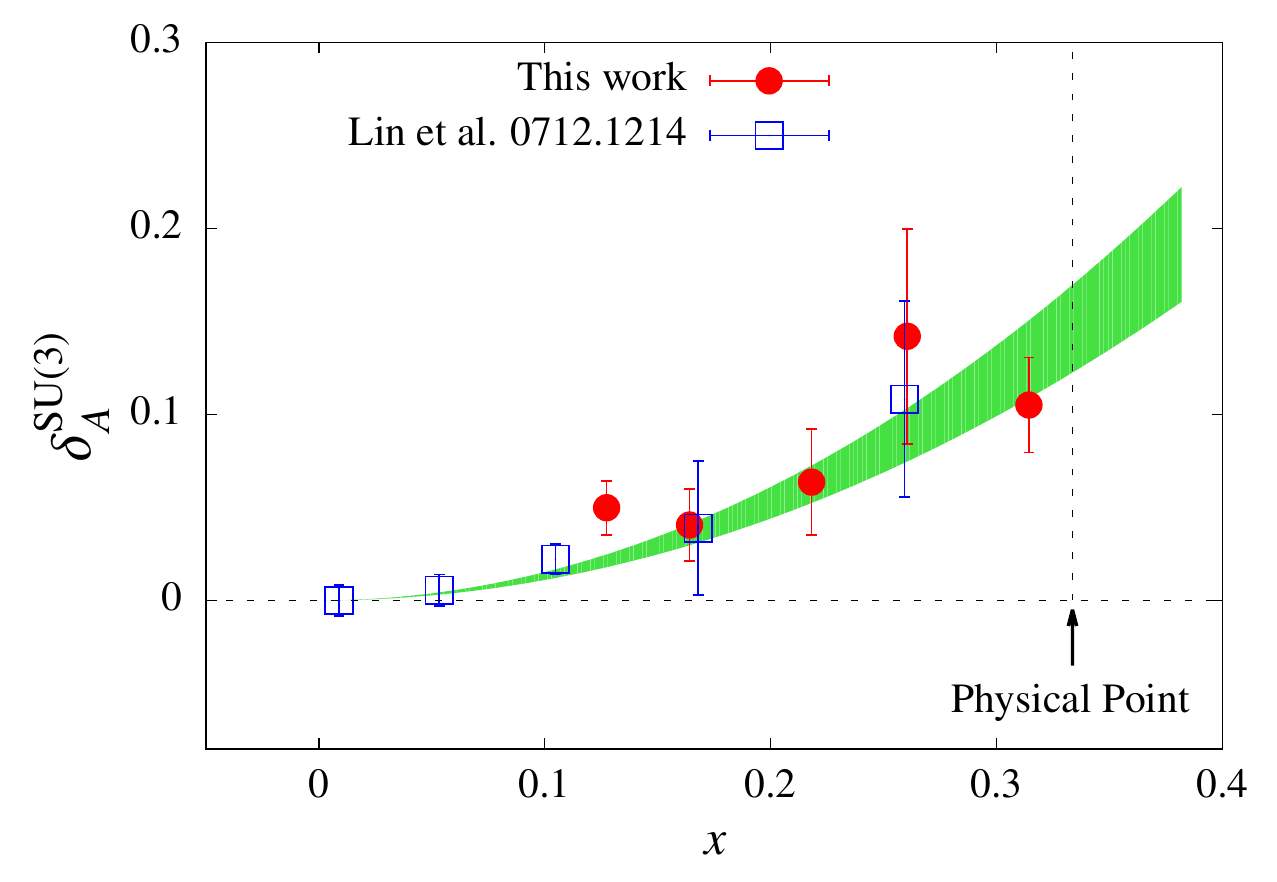}
\end{minipage}
\hfill
\begin{minipage}[t]{0.49\linewidth}
\center
\includegraphics[width=\linewidth]{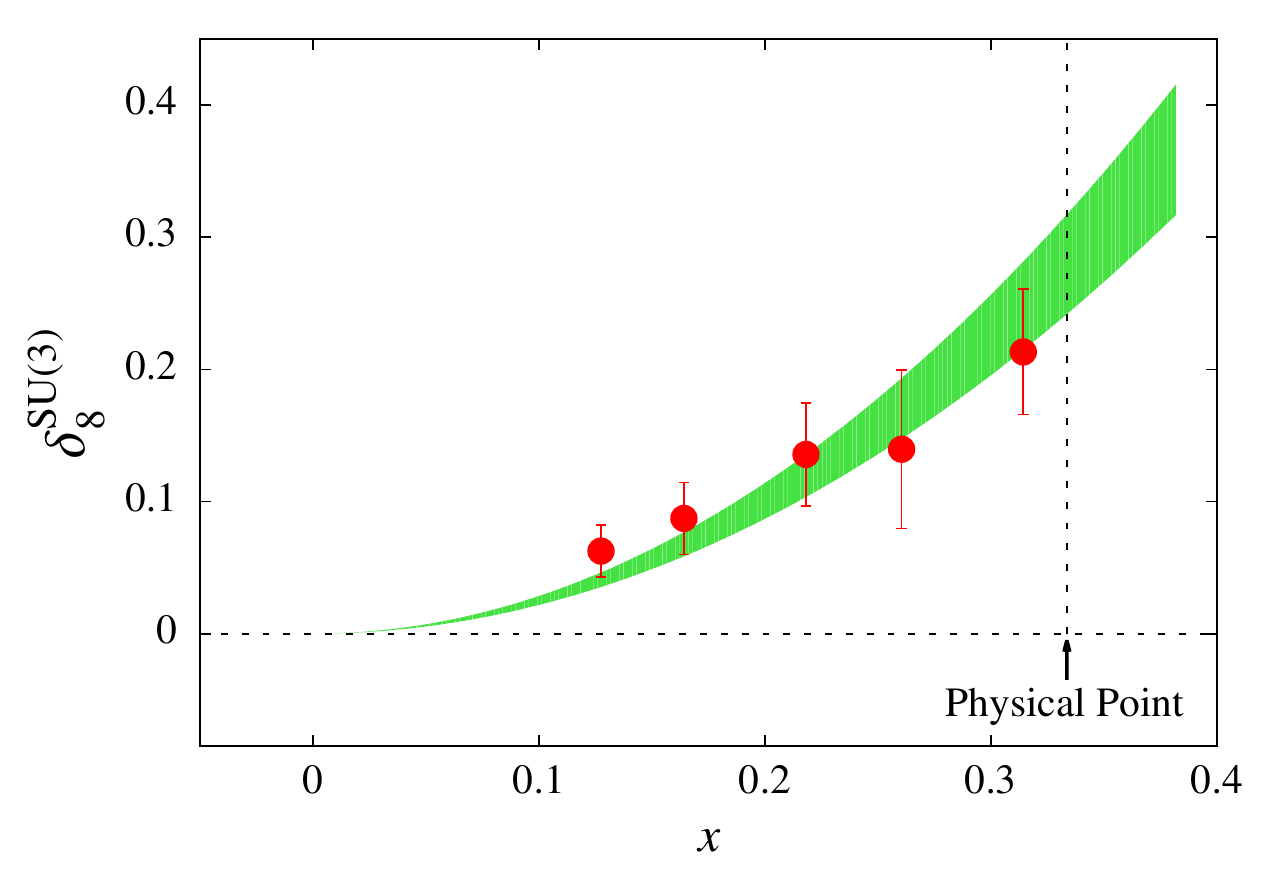}
\end{minipage}
\caption{\small Left: The $SU(3)$ flavour symmetry breaking for the octet as a function of the breaking parameter $x$ from our results (red circles). Results from Ref.~\cite{Lin:2007ap} are also shown for comparison in open blue squares. Right: The $SU(3)$ flavour symmetry breaking as a function of the breaking parameter $x$ using \eq{eq:su3_breaking_eta2}. In both cases, the green band represents a quadratic fit to the data of this work. }
\label{fig:su3_octet}
\end{figure}

Having results for different pion masses enables us to  examine SU(3) flavour symmetry breaking effects as a function of the breaking parameter $x = (m_K^2 - m^2_\pi)/(4\pi^2 f^2_\pi)$. 
The tree-level relations in terms of the low-energy constants (LEC) $D$ and $F$ can be
written as
\begin{equation}\label{eq:lec_octet}
 g_A^N = F + D + \sum_n C^{(n)}_N x^n, \;\;\; g_A^\Sigma = 2 F + \sum_n C_{\Sigma}^{(n)} x^n, \;\;\; g_A^\Xi = F - D + \sum_n C^{(n)}_\Xi x^n
\end{equation}
in correspondence with~\eq{eq:ax_param}. We define $\delta_A^{SU(3)}$ to be the quantity measuring $SU(3)$ symmetry breaking~\citep{Tiburzi:2008bk,Lin:2007ap}
\begin{equation}
\delta_A^{SU(3)} = g_A^N - g_A^\Sigma + g_A^\Xi  = \sum_n c_n x^n\;.
\label{eq:su3_breaking}
\end{equation}
In the SU(3) limit \eq{eq:lec_octet} reduces to \eq{eq:ax_param} and $\delta_A^{SU(3)}\rightarrow 0$. In \fig{fig:su3_octet} we show our results for $\delta_A^{SU(3)}$. Our data as well as the data from Ref.~\cite{Lin:2007ap}, also shown in the plot, suggest that $ \delta_A^{SU(3)}\sim x^2$. After fitting and extrapolating to the physical point, we find that the SU(3) breaking effects in the octet at the physical pion mass  amount to $(14.7\pm 2.4)\%$. In a recent study~\cite{Carrillo-Serrano:2014zta} using the Nambu-Jona-Lasinio model, the values of the  LECs are found to be $F_\Sigma = 0.441$, $D_\Sigma = 0.829$ and $F_\Xi = 0.496$, $D_\Xi = 0.774$, suggesting SU(3) breaking effects of around $10\%$, which is consistent in fact with our findings.

Similarly one can expand the $\lambda_8$ couplings in a terms of $x$, in correspondence with~\eq{eta-baryon}, as follows
\begin{eqnarray}
\label{eq:su3_breaking_eta}
&&g_8^N = - \frac{1}{\sqrt{3}} (D + 3F) + \sum_n C_N^{'(n)} x^n, \;\;\; g_8^\Lambda = - \frac{2}{\sqrt{3}} D  + \sum_n C_\Lambda^{'(n)} x^n, \nonumber \\
&& g_8^\Sigma = \frac{2}{\sqrt{3}} D  + \sum_n C_\Sigma^{'(n)} x^n, \;\;\; g_8^\Xi = - \frac{1}{\sqrt{3}} (D - 3F) + \sum_n C_\Xi^{'(n)} x^n,
\end{eqnarray}
where again the $SU(3)$ flavour symmetry is recovered as $x$ goes to zero. The corresponding SU(3) breaking can be probed via 
\begin{equation}
\label{eq:su3_breaking_eta2}
\delta_8^{SU(3)} = g_8^N + g_8^\Xi - \frac{g_8^\Lambda}{2} + \frac{g_8^\Sigma}{2} = \sum_n c^{\prime}_n x^n .
\end{equation}
We show in \fig{fig:su3_octet} the value of $\delta_8^{SU(3)}$ as a function of $x$. As can be seen, we observe larger SU(3) breaking affects for all values of $x$ up to the physical point where we find a value of $(28.2\pm 3.8)\%$, i.e. twice as large the result of the SU(3) breaking for the pion-baryon axial couplings. This is to be expected since in the $\lambda_8$ couplings the strange quark enters. 

\begin{figure}[!ht]
\begin{minipage}[t]{0.49\linewidth}
\includegraphics[width=\linewidth]{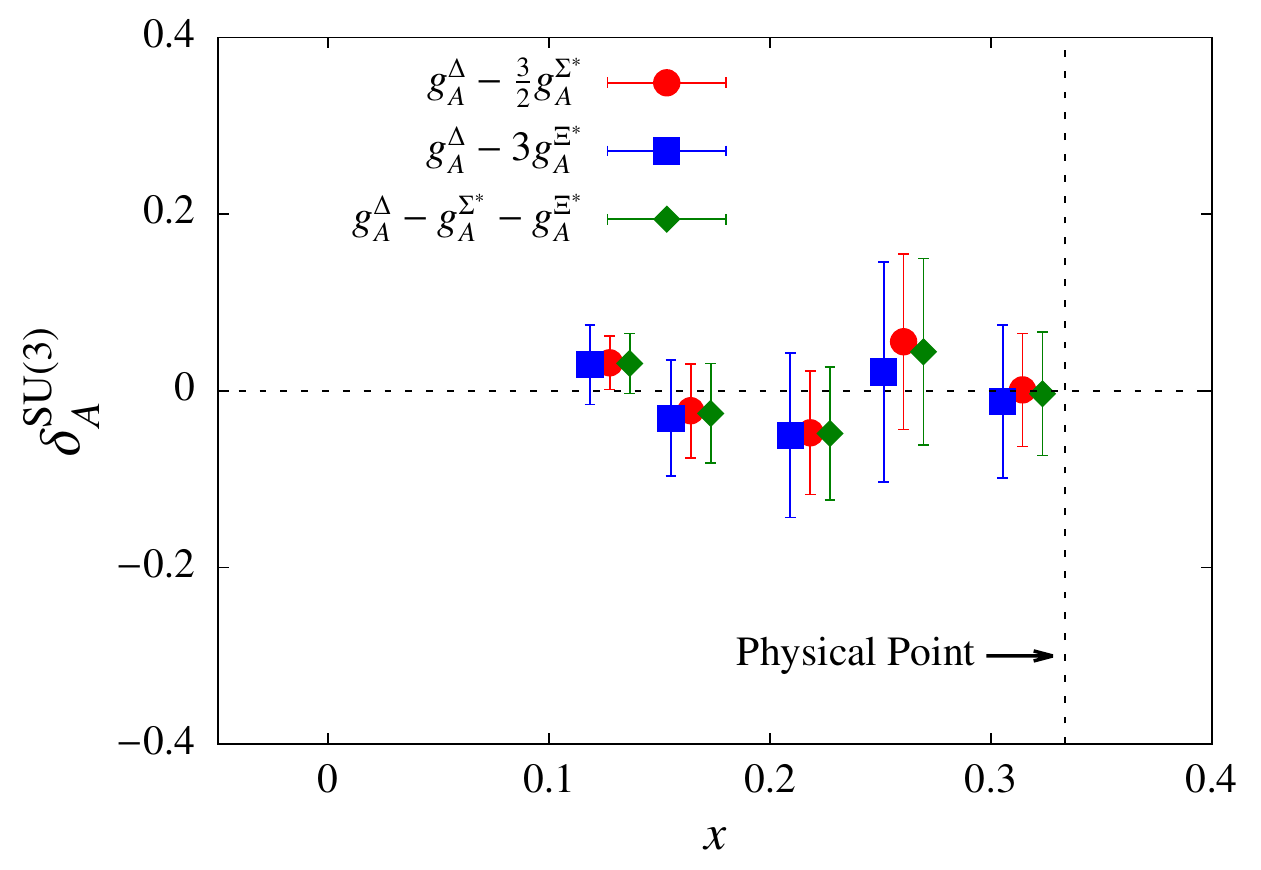}
\end{minipage}
\hfill
\begin{minipage}[t]{0.49\linewidth}
\center
\includegraphics[width=\linewidth]{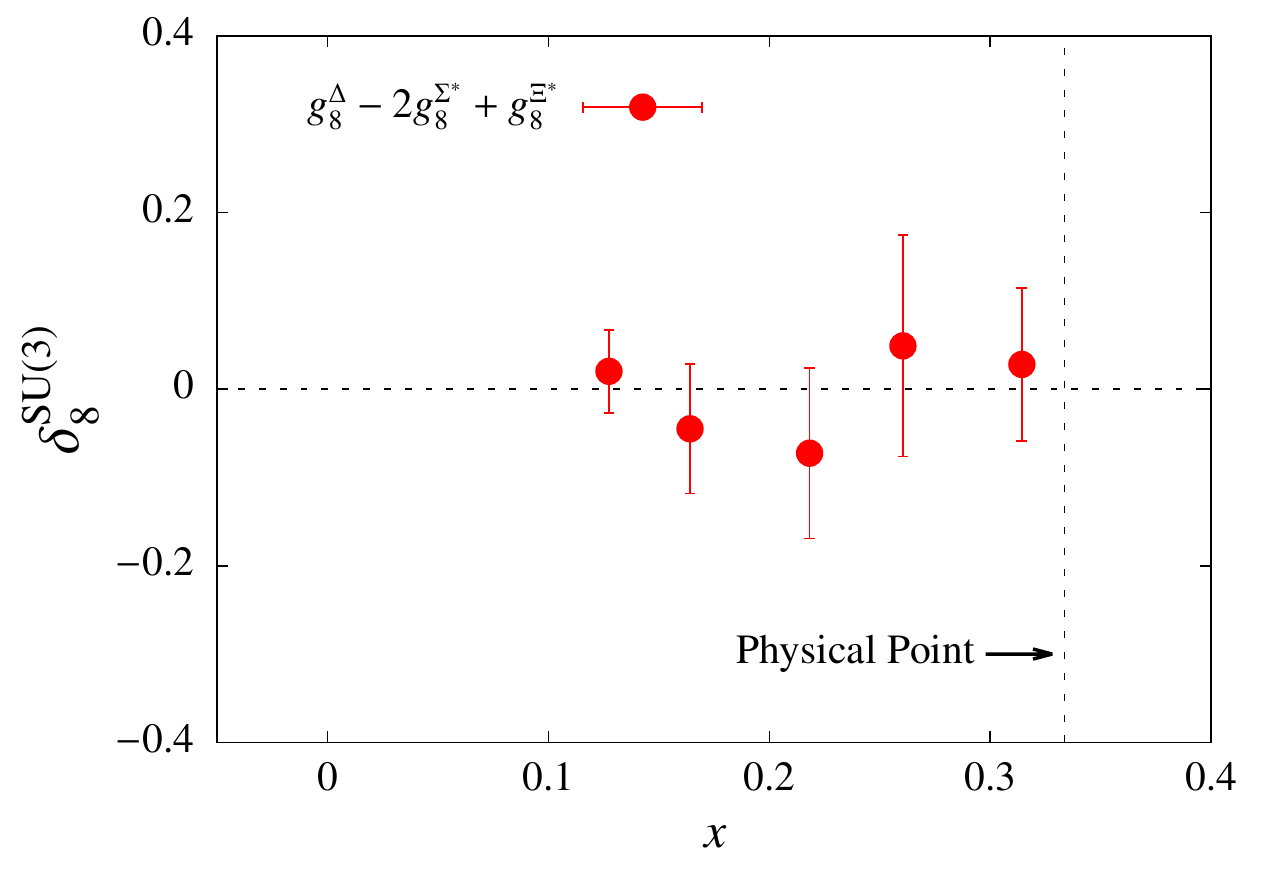}
\end{minipage}
\caption{The $SU(3)$ flavour symmetry breaking parameter as a function of the breaking parameter $x$ for the decuplet using~\eq{eq:su3_dec_1} (red circles),~\eq{eq:su3_dec_2} (blue squares) and~\eq{eq:su3_dec_3} (green diamonds) (left) and using~\eq{eq:su3_dec_eta8} (right).}
\label{fig:su3_flavour_breaking_decuplet}
\end{figure}

The same study can be carried out for the decuplet. The flavour symmetry breaking is given now from a combination of $\Delta$, $\Sigma^*$ and $\Xi^*$. Using chiral perturbation theory, the couplings of these baryons can be expressed in terms of a single LEC as \citep{Tiburzi:2008bk},
\begin{equation}
g_A^\Delta = H + \sum_n C^{(n)}_\Delta x^n, \;\;\;\; g_A^{\Sigma^*} = \frac{2}{3} H + \sum_n C^{(n)}_{\Sigma^*} x^n, \;\;\;\; g_A^{\Xi^*} = \frac{1}{3} H + \sum_n C^{(n)}_{\Xi^*} x^n \;.
\end{equation} 
Using the above relations we can construct the following 3 expressions

\begin{eqnarray}  
\delta_A^{SU(3)} &=& \sum_n c^{\prime\prime}_n x^n \nonumber\\
&=&  g_A^\Delta - \frac{3}{2} g_A^{\Sigma^*} \label{eq:su3_dec_1}\\
&=&  g_A^\Delta - 3 g_A^{\Xi^*}              \label{eq:su3_dec_2}\\ 
&=& g_A^\Delta - g_A^{\Sigma^*} - g_A^{\Xi^*}\;,  \label{eq:su3_dec_3}
\end{eqnarray}
which hold at the SU(3) limit. Another expression involving the $\lambda_8$ couplings of the decuplet can be inferred by our results, which reads
\be\label{eq:su3_dec_eta8}
\delta_8^{SU(3)} = g_8^\Delta - 2g_8^{\Sigma^*} + g_8^{\Xi^*}\;.
\ee
In \fig{fig:su3_flavour_breaking_decuplet} we plot the SU(3) breaking for the decuplet. As one can see, the breaking effects for the decuplet are consistent with zero across the range of $x$ for all three expressions involving $g_A$, as well as for~\eq{eq:su3_dec_eta8} involving the $\lambda_8$ coupling of the decuplet baryons, which is an interesting result.

Given the large charm quark mass SU(3) symmetry is not expected to be well satisfied for charmed baryons.
Interchanging the strange with the charm quark in \eq{eq:su3_breaking} one finds
\be
g_A^N-g_A^{\Sigma_c}+g_A^{\Xi_{cc}}=0 
\label{charm SU3}
\ee
As expected the breaking in this case is larger and we find that this relation is broken by $(36.6\pm 3.3)$\% at the physical point.


\section{Conclusions}
In this work, we present the calculation of all  axial charges of the
nucleon, $\Delta$, the hyperons and charmed baryons. The complete set of results are given in Appendices B and C.  We consider the axial vector currents with flavor combinations corresponding to three diagonal generators of SU(4) for which  disconnected contributions vanish in the mass symmetric case. In addition, we consider the isoscalar combination neglecting the disconnected contributions, which are smaller compared to the connected ones. Having these four combinations one can extract all four quark axial couplings $g_A^q$ of all forty particles. Comparing results of the B-ensembles with those of the D-ensemble with smaller lattice spacing we found no detectable cut-off effects. Agreement of results among isospin doublets also corroborates that for these lattice spacings 
finite lattice spacing effects are small. This enables us to use all the lattice QCD data 
to make an extrapolation to the physical value of the pion mass. We have found that a linear fit in terms of $m_\pi^2$ describes well most of our data for the axial charge of hyperons and charmed baryons allowing us to provide estimates of the
axial charges at the physical point.
  
Having all the  axial couplings for a range of pion masses  we are able to check SU(3) breaking effects. We found that the pion axial couplings for the octet baryons exhibit a breaking of $(13.6\pm 2.4)$\% at the physical point while for the $\eta_8$ couplings this increases to $(26.8\pm 3.8)$. In the decuplet, on the other hand, the isospin splitting is found to be consistent with zero within our current statistics. For singly charmed baryons one can examine
similar relations by replacing the strange by the charm quark. As expected
a larger  SU(3) breaking is exhibited.

\section{ Acknowledgments}

This work was supported by a grant from the Swiss National Supercomputing Centre (CSCS) under project ID s540 and in addition used computational resources  from
the John von Neumann-Institute for Computing on the JUROPA system and
the BlueGene/Q system Juqueen at the research center in J\"ulich, and
Cy-Tera at the Cyprus Institute. 
We
thank the staff members at all sites for their kind and sustained
support. This work is supported in part by funding
received from the Cyprus Research Promotion Foundation under contracts
NEA Y$\Pi$O$\Delta$OMH/$\Sigma$TPATH/0308/31) co-financed by the
European Regional Development Fund. K.H. and Ch. K. acknowledge
support from the Cyprus Research Promotion Foundation under contract
T$\Pi$E/$\Pi\Lambda$HPO/0311(BIE)/09. 
\clearpage


\clearpage
\bibliographystyle{unsrtnat}
\bibliography{refs}

\clearpage
\appendix
\begin{center}
{\bf Appendix A: Interpolating fields for baryons}\label{app:int_fields}
\end{center}

In the following tables we give the interpolating fields for baryons used in this work.
The sorting is in correspondence with \fig{fig:spin12_32}.
Throughout, $C$ denotes the charge conjugation matrix and the transposition sign refers to spinor indices which are suppressed.

%
%
\begin{table}[!ht]
\begin{center}
\renewcommand{\arraystretch}{1.2}
\renewcommand{\tabcolsep}{5.5pt}
\makebox[\textwidth]{%
\begin{tabular}{c|c|c c c c c}
\hline
\hline
\multirow{2}{*}{Charm} & \multirow{2}{*}{Strange} &  \multirow{2}{*}{Baryon} & Quark & \multirow{2}{*}{Interpolating field}  & \multirow{2}{*}{$I$} & 	\multirow{2}{*}{$I_z$}	 \\
	                     &                         &                          & content  &                                     &                      &           \\
\hline
\renewcommand{\arraystretch}{1.6}
\renewcommand{\tabcolsep}{5.5pt} 
\multirow{3}{*}{$c=2$}&	  \multirow{2}{*}{$s=0$}		&$\Xi_{cc}^{++}$        & ucc &   $    \cone{c}{u}{c} $ & 1/2 & +1/2 \\
		    		&	                     &$\Xi_{cc}^{+}$        &  dcc & $    \cone{c}{d}{c} $ & 1/2 & -1/2 \\ 
\cline{2-7}
		         &$s=1$			&$\Omega_{cc}^{+}$        & scc &   $    \cone{c}{s}{c} $  & 0 & 0\\ 
\hline\hline
\multirow{9}{*}{$c=1$}&	\multirow{3}{*}{$s=0$}	 	&$\Sigma_c^{++}$  &  uuc & $    \cone{u}{c}{u} $ & 1 & +1 \\
		&	  	&$\Sigma_c^{+}$   & udc & $   \reci{\sqrt{2}} \eps \left[ \con{u}{c}{d} + \con{d}{c}{u} \right] $ & 1 & 0 \\
		    	&	     	&$\Sigma_c^{0}$  &   ddc & $    \cone{d}{c}{d} $ & 1 & -1 \\						
\cline{2-7}		 
	& \multirow{2}{*}{$s=1$}	 &$\Xi_c^{\prime +}$  &  usc & $   \reci{\sqrt{2}} \eps \left[ \con{u}{c}{s} + \con{s}{c}{u} \right] $ & 1/2 & +1/2 \\				
					 &     		&$\Xi_c^{\prime 0}$  &   dsc &$   \reci{\sqrt{2}} \eps \left[ \con{d}{c}{s} + \con{s}{c}{d} \right] $ & 1/2 & -1/2 \\
\cline{2-7}
		  & $s=2$	&$\Omega_c^{0}$        & ssc &   $    \cone{s}{c}{s} $ & 0 & 0\\
\cline{2-7}
 & $s=0$ & $\Lambda_c^+$  &	udc	 & $   \reci{\sqrt{6}} \eps \left[ 2\con{u}{d}{c} + \con{u}{c}{d} - \con{d}{c}{u} \right] $   & 0 & 0   \\
\cline{2-7}
& \multirow{2}{*}{$s=1$}	 	&$\Xi_c^+$      & usc &	$   \reci{\sqrt{6}} \eps \left[ 2\con{s}{u}{c} + \con{s}{c}{u} - \con{u}{c}{s} \right] $ & 1/2 & +1/2 \\
 &	  &$\Xi_c^0$   & dsc & $   \reci{\sqrt{6}} \eps \left[ 2\con{s}{d}{c} + \con{s}{c}{d} - \con{d}{c}{s} \right] $ & 1/2 & -1/2 \\
\hline\hline
\multirow{8}{*}{$c=0$}&	\multirow{2}{*}{$s=0$}	   & \emph{p}     &	uud	      &    $\cone{u}{d}{u} $      & 1/2 & +1/2     \\
	&	 	& \emph{n}     & udd	   & $   \cone{d}{u}{d} $      & 1/2 & -1/2      \\
\cline{2-7}	 
& \multirow{4}{*}{$s=1$}		&$\Lambda$     & uds     &	$   \reci{\sqrt{6}} \eps \left[ 2\con{u}{d}{s} + \con{u}{s}{d} - \con{d}{s}{u} \right] $ & 0 & 0 \\
\cline{3-7}
	   	&	&$\Sigma^{+}$  &  uus    & $    \cone{u}{s}{u} $     & 1 & +1\\
      & 	&$\Sigma^{0}$  & uds     & $   \reci{\sqrt{2}} \eps \left[ \con{u}{s}{d} + \con{d}{s}{u} \right] $& 1 & 0 \\
	   &   &$\Sigma^{-}$  &   dds   & $    \cone{d}{s}{d} $     & 1 & -1\\
\cline{2-7}
& \multirow{2}{*}{$s=2$}    &$\Xi^{0}$      & uss    &   $    \cone{s}{u}{s} $    & 1/2 & +1/2 \\
		&   &$\Xi^{-}$     &  dss   & $    \cone{s}{d}{s} $       & 1/2 & -1/2\\
\hline\hline
\end{tabular}
}
\end{center}
\caption{Interpolating fields and quantum numbers for the $20^\prime$-plet of spin-1/2 baryons.}
\label{spin12_tab}
\end{table}
%
\clearpage
\begin{table}[!ht]
\begin{center}
\renewcommand{\arraystretch}{1.2}
\renewcommand{\tabcolsep}{5.5pt}
\makebox[\textwidth]{%
\begin{tabular}{c|c|c c c c c}
\hline
\hline
\multirow{2}{*}{Charm} & \multirow{2}{*}{Strange} &  \multirow{2}{*}{Baryon} & Quark & \multirow{2}{*}{Interpolating field}  & \multirow{2}{*}{$I$} & 	\multirow{2}{*}{$I_z$}	 \\
	                     &                         &                          & content  &                                     &                      &           \\
\hline
\renewcommand{\arraystretch}{1.6}
\renewcommand{\tabcolsep}{5.5pt}
	$c=3$ &	 $s=0$	&$\Omega_{ccc}^{++}$        & ccc &   $     \conme{c}{c}{c} $ & 0 & 0 \\
\hline\hline
 \multirow{3}{*}{$c=2$}    & \multirow{2}{*}{$s=0$}	&$\Xi_{cc}^{\star ++}$        & ucc &  $   \reci{\sqrt{3}} \eps \left[ 2\conm{c}{u}{c} + \conm{c}{c}{u} \right] $ & 1/2 & +1/2\\
		    		&           	&$\Xi_{cc}^{\star +}$        &  dcc & $   \reci{\sqrt{3}} \eps \left[ 2\conm{c}{d}{c} + \conm{c}{c}{d} \right] $ & 1/2 & -1/2\\ 
\cline{2-7}
		  		&$s=1$ 	&$\Omega_{cc}^{\star +}$        & scc &   $   \reci{\sqrt{3}} \eps \left[ 2\conm{c}{s}{c} + \conm{c}{c}{s} \right] $ & 0 & 0\\ 
\hline\hline
\multirow{6}{*}{$c=1$}&	  \multirow{3}{*}{$s=0$}	&$\Sigma_c^{\star ++}$  & uuc &   $   \reci{\sqrt{3}} \eps \left[ \conm{u}{u}{c} + 2\conm{c}{u}{u} \right] $ & 1 & +1 \\
			 &			    &$\Sigma_c^{\star +}$   		   & udc   & $   \sqrt{\frac{2}{3}} \eps \left[ \conm{u}{d}{c} + \conm{d}{c}{u}+ \conm{c}{u}{d} \right] $ & 1 & 0 \\
		    	&		        &$\Sigma_c^{\star 0}$  &   ddc        &   $     \reci{\sqrt{3}} \eps \left[ \conm{d}{d}{c} + 2\conm{c}{d}{d} \right] $ & 1 & -1 \\					
\cline{2-7}								 
       &	  \multirow{2}{*}{$s=1$}	&$\Xi_c^{\star +}$      & usc &	$   \sqrt{\frac{2}{3}} \eps \left[ \conm{u}{s}{c} + \conm{s}{c}{u}+ \conm{c}{u}{s} \right] $  & 1/2 & +1/2\\
		 &		      	&$\Xi_c^{\star 0}$   & dsc & $   \sqrt{\frac{2}{3}} \eps \left[ \conm{d}{s}{c} + \conm{s}{c}{d}+ \conm{c}{d}{s} \right] $  & 1/2 & -1/2\\
\cline{2-7}
		  &$s=2$	&$\Omega_c^{\star 0}$        & ssc & $   \reci{\sqrt{3}} \eps \left[ 2\conm{s}{c}{s} + \conm{s}{s}{c} \right] $ & 0 & 0  \\
\hline\hline
\multirow{10}{*}{$c=0$}&   \multirow{4}{*}{$s=0$}		& $\Delta^{++}$  & uuu & $    \conme{u}{u}{u} $   & 3/2 & +3/2         \\
	&	& $\Delta^{+}  $  &  uud & $   \reci{\sqrt{3}} \eps \left[ 2\conm{u}{d}{u} + \conm{u}{u}{d} \right] $& 3/2 & +1/2         \\
	&	& $\Delta^{0}  $  &  udd & $   \reci{\sqrt{3}} \eps \left[ 2\conm{d}{u}{d} + \conm{d}{d}{u} \right] $ & 3/2 & -1/2        \\						
 	&	& $\Delta^{-}$  & 	ddd & $    \conme{d}{d}{d} $   & 3/2 & -3/2         \\						
\cline{2-7}									 
   & \multirow{3}{*}{$s=1$} 		 &$\Sigma^{\star +}$  & uus &   $   \reci{\sqrt{3}} \eps \left[ \conm{u}{u}{s} + 2\conm{s}{u}{u} \right] $ & 1 & +1 \\
    &	&$\Sigma^{\star 0}$   		   & uds   & $   \sqrt{\frac{2}{3}} \eps \left[ \conm{u}{d}{s} + \conm{d}{s}{u}+ \conm{s}{u}{d} \right] $& 1 & 0 \\
	 & &$\Sigma^{\star -}$  &   dds        &   $     \reci{\sqrt{3}} \eps \left[ \conm{d}{d}{s} + 2\conm{s}{d}{d} \right] $ & 1 & -1\\
\cline{2-7}
& \multirow{2}{*}{$s=2$}		&$\Xi^{\star 0}$        & uss &  $   \reci{\sqrt{3}} \eps \left[ 2\conm{s}{u}{s} + \conm{s}{s}{u} \right] $ & 1/2 & +1/2\\
 		&	&$\Xi^{\star -}$        &  dss & $   \reci{\sqrt{3}} \eps \left[ 2\conm{s}{d}{s} + \conm{s}{s}{d} \right] $ & 1/2 & -1/2\\
\cline{2-7}
& $s=3$		& $\Omega^-$  & sss &$   \conme{s}{s}{s}$ &0&0 \\
\hline\hline
\end{tabular}
}
\end{center}
\caption{Interpolating fields and quantum numbers for the 20-plet of spin-3/2 baryons.}
\label{spin32_tab}
\end{table}


\clearpage
\appendix
\begin{center}
{\bf Appendix B: Lattice results on the axial couplings}\label{app:results}
\end{center}

Here we present the lattice results on the axial couplings for all the baryons considered in this work. All errors given in the tables are jack-knife errors. Although the  individual flavour components can be deduced from the Tables~(\ref{Table:axial_IV}-\ref{Table:axial_IS}) we also tabulate the $ \bar{q}_f\gamma_\mu\gamma_5 q_f$ components of the current for baryons that contain at least one valence $q_f$-quark. This ensures that the correct statistical errors for these components are listed. The individual $q_f$ components as well as the $\lambda_8$, $\lambda_{15}$ and isovector combinations which are purely disconnected for a given baryon are excluded from the tables.

\begin{table}[!ht]
\renewcommand*{\arraystretch}{1.4}
	\begin{center}
		\begin{tabular}{l|c|c|c|c|c}
Baryon & D15.48 & B25.32 & B35.32 & B55.32 & B75.32 \\
		\hline\hline
			                   $N$       &   1.1442(349)       &   1.1069(467)       &   1.1247(378)       &   1.1279(277)       &   1.1944(210)     \\ 
             $\Lambda$       &   0.0782(157)       &   0.1396(236)       &   0.0892(156)       &   0.0846(141)       &   0.1088(112)     \\ 
            $\Sigma^+$       &   0.7737(321)       &   0.8126(467)       &   0.8134(400)       &   0.8302(248)       &   0.8724(257)     \\ 
            $\Sigma^0$       &   0.0525(242)       &   0.1389(264)       &   0.1006(247)       &   0.1424(171)       &   0.1207(148)     \\ 
            $\Sigma^-$      &   -0.8140(282)      &   -0.7551(379)      &   -0.7862(286)      &   -0.8508(203)      &   -0.8826(179)     \\ 
               $\Xi^0$      &   -0.2384(123)      &   -0.2490(159)      &   -0.2680(161)      &   -0.2508(117)      &   -0.2546(104)     \\ 
               $\Xi^-$       &   0.2522(104)       &   0.2707(116)       &   0.2545(106)       &   0.2427(109)       &   0.2794(115)     \\ 
         $\Delta^{++}$      &   1.9777(1458)      &   1.6956(1897)      &   1.9574(1552)      &   1.7602(1035)       &   1.8520(875)     \\ 
          $\Delta^{+}$       &   0.5181(981)      &   0.5670(1479)       &   0.6374(976)       &   0.5215(639)       &   0.6129(478)     \\ 
          $\Delta^{0}$      &   -0.6499(973)     &   -0.5929(1167)     &   -0.4798(1063)      &   -0.5676(635)      &   -0.5949(489)     \\ 
          $\Delta^{-}$      &   1.7090(1422)      &   1.7322(1718)      &   1.4374(1331)      &   1.5872(1270)       &   1.8108(868)     \\ 
         $\Sigma^{*+}$       &   1.1929(521)       &   1.1462(720)       &   1.2839(636)       &   1.1478(558)       &   1.2228(473)     \\ 
         $\Sigma^{*0}$      &   -0.1367(685)       &   0.0148(542)       &   0.0654(444)      &   -0.0130(323)       &   0.0124(244)     \\ 
         $\Sigma^{*-}$      &   -1.2633(516)      &   -1.0646(661)      &   -1.0423(619)      &   -1.1139(485)      &   -1.1765(450)     \\ 
            $\Xi^{*0}$       &   0.5869(216)       &   0.5785(278)       &   0.6204(256)       &   0.5741(243)       &   0.6059(213)     \\ 
            $\Xi^{*-}$      &   -0.6682(382)      &   -0.5424(303)      &   -0.5459(299)      &   -0.5702(230)      &   -0.5885(223)     \\ 
         $\Lambda_c^+$        &   0.0059(75)       &   0.0330(116)        &   0.0166(86)        &   0.0187(83)        &   0.0192(55)     \\ 
             $\Xi_c^+$       &   -0.0158(45)       &   -0.0148(54)       &   -0.0268(63)       &   -0.0134(49)       &   -0.0112(36)     \\ 
             $\Xi_c^0$        &   0.0205(42)        &   0.0254(44)        &   0.0273(51)        &   0.0260(41)        &   0.0211(34)     \\ 
       $\Sigma_c^{++}$       &   0.7344(261)       &   0.7892(306)       &   0.7769(309)       &   0.7613(215)       &   0.8276(210)     \\ 
        $\Sigma_c^{+}$      &   -0.0161(204)       &   0.0404(236)       &   0.0472(207)       &   0.0408(162)       &   0.0065(115)     \\ 
        $\Sigma_c^{0}$      &   -0.7695(263)      &   -0.6563(344)      &   -0.7101(279)      &   -0.7787(201)      &   -0.8449(191)     \\ 
     $\Xi_c^{\prime+}$       &   0.3500(116)       &   0.3778(127)       &   0.3733(125)        &   0.3806(95)        &   0.4016(95)     \\ 
     $\Xi_c^{\prime0}$      &   -0.3764(122)      &   -0.3373(130)      &   -0.3351(113)      &   -0.3603(100)       &   -0.3996(89)     \\ 
      $\Sigma_c^{*++}$       &   1.1285(424)       &   1.1981(503)       &   1.1473(529)       &   1.1376(403)       &   1.2161(360)     \\ 
       $\Sigma_c^{*+}$      &   -0.0460(343)       &   0.0311(377)       &   0.0448(320)       &   0.0176(245)      &   -0.0355(175)     \\ 
       $\Sigma_c^{*0}$      &   -1.1596(423)      &   -1.0559(565)      &   -0.9528(448)      &   -1.1561(369)      &   -1.1823(317)     \\ 
          $\Xi_c^{*+}$       &   0.5499(207)       &   0.5810(209)       &   0.5574(214)       &   0.5745(168)       &   0.5925(167)     \\ 
          $\Xi_c^{*0}$      &   -0.5985(226)      &   -0.5467(243)      &   -0.5210(215)      &   -0.5672(178)      &   -0.6108(161)     \\ 
       $\Xi_{cc}^{++}$       &   -0.1853(70)       &   -0.2016(84)       &   -0.2034(78)       &   -0.2104(70)       &   -0.1972(66)     \\ 
        $\Xi_{cc}^{+}$        &   0.1895(67)        &   0.1983(84)        &   0.1928(74)        &   0.2126(73)        &   0.2062(57)     \\ 
      $\Xi_{cc}^{*++}$       &   0.5435(211)       &   0.5864(215)       &   0.5525(196)       &   0.5702(157)       &   0.5833(148)     \\ 
       $\Xi_{cc}^{*+}$      &   -0.5405(199)      &   -0.5293(236)      &   -0.5149(187)      &   -0.5702(185)      &   -0.5844(141)     \\ 

			\hline
		\end{tabular}
	\end{center}
	\caption{The isovector combination for all ensembles considered in this work. Baryons for which this combination is purely disconnected are not included in the table.}
	\label{Table:axial_IV}
\end{table}

\begin{table}[!ht]
\renewcommand*{\arraystretch}{1.4}
	\begin{center}
		\begin{tabular}{l|c|c|c|c|c}
Baryon & D15.48 & B25.32 & B35.32 & B55.32 & B75.32 \\
		\hline\hline
			                   $N$       &   0.5056(387)       &   0.5366(484)       &   0.5588(284)       &   0.5802(236)       &   0.5980(195)     \\ 
             $\Lambda$      &   -1.5117(289)      &   -1.5410(277)      &   -1.4811(270)      &   -1.4792(230)      &   -1.5011(201)     \\ 
            $\Sigma^+$       &   1.2945(437)       &   1.3605(590)       &   1.3423(576)       &   1.3435(346)       &   1.4020(299)     \\ 
            $\Sigma^0$       &   1.3478(285)       &   1.3016(462)       &   1.2825(362)       &   1.3840(289)       &   1.4078(232)     \\ 
            $\Sigma^-$       &   1.3372(366)       &   1.3004(533)       &   1.3544(418)       &   1.4050(310)       &   1.4665(300)     \\ 
               $\Xi^0$      &   -2.1024(270)      &   -2.0865(447)      &   -2.1251(412)      &   -2.1124(345)      &   -2.1144(354)     \\ 
               $\Xi^-$      &   -2.1283(264)      &   -2.1373(269)      &   -2.0933(338)      &   -2.0760(312)      &   -2.1419(268)     \\ 
         $\Delta^{++}$      &   1.9777(1458)      &   1.6956(1897)      &   1.9574(1552)      &   1.7602(1035)       &   1.8520(875)     \\ 
          $\Delta^{+}$      &   1.9793(1120)      &   1.6860(1388)      &   1.6936(1209)       &   1.6694(854)       &   1.7720(700)     \\ 
          $\Delta^{0}$       &   1.8431(985)      &   1.5873(1292)      &   1.5251(1304)      &   1.4883(1394)       &   1.7566(724)     \\ 
          $\Delta^{-}$     &   -1.7090(1422)     &   -1.7322(1718)     &   -1.4374(1331)     &   -1.5872(1270)      &   -1.8108(868)     \\ 
         $\Sigma^{*+}$      &   -0.2109(437)      &   -0.1557(628)      &   -0.1013(485)      &   -0.1497(357)      &   -0.0886(220)     \\ 
         $\Sigma^{*0}$      &   -0.1119(481)      &   -0.1982(528)      &   -0.1573(467)      &   -0.1137(321)      &   -0.1113(247)     \\ 
         $\Sigma^{*-}$      &   -0.1211(614)      &   -0.2118(663)      &   -0.2314(716)      &   -0.0663(504)      &   -0.1107(432)     \\ 
            $\Xi^{*0}$      &   -2.1563(464)      &   -2.0341(617)      &   -2.0968(605)      &   -1.9758(554)      &   -2.0025(515)     \\ 
            $\Xi^{*-}$      &   -2.1304(537)      &   -2.0179(601)      &   -2.0973(646)      &   -1.8916(735)      &   -2.0101(585)     \\ 
            $\Omega^-$      &   -4.0731(606)      &   -3.9212(752)      &   -4.0431(883)      &   -3.8087(877)      &   -3.9125(915)     \\ 
         $\Lambda_c^+$       &   -0.0293(88)      &   -0.0221(114)       &   -0.0238(93)       &   -0.0231(75)       &   -0.0152(67)     \\ 
             $\Xi_c^+$       &   -0.0008(62)       &   -0.0040(79)       &   -0.0225(73)       &   -0.0076(55)       &   -0.0015(43)     \\ 
             $\Xi_c^0$       &   -0.0215(79)      &   -0.0319(101)       &   -0.0449(93)       &   -0.0392(96)       &   -0.0343(71)     \\ 
       $\Sigma_c^{++}$       &   0.7344(261)       &   0.7892(306)       &   0.7769(309)       &   0.7613(215)       &   0.8276(210)     \\ 
        $\Sigma_c^{+}$       &   0.7837(234)       &   0.7437(288)       &   0.7202(230)       &   0.7638(176)       &   0.8205(164)     \\ 
        $\Sigma_c^{0}$       &   0.7695(263)       &   0.6563(344)       &   0.7101(279)       &   0.7787(201)       &   0.8449(191)     \\ 
     $\Xi_c^{\prime+}$      &   -0.5387(123)      &   -0.5242(136)      &   -0.5045(136)       &   -0.4827(90)       &   -0.4770(81)     \\ 
     $\Xi_c^{\prime0}$      &   -0.5327(165)      &   -0.5727(171)      &   -0.5685(188)      &   -0.5315(156)      &   -0.4926(148)     \\ 
          $\Omega_c^0$      &   -1.6921(306)      &   -1.7293(221)      &   -1.7074(258)      &   -1.7091(240)      &   -1.7189(265)     \\ 
      $\Sigma_c^{*++}$       &   1.1285(424)       &   1.1981(503)       &   1.1473(529)       &   1.1376(403)       &   1.2161(360)     \\ 
       $\Sigma_c^{*+}$       &   1.1856(444)       &   1.1043(462)       &   1.0383(358)       &   1.1112(316)       &   1.2014(292)     \\ 
       $\Sigma_c^{*0}$       &   1.1596(423)       &   1.0559(565)       &   0.9528(448)       &   1.1561(369)       &   1.1823(317)     \\ 
          $\Xi_c^{*+}$      &   -0.7501(224)      &   -0.7658(320)      &   -0.6976(226)      &   -0.6748(165)      &   -0.6925(134)     \\ 
          $\Xi_c^{*0}$      &   -0.7455(288)      &   -0.7199(324)      &   -0.7143(399)      &   -0.6964(257)      &   -0.6548(233)     \\ 
       $\Omega_c^{*0}$      &   -2.5198(589)      &   -2.6147(360)      &   -2.5606(466)      &   -2.5603(431)      &   -2.5739(451)     \\ 
       $\Xi_{cc}^{++}$       &   -0.1853(70)       &   -0.2016(84)       &   -0.2034(78)       &   -0.2104(70)       &   -0.1972(66)     \\ 
        $\Xi_{cc}^{+}$       &   -0.1895(67)       &   -0.1983(84)       &   -0.1928(74)       &   -0.2126(73)       &   -0.2062(57)     \\ 
       $\Omega_{cc}^+$        &   0.4306(72)        &   0.4476(79)        &   0.4397(84)        &   0.4420(86)        &   0.4281(96)     \\ 
      $\Xi_{cc}^{*++}$       &   0.5435(211)       &   0.5864(215)       &   0.5525(196)       &   0.5702(157)       &   0.5833(148)     \\ 
       $\Xi_{cc}^{*+}$       &   0.5405(199)       &   0.5293(236)       &   0.5149(187)       &   0.5702(185)       &   0.5844(141)     \\ 
    $\Omega_{cc}^{*+}$      &   -1.2739(155)      &   -1.2827(164)      &   -1.2365(212)      &   -1.2767(203)      &   -1.2570(193)     \\ 

			\hline
		\end{tabular}
	\end{center}
	\caption{The $\lambda_8$ combination for all ensembles considered in this work. This combination is purely disconnected only for the triply charmed $\Omega_{ccc}^{++}$.}
	\label{Table:axial_eta8}
\end{table}

\begin{table}[!ht]
\renewcommand*{\arraystretch}{1.4}
	\begin{center}
		\begin{tabular}{l|c|c|c|c|c}
Baryon & D15.48 & B25.32 & B35.32 & B55.32 & B75.32 \\
		\hline\hline
			                   $N$       &   0.5056(387)       &   0.5366(484)       &   0.5588(284)       &   0.5802(236)       &   0.5980(195)     \\ 
             $\Lambda$       &   0.6422(200)       &   0.6341(246)       &   0.6380(177)       &   0.6414(186)       &   0.6467(180)     \\ 
            $\Sigma^+$       &   0.5112(359)       &   0.5364(535)       &   0.5499(414)       &   0.5722(301)       &   0.6046(291)     \\ 
            $\Sigma^0$       &   0.5372(273)       &   0.5143(304)       &   0.4967(264)       &   0.5779(233)       &   0.6029(213)     \\ 
            $\Sigma^-$       &   0.5533(322)       &   0.4866(370)       &   0.5028(295)       &   0.5730(215)       &   0.5900(192)     \\ 
               $\Xi^0$       &   0.6903(190)       &   0.6729(235)       &   0.6607(218)       &   0.6791(200)       &   0.6741(198)     \\ 
               $\Xi^-$       &   0.6758(199)       &   0.6308(326)       &   0.6660(144)       &   0.6720(162)       &   0.6503(224)     \\ 
         $\Delta^{++}$      &   1.9777(1458)      &   1.6956(1897)      &   1.9574(1552)      &   1.7602(1035)       &   1.8520(875)     \\ 
          $\Delta^{+}$      &   1.9793(1120)      &   1.6860(1388)      &   1.6936(1209)       &   1.6694(854)       &   1.7720(700)     \\ 
          $\Delta^{0}$       &   1.8431(985)      &   1.5873(1292)      &   1.5251(1304)      &   1.4883(1394)       &   1.7566(724)     \\ 
          $\Delta^{-}$     &   -1.7090(1422)     &   -1.7322(1718)     &   -1.4374(1331)     &   -1.5872(1270)      &   -1.8108(868)     \\ 
         $\Sigma^{*+}$       &   1.8981(662)       &   1.7977(892)       &   1.9777(831)       &   1.7980(759)       &   1.8785(665)     \\ 
         $\Sigma^{*0}$       &   1.9754(624)       &   1.7536(691)       &   1.7624(669)       &   1.7243(612)       &   1.8342(571)     \\ 
         $\Sigma^{*-}$       &   1.9556(612)       &   1.7090(790)       &   1.6782(736)       &   1.6640(815)       &   1.8205(613)     \\ 
            $\Xi^{*0}$       &   1.9558(429)       &   1.8849(530)       &   1.9802(562)       &   1.8454(576)       &   1.9113(537)     \\ 
            $\Xi^{*-}$       &   2.0431(439)       &   1.8230(490)       &   1.8692(502)       &   1.8045(483)       &   1.8852(492)     \\ 
            $\Omega^-$       &   2.0365(303)       &   1.9606(376)       &   2.0215(441)       &   1.9044(439)       &   1.9562(457)     \\ 
         $\Lambda_c^+$      &   -2.7802(185)      &   -2.7781(245)      &   -2.7871(210)      &   -2.8369(211)      &   -2.8168(184)     \\ 
             $\Xi_c^+$      &   -2.7747(139)      &   -2.7851(225)      &   -2.7965(223)      &   -2.8216(222)      &   -2.8030(174)     \\ 
             $\Xi_c^0$      &   -2.7565(124)      &   -2.7833(175)      &   -2.7816(168)      &   -2.8183(181)      &   -2.8117(163)     \\ 
       $\Sigma_c^{++}$       &   1.6358(578)       &   1.7597(720)       &   1.7284(673)       &   1.6725(584)       &   1.8153(504)     \\ 
        $\Sigma_c^{+}$       &   1.7262(456)       &   1.7231(695)       &   1.6897(550)       &   1.6787(500)       &   1.7664(416)     \\ 
        $\Sigma_c^{0}$       &   1.6519(513)       &   1.5627(782)       &   1.6470(663)       &   1.6883(497)       &   1.8175(533)     \\ 
     $\Xi_c^{\prime+}$       &   1.7414(309)       &   1.7804(433)       &   1.7579(420)       &   1.7342(405)       &   1.8216(373)     \\ 
     $\Xi_c^{\prime0}$       &   1.7655(281)       &   1.7720(386)       &   1.7271(363)       &   1.7187(336)       &   1.7858(332)     \\ 
          $\Omega_c^0$       &   1.8074(205)       &   1.8165(275)       &   1.7998(297)       &   1.7813(296)       &   1.8299(293)     \\ 
      $\Sigma_c^{*++}$      &   -1.5979(486)      &   -1.5023(595)      &   -1.5675(609)      &   -1.5945(447)      &   -1.5164(415)     \\ 
       $\Sigma_c^{*+}$      &   -1.5389(466)      &   -1.6453(488)      &   -1.6626(432)      &   -1.6326(402)      &   -1.5555(346)     \\ 
       $\Sigma_c^{*0}$      &   -1.5402(500)      &   -1.6804(630)      &   -1.7861(576)      &   -1.5990(480)      &   -1.5669(386)     \\ 
          $\Xi_c^{*+}$      &   -1.5298(315)      &   -1.4556(360)      &   -1.5248(387)      &   -1.5425(347)      &   -1.4775(330)     \\ 
          $\Xi_c^{*0}$      &   -1.4557(297)      &   -1.5496(365)      &   -1.5597(321)      &   -1.5632(340)      &   -1.4870(296)     \\ 
       $\Omega_c^{*0}$      &   -1.4057(175)      &   -1.4233(216)      &   -1.4268(266)      &   -1.4684(267)      &   -1.4279(276)     \\ 
       $\Xi_{cc}^{++}$      &   -3.7969(184)      &   -3.7925(298)      &   -3.8385(268)      &   -3.9084(243)      &   -3.7916(249)     \\ 
        $\Xi_{cc}^{+}$      &   -3.7909(187)      &   -3.7659(281)      &   -3.8358(266)      &   -3.9119(257)      &   -3.8262(238)     \\ 
       $\Omega_{cc}^+$      &   -3.7608(197)      &   -3.7711(321)      &   -3.8175(236)      &   -3.8319(323)      &   -3.7578(365)     \\ 
      $\Xi_{cc}^{*++}$      &   -4.8413(424)      &   -4.7821(450)      &   -4.7476(663)      &   -4.8364(476)      &   -4.7348(544)     \\ 
       $\Xi_{cc}^{*+}$      &   -4.8291(460)      &   -4.8804(468)      &   -4.7705(610)      &   -4.9335(472)      &   -4.6946(728)     \\ 
    $\Omega_{cc}^{*+}$      &   -4.7330(244)      &   -4.7277(323)      &   -4.6907(392)      &   -4.8042(377)      &   -4.7006(381)     \\ 
   $\Omega_{ccc}^{++}$      &   -7.8548(480)      &   -7.8496(740)      &   -7.7605(801)      &   -8.0074(693)      &   -7.8314(666)     \\ 

			\hline
		\end{tabular}
	\end{center}
	\caption{The $\lambda_{15}$ combination for all ensembles considered in this work.}
	\label{Table:axial_eta15}
\end{table}

\begin{table}[!ht]
\renewcommand*{\arraystretch}{1.4}
	\begin{center}
		\begin{tabular}{l|c|c|c|c|c}
Baryon & D15.48 & B25.32 & B35.32 & B55.32 & B75.32 \\
		\hline\hline
			                   $N$       &   0.5056(387)       &   0.5366(484)       &   0.5588(284)       &   0.5802(236)       &   0.5980(195)     \\ 
             $\Lambda$       &   0.6422(200)       &   0.6341(246)       &   0.6380(177)       &   0.6414(186)       &   0.6467(180)     \\ 
            $\Sigma^+$       &   0.5112(359)       &   0.5364(535)       &   0.5499(414)       &   0.5722(301)       &   0.6046(291)     \\ 
            $\Sigma^0$       &   0.5372(273)       &   0.5143(304)       &   0.4967(264)       &   0.5779(233)       &   0.6029(213)     \\ 
            $\Sigma^-$       &   0.5533(322)       &   0.4866(370)       &   0.5028(295)       &   0.5730(215)       &   0.5900(192)     \\ 
               $\Xi^0$       &   0.6903(190)       &   0.6729(235)       &   0.6607(218)       &   0.6791(200)       &   0.6741(198)     \\ 
               $\Xi^-$       &   0.6758(199)       &   0.6308(326)       &   0.6660(144)       &   0.6720(162)       &   0.6503(224)     \\ 
         $\Delta^{++}$      &   1.9777(1458)      &   1.6956(1897)      &   1.9574(1552)      &   1.7602(1035)       &   1.8520(875)     \\ 
          $\Delta^{+}$      &   1.9793(1120)      &   1.6860(1388)      &   1.6936(1209)       &   1.6694(854)       &   1.7720(700)     \\ 
          $\Delta^{0}$       &   1.8431(985)      &   1.5873(1292)      &   1.5251(1304)      &   1.4883(1394)       &   1.7566(724)     \\ 
          $\Delta^{-}$     &   -1.7090(1422)     &   -1.7322(1718)     &   -1.4374(1331)     &   -1.5872(1270)      &   -1.8108(868)     \\ 
         $\Sigma^{*+}$       &   1.8981(662)       &   1.7977(892)       &   1.9777(831)       &   1.7980(759)       &   1.8785(665)     \\ 
         $\Sigma^{*0}$       &   1.9754(624)       &   1.7536(691)       &   1.7624(669)       &   1.7243(612)       &   1.8342(571)     \\ 
         $\Sigma^{*-}$       &   1.9556(612)       &   1.7090(790)       &   1.6782(736)       &   1.6640(815)       &   1.8205(613)     \\ 
            $\Xi^{*0}$       &   1.9558(429)       &   1.8849(530)       &   1.9802(562)       &   1.8454(576)       &   1.9113(537)     \\ 
            $\Xi^{*-}$       &   2.0431(439)       &   1.8230(490)       &   1.8692(502)       &   1.8045(483)       &   1.8852(492)     \\ 
            $\Omega^-$       &   2.0365(303)       &   1.9606(376)       &   2.0215(441)       &   1.9044(439)       &   1.9562(457)     \\ 
         $\Lambda_c^+$        &   0.8883(89)       &   0.8961(140)       &   0.8977(107)        &   0.9156(95)        &   0.9195(90)     \\ 
             $\Xi_c^+$        &   0.8947(70)       &   0.9014(100)       &   0.8947(111)        &   0.9190(94)        &   0.9135(92)     \\ 
             $\Xi_c^0$        &   0.8930(55)        &   0.8986(76)        &   0.9026(90)        &   0.9144(77)        &   0.9178(75)     \\ 
       $\Sigma_c^{++}$       &   0.4367(267)       &   0.4644(321)       &   0.4615(322)       &   0.4577(259)       &   0.4968(233)     \\ 
        $\Sigma_c^{+}$       &   0.4696(243)       &   0.4181(276)       &   0.3975(229)       &   0.4595(196)       &   0.5050(188)     \\ 
        $\Sigma_c^{0}$       &   0.4776(301)       &   0.3649(319)       &   0.4006(293)       &   0.4757(229)       &   0.5211(210)     \\ 
     $\Xi_c^{\prime+}$       &   0.4838(152)       &   0.5121(183)       &   0.4972(186)       &   0.5038(185)       &   0.5130(186)     \\ 
     $\Xi_c^{\prime0}$       &   0.5236(161)       &   0.4692(167)       &   0.4735(158)       &   0.5026(147)       &   0.5319(158)     \\ 
          $\Omega_c^0$       &   0.5349(147)       &   0.5474(112)       &   0.5388(128)       &   0.5454(140)       &   0.5360(150)     \\ 
      $\Sigma_c^{*++}$       &   2.0368(460)       &   2.0984(577)       &   2.0502(612)       &   2.0482(489)       &   2.1239(418)     \\ 
       $\Sigma_c^{*+}$       &   2.0928(477)       &   2.0216(523)       &   1.9377(434)       &   2.0254(398)       &   2.1186(349)     \\ 
       $\Sigma_c^{*0}$       &   2.0594(446)       &   1.9759(629)       &   1.8671(521)       &   2.0731(438)       &   2.0987(378)     \\ 
          $\Xi_c^{*+}$       &   2.1090(307)       &   2.1507(349)       &   2.0829(421)       &   2.1149(352)       &   2.1363(335)     \\ 
          $\Xi_c^{*0}$       &   2.1797(304)       &   2.0980(339)       &   2.0450(368)       &   2.1201(330)       &   2.1533(300)     \\ 
       $\Omega_c^{*0}$       &   2.1582(322)       &   2.2186(232)       &   2.1810(298)       &   2.1967(282)       &   2.1738(380)     \\ 
       $\Xi_{cc}^{++}$        &   1.0178(78)        &   0.9983(99)       &   1.0084(102)        &   1.0218(87)        &   1.0016(80)     \\ 
        $\Xi_{cc}^{+}$        &   1.0102(79)       &   0.9935(101)       &   1.0214(107)        &   1.0205(87)        &   1.0002(81)     \\ 
       $\Omega_{cc}^+$        &   0.9724(51)       &   0.9506(109)        &   0.9790(80)        &   0.9901(83)        &   0.9771(86)     \\ 
      $\Xi_{cc}^{*++}$       &   2.2823(437)       &   2.3750(266)       &   2.3151(357)       &   2.3654(307)       &   2.3473(319)     \\ 
       $\Xi_{cc}^{*+}$       &   2.3268(253)       &   2.3341(280)       &   2.2792(330)       &   2.4044(267)       &   2.3394(346)     \\ 
    $\Omega_{cc}^{*+}$       &   2.4253(136)       &   2.4309(164)       &   2.3882(204)       &   2.4526(198)       &   2.4045(185)     \\ 
   $\Omega_{ccc}^{++}$       &   2.6183(160)       &   2.6165(247)       &   2.5868(267)       &   2.6691(231)       &   2.6105(222)     \\ 

			\hline
		\end{tabular}
	\end{center}
	\caption{The isoscalar combination for all ensembles considered in this work.}
	\label{Table:axial_IS}
\end{table}

\begin{table}[!ht]
\renewcommand*{\arraystretch}{1.4}
	\begin{center}
		\begin{tabular}{l|c|c|c|c|c}
Baryon & D15.48 & B25.32 & B35.32 & B55.32 & B75.32 \\
		\hline\hline
			                   $N$       &   0.8251(310)       &   0.8215(402)       &   0.8409(284)       &   0.8543(200)       &   0.8939(165)     \\ 
             $\Lambda$       &   0.0006(115)       &   0.0248(168)       &   0.0107(108)        &   0.0091(96)        &   0.0195(86)     \\ 
            $\Sigma^+$       &   0.7737(321)       &   0.8126(467)       &   0.8134(400)       &   0.8302(248)       &   0.8724(257)     \\ 
            $\Sigma^0$       &   0.4296(169)       &   0.4589(223)       &   0.4296(197)       &   0.4944(140)       &   0.4963(124)     \\ 
               $\Xi^0$      &   -0.2384(123)      &   -0.2490(159)      &   -0.2680(161)      &   -0.2508(117)      &   -0.2546(104)     \\ 
         $\Delta^{++}$      &   1.9777(1458)      &   1.6956(1897)      &   1.9574(1552)      &   1.7602(1035)       &   1.8520(875)     \\ 
          $\Delta^{+}$       &   1.2473(901)      &   1.1243(1211)       &   1.1703(924)       &   1.0947(633)       &   1.1963(519)     \\ 
          $\Delta^{0}$       &   0.6001(560)       &   0.4982(639)       &   0.5277(691)       &   0.4855(553)       &   0.5820(331)     \\ 
         $\Sigma^{*+}$       &   1.1929(521)       &   1.1462(720)       &   1.2839(636)       &   1.1478(558)       &   1.2228(473)     \\ 
         $\Sigma^{*0}$       &   0.5988(312)       &   0.5580(354)       &   0.5954(314)       &   0.5502(269)       &   0.6030(228)     \\ 
            $\Xi^{*0}$       &   0.5869(216)       &   0.5785(278)       &   0.6204(256)       &   0.5741(243)       &   0.6059(213)     \\ 
         $\Lambda_c^+$       &   -0.0116(57)        &   0.0049(83)       &   -0.0036(64)       &   -0.0019(58)        &   0.0025(46)     \\ 
             $\Xi_c^+$       &   -0.0158(45)       &   -0.0148(54)       &   -0.0268(63)       &   -0.0134(49)       &   -0.0112(36)     \\ 
       $\Sigma_c^{++}$       &   0.7344(261)       &   0.7892(306)       &   0.7769(309)       &   0.7613(215)       &   0.8276(210)     \\ 
        $\Sigma_c^{+}$       &   0.3837(157)       &   0.3926(189)       &   0.3843(159)       &   0.4028(117)       &   0.4130(101)     \\ 
     $\Xi_c^{\prime+}$       &   0.3500(116)       &   0.3778(127)       &   0.3733(125)        &   0.3806(95)        &   0.4016(95)     \\ 
      $\Sigma_c^{*++}$       &   1.1285(424)       &   1.1981(503)       &   1.1473(529)       &   1.1376(403)       &   1.2161(360)     \\ 
       $\Sigma_c^{*+}$       &   0.5719(277)       &   0.5679(290)       &   0.5428(239)       &   0.5642(207)       &   0.5827(174)     \\ 
          $\Xi_c^{*+}$       &   0.5499(207)       &   0.5810(209)       &   0.5574(214)       &   0.5745(168)       &   0.5925(167)     \\ 
       $\Xi_{cc}^{++}$       &   -0.1853(70)       &   -0.2016(84)       &   -0.2034(78)       &   -0.2104(70)       &   -0.1972(66)     \\ 
      $\Xi_{cc}^{*++}$       &   0.5435(211)       &   0.5864(215)       &   0.5525(196)       &   0.5702(157)       &   0.5833(148)     \\ 

			\hline
		\end{tabular}
	\end{center}
	\caption{The component $\bar{u}\gamma_\mu\gamma_5 u$ of the axial current for all ensembles considered in this work. Only baryons with an up quark are included in the table.}
	\label{Table:axial_up}
\end{table}

\begin{table}[!ht]
\renewcommand*{\arraystretch}{1.4}
	\begin{center}
		\begin{tabular}{l|c|c|c|c|c}
Baryon & D15.48 & B25.32 & B35.32 & B55.32 & B75.32 \\
		\hline\hline
			                   $N$      &   -0.3230(209)      &   -0.2886(252)      &   -0.2828(174)      &   -0.2711(123)      &   -0.3005(120)     \\ 
             $\Lambda$      &   -0.0780(125)      &   -0.1146(142)      &   -0.0788(105)      &   -0.0754(107)       &   -0.0893(90)     \\ 
            $\Sigma^0$       &   0.3766(166)       &   0.3181(175)       &   0.3296(154)       &   0.3517(132)       &   0.3755(113)     \\ 
            $\Sigma^-$       &   0.8140(282)       &   0.7551(379)       &   0.7862(286)       &   0.8508(203)       &   0.8826(179)     \\ 
               $\Xi^-$      &   -0.2522(104)      &   -0.2707(116)      &   -0.2545(106)      &   -0.2427(109)      &   -0.2794(115)     \\ 
          $\Delta^{+}$      &   0.8839(1379)       &   0.5677(766)       &   0.5295(593)       &   0.5770(403)       &   0.5800(295)     \\ 
          $\Delta^{0}$       &   1.2459(815)      &   1.0885(1059)       &   1.0035(969)       &   1.0740(597)       &   1.1767(524)     \\ 
          $\Delta^{-}$     &   -1.7090(1422)     &   -1.7322(1718)     &   -1.4374(1331)     &   -1.5872(1270)      &   -1.8108(868)     \\ 
         $\Sigma^{*0}$       &   0.7156(530)       &   0.5451(420)       &   0.5287(370)       &   0.5656(283)       &   0.5854(243)     \\ 
         $\Sigma^{*-}$       &   1.2633(516)       &   1.0646(661)       &   1.0423(619)       &   1.1139(485)       &   1.1765(450)     \\ 
            $\Xi^{*-}$       &   0.6682(382)       &   0.5424(303)       &   0.5459(299)       &   0.5702(230)       &   0.5885(223)     \\ 
         $\Lambda_c^+$       &   -0.0174(58)       &   -0.0276(80)       &   -0.0200(63)       &   -0.0208(54)       &   -0.0172(41)     \\ 
             $\Xi_c^0$       &   -0.0205(42)       &   -0.0254(44)       &   -0.0273(51)       &   -0.0260(41)       &   -0.0211(34)     \\ 
        $\Sigma_c^{+}$       &   0.3998(154)       &   0.3498(182)       &   0.3370(151)       &   0.3609(122)        &   0.4071(99)     \\ 
        $\Sigma_c^{0}$       &   0.7695(263)       &   0.6563(344)       &   0.7101(279)       &   0.7787(201)       &   0.8449(191)     \\ 
     $\Xi_c^{\prime0}$       &   0.3764(122)       &   0.3373(130)       &   0.3351(113)       &   0.3603(100)        &   0.3996(89)     \\ 
       $\Sigma_c^{*+}$       &   0.6153(282)       &   0.5371(307)       &   0.4968(241)       &   0.5480(193)       &   0.6172(166)     \\ 
       $\Sigma_c^{*0}$       &   1.1596(423)       &   1.0559(565)       &   0.9528(448)       &   1.1561(369)       &   1.1823(317)     \\ 
          $\Xi_c^{*0}$       &   0.5985(226)       &   0.5467(243)       &   0.5210(215)       &   0.5672(178)       &   0.6108(161)     \\ 
        $\Xi_{cc}^{+}$       &   -0.1895(67)       &   -0.1983(84)       &   -0.1928(74)       &   -0.2126(73)       &   -0.2062(57)     \\ 
       $\Xi_{cc}^{*+}$       &   0.5405(199)       &   0.5293(236)       &   0.5149(187)       &   0.5702(185)       &   0.5844(141)     \\ 

			\hline
		\end{tabular}
	\end{center}
	\caption{The component $\bar{d}\gamma_\mu\gamma_5 d$ of the axial current for all ensembles considered in this work. Only baryons with a down quark are included in the table.}
	\label{Table:axial_down}
\end{table}

\begin{table}[!ht]
\renewcommand*{\arraystretch}{1.4}
	\begin{center}
		\begin{tabular}{l|c|c|c|c|c}
Baryon & D15.48 & B25.32 & B35.32 & B55.32 & B75.32 \\
		\hline\hline
			             $\Lambda$       &   0.7184(102)       &   0.7251(112)       &   0.7068(110)        &   0.7075(97)        &   0.7159(91)     \\ 
            $\Sigma^+$      &   -0.2616(153)      &   -0.2746(210)      &   -0.2647(178)      &   -0.2567(140)      &   -0.2672(103)     \\ 
            $\Sigma^0$      &   -0.2692(109)      &   -0.2623(144)      &   -0.2620(119)       &   -0.2686(98)       &   -0.2692(83)     \\ 
            $\Sigma^-$      &   -0.2613(131)      &   -0.2730(148)      &   -0.2848(132)      &   -0.2771(104)      &   -0.2925(106)     \\ 
               $\Xi^0$       &   0.9243(170)       &   0.9181(205)       &   0.9282(176)       &   0.9303(162)       &   0.9296(169)     \\ 
               $\Xi^-$       &   0.9390(113)       &   0.9290(155)       &   0.9177(175)       &   0.9171(137)       &   0.9335(127)     \\ 
         $\Sigma^{*+}$       &   0.7040(207)       &   0.6522(267)       &   0.6938(260)       &   0.6510(238)       &   0.6548(209)     \\ 
         $\Sigma^{*0}$       &   0.6977(190)       &   0.6523(225)       &   0.6417(230)       &   0.6158(202)       &   0.6491(191)     \\ 
         $\Sigma^{*-}$       &   0.6930(232)       &   0.6418(252)       &   0.6393(277)       &   0.5614(427)       &   0.6452(235)     \\ 
            $\Xi^{*0}$       &   1.3712(272)       &   1.3065(344)       &   1.3597(365)       &   1.2737(363)       &   1.3047(343)     \\ 
            $\Xi^{*-}$       &   1.3909(276)       &   1.2804(307)       &   1.3214(333)       &   1.2222(399)       &   1.2993(335)     \\ 
            $\Omega^-$       &   2.0365(303)       &   1.9606(376)       &   2.0215(441)       &   1.9044(439)       &   1.9562(457)     \\ 
             $\Xi_c^+$       &   -0.0076(30)       &   -0.0056(43)       &   -0.0014(42)       &   -0.0031(37)       &   -0.0049(34)     \\ 
             $\Xi_c^0$        &   0.0004(31)        &   0.0029(42)        &   0.0087(40)        &   0.0067(41)        &   0.0067(34)     \\ 
     $\Xi_c^{\prime+}$        &   0.4462(59)        &   0.4504(74)        &   0.4386(80)        &   0.4311(69)        &   0.4390(75)     \\ 
     $\Xi_c^{\prime0}$        &   0.4552(66)        &   0.4573(75)        &   0.4520(88)        &   0.4465(69)        &   0.4458(77)     \\ 
          $\Omega_c^0$       &   0.8461(153)       &   0.8646(110)       &   0.8537(129)       &   0.8546(120)       &   0.8594(133)     \\ 
          $\Xi_c^{*+}$       &   0.6506(113)       &   0.6774(232)       &   0.6260(156)       &   0.6252(129)       &   0.6416(127)     \\ 
          $\Xi_c^{*0}$       &   0.6716(114)       &   0.6366(135)       &   0.6241(145)       &   0.6316(134)       &   0.6326(130)     \\ 
       $\Omega_c^{*0}$       &   1.2599(295)       &   1.3073(180)       &   1.2803(233)       &   1.2801(216)       &   1.2870(225)     \\ 
       $\Omega_{cc}^+$       &   -0.2153(36)       &   -0.2238(40)       &   -0.2198(42)       &   -0.2210(43)       &   -0.2141(48)     \\ 
    $\Omega_{cc}^{*+}$        &   0.6370(78)        &   0.6414(82)       &   0.6182(106)       &   0.6383(102)        &   0.6285(97)     \\ 

			\hline
		\end{tabular}
	\end{center}
	\caption{The component $\bar{s}\gamma_\mu\gamma_5 s$ of the axial current for all ensembles considered in this work. Only baryons with a strange quark are included in the table.}
	\label{Table:axial_strange}
\end{table}

\begin{table}[!ht]
\renewcommand*{\arraystretch}{1.4}
	\begin{center}
		\begin{tabular}{l|c|c|c|c|c}
Baryon & D15.48 & B25.32 & B35.32 & B55.32 & B75.32 \\
		\hline\hline
			         $\Lambda_c^+$        &   0.9168(47)        &   0.9179(75)        &   0.9212(60)        &   0.9383(64)        &   0.9339(58)     \\ 
             $\Xi_c^+$        &   0.9171(40)        &   0.9205(66)        &   0.9230(66)        &   0.9352(65)        &   0.9288(56)     \\ 
             $\Xi_c^0$        &   0.9120(36)        &   0.9192(54)        &   0.9207(53)        &   0.9332(57)        &   0.9321(52)     \\ 
       $\Sigma_c^{++}$      &   -0.3000(152)      &   -0.3247(196)      &   -0.3165(177)      &   -0.3037(173)      &   -0.3297(141)     \\ 
        $\Sigma_c^{+}$      &   -0.3146(119)      &   -0.3262(178)      &   -0.3233(144)      &   -0.3049(142)      &   -0.3153(119)     \\ 
        $\Sigma_c^{0}$      &   -0.2941(145)      &   -0.2987(191)      &   -0.3131(178)      &   -0.3033(143)      &   -0.3243(150)     \\ 
     $\Xi_c^{\prime+}$       &   -0.3140(78)      &   -0.3175(115)      &   -0.3151(110)      &   -0.3076(119)      &   -0.3273(105)     \\ 
     $\Xi_c^{\prime0}$       &   -0.3103(74)      &   -0.3250(102)       &   -0.3136(94)       &   -0.3039(96)       &   -0.3135(96)     \\ 
          $\Omega_c^0$       &   -0.3133(51)       &   -0.3172(73)       &   -0.3152(77)       &   -0.3090(86)       &   -0.3234(83)     \\ 
      $\Sigma_c^{*++}$       &   0.9094(113)       &   0.9004(168)       &   0.9039(177)       &   0.9110(148)       &   0.9093(121)     \\ 
       $\Sigma_c^{*+}$        &   0.9087(97)       &   0.9179(132)       &   0.9016(142)       &   0.9156(141)       &   0.9179(109)     \\ 
       $\Sigma_c^{*0}$       &   0.9009(104)       &   0.9129(158)       &   0.9165(170)       &   0.9172(148)       &   0.9162(121)     \\ 
          $\Xi_c^{*+}$        &   0.9111(78)       &   0.9011(105)       &   0.9017(135)       &   0.9146(121)       &   0.8833(185)     \\ 
          $\Xi_c^{*0}$        &   0.9091(75)        &   0.9110(99)       &   0.9011(118)       &   0.9210(120)       &   0.9052(113)     \\ 
       $\Omega_c^{*0}$       &   0.8935(102)        &   0.9104(81)        &   0.9010(99)       &   0.9167(102)       &   0.8903(157)     \\ 
       $\Xi_{cc}^{++}$        &   1.2040(52)        &   1.1970(87)        &   1.2118(81)        &   1.2327(72)        &   1.1985(73)     \\ 
        $\Xi_{cc}^{+}$        &   1.2008(54)        &   1.1902(83)        &   1.2141(84)        &   1.2332(75)        &   1.2067(73)     \\ 
       $\Omega_{cc}^+$        &   1.1831(59)        &   1.1804(98)        &   1.1991(73)        &   1.2049(97)       &   1.1828(111)     \\ 
      $\Xi_{cc}^{*++}$       &   1.7903(135)       &   1.7881(169)       &   1.7655(223)       &   1.7986(197)       &   1.7629(222)     \\ 
       $\Xi_{cc}^{*+}$       &   1.7867(150)       &   1.8046(140)       &   1.7620(202)       &   1.8343(158)       &   1.7595(239)     \\ 
    $\Omega_{cc}^{*+}$        &   1.7896(87)       &   1.7896(115)       &   1.7695(140)       &   1.8142(135)       &   1.7761(133)     \\ 
   $\Omega_{ccc}^{++}$       &   2.6183(160)       &   2.6165(247)       &   2.5868(267)       &   2.6691(231)       &   2.6105(222)     \\ 

			\hline
		\end{tabular}
	\end{center}
	\caption{The component $\bar{c}\gamma_\mu\gamma_5 c$ of the axial current for all ensembles considered in this work. Only baryons with a charm quark are included in the table.}
	\label{Table:axial_charm}
\end{table}


\clearpage
\appendix
\begin{center}
{\bf Appendix C: Axial couplings at the physical pion mass}\label{app:results}
\end{center}

Here we tabulate the values of the axial couplings at the physical pion mass with their jackknife error, estimated from the chiral extrapolations we perform on our lattice data. As already stated, we average over the various isospin partners.

\begin{table}[!ht]
\renewcommand*{\arraystretch}{1.4}
	\begin{center}
		\begin{tabular}{l|c|c|c|c|c|c|c|c}
Baryon & $\bar{u}\gamma_\mu\gamma_5 u$ & $\bar{d}\gamma_\mu\gamma_5 d$ & $\bar{s}\gamma_\mu\gamma_5 s$ & $\bar{c}\gamma_\mu\gamma_5 c$ & $\lambda_3$ & $\lambda_8$ & $\lambda_{15}$ & isoscalar \\
		\hline\hline
			             $\Lambda$       &   0.0035(105)      &   -0.0861(106)        &   0.7185(92)                 &   -       &   0.0851(145)      &   -1.5169(238)       &   0.6361(180)       &   0.6361(180)     \\ 
              $\Sigma$       &   0.7629(218)                 &   -      &   -0.2634(101)                 &   -       &   0.7629(218)       &   1.2885(288)       &   0.4984(244)       &   0.4984(244)     \\ 
                 $\Xi$       &   -0.2479(87)                 &   -       &   0.9266(121)                 &   -       &   -0.2479(87)      &   -2.1092(236)       &   0.6735(162)       &   0.6735(162)     \\ 
          $\Sigma^{*}$       &   1.1740(380)                 &   -       &   0.6852(171)                 &   -       &   1.1740(380)      &   -0.1925(336)       &   1.8616(498)       &   1.8616(498)     \\ 
             $\Xi^{*}$       &   0.5891(198)                 &   -       &   1.3637(245)                 &   -       &   0.5891(198)      &   -2.1321(415)       &   1.9571(379)       &   1.9571(379)     \\ 
            $\Omega^-$                 &   -                 &   -       &   2.0338(310)                 &   -                 &   -      &   -4.0677(620)       &   2.0338(310)       &   2.0338(310)     \\ 
         $\Lambda_c^+$       &   -0.0092(54)       &   -0.0217(54)                 &   -        &   0.9128(49)        &   0.0120(72)       &   -0.0304(80)      &   -2.7699(177)        &   0.8832(89)     \\ 
               $\Xi_c$       &   -0.0217(32)                 &   -       &   -0.0026(26)        &   0.9124(37)       &   -0.0217(32)       &   -0.0189(54)      &   -2.7626(125)        &   0.8901(57)     \\ 
            $\Sigma_c$       &   0.7055(191)                 &   -                 &   -      &   -0.2970(113)       &   0.7055(191)       &   0.7055(191)       &   1.6027(422)       &   0.4094(199)     \\ 
        $\Xi_c^\prime$        &   0.3433(85)                 &   -        &   0.4539(55)       &   -0.3133(69)        &   0.3433(85)       &   -0.5596(99)       &   1.7440(266)       &   0.4872(127)     \\ 
          $\Omega_c^0$                 &   -                 &   -       &   0.8554(117)       &   -0.3125(54)                 &   -      &   -1.7108(233)       &   1.8042(211)       &   0.5428(118)     \\ 
        $\Sigma_c^{*}$       &   1.0899(308)                 &   -                 &   -        &   0.9043(90)       &   1.0899(308)       &   1.0899(308)      &   -1.6170(349)       &   2.0004(346)     \\ 
           $\Xi_c^{*}$       &   0.5466(150)                 &   -       &   0.6587(104)        &   0.9103(75)       &   0.5466(150)      &   -0.7581(183)      &   -1.5124(251)       &   2.1192(254)     \\ 
       $\Omega_c^{*0}$                 &   -                 &   -       &   1.2909(204)        &   0.9026(90)                 &   -      &   -2.5817(408)      &   -1.4060(181)       &   2.1961(261)     \\ 
            $\Xi_{cc}$       &   -0.1912(53)                 &   -                 &   -        &   1.2010(51)       &   -0.1912(53)       &   -0.1912(53)      &   -3.7911(175)        &   1.0112(65)     \\ 
       $\Omega_{cc}^+$                 &   -                 &   -       &   -0.2199(35)        &   1.1840(65)                 &   -        &   0.4398(69)      &   -3.7702(216)        &   0.9681(58)     \\ 
        $\Xi_{cc}^{*}$       &   0.5290(142)                 &   -                 &   -       &   1.7928(127)       &   0.5290(142)       &   0.5290(142)      &   -4.8477(375)       &   2.3176(236)     \\ 
    $\Omega_{cc}^{*+}$                 &   -                 &   -        &   0.6383(74)        &   1.7891(91)                 &   -      &   -1.2765(148)      &   -4.7297(255)       &   2.4265(137)     \\ 
   $\Omega_{ccc}^{++}$                 &   -                 &   -                 &   -       &   2.6141(170)                 &   -                 &   -      &   -7.8423(510)       &   2.6141(170)     \\ 

			\hline
		\end{tabular}
	\end{center}
	\label{Table:extrap_values}
	\caption{The extrapolated values for the axial couplings of hyperons and charmed baryons at the physical pion mass. Purely disconnected contributions are omitted.}
\end{table}

\end{document}